\documentclass[11pt,eqsecnum,tightenlines,floatfix]{revtex4}

\usepackage{fancyhdr}
\usepackage{amsfonts}
\usepackage{amsmath}
\usepackage{amssymb}
\usepackage{subfigure}
\usepackage{graphicx}
\usepackage{bm}
\usepackage{sidecap}
\usepackage{epsfig}
\usepackage[OT1,T1]{fontenc}
\usepackage{setspace}
\usepackage{multirow}
\usepackage{array}
\usepackage{tabulary}
\usepackage{fancyhdr}
\usepackage{wrapfig}
\usepackage{url}
\usepackage[ruled,noline,longend]{algorithm2e}

\SetArgSty{textrm}
\DontPrintSemicolon

\newcommand{\linfnorm}[1]{\left|\left|#1\right|\right|_0}
\newcommand{\us}{|\hspace{-0.04in}\uparrow\rangle}
\newcommand{\ds}{|\hspace{-0.04in}\downarrow\rangle}

\begin{document}

\title{A Near-Term Quantum Computing Approach for
Hard Computational Problems in Space Exploration}

\author{Vadim N. Smelyanskiy}
\thanks{To whom correspondence should be addressed. \\
Electronic address: Vadim.N.Smelyanskiy@nasa.gov}
\author{Eleanor G. Rieffel}
\author{Sergey I. Knysh}
\affiliation{NASA Ames Research Center, Mail Stop 269-3, Moffett Field, CA 94035}
\author{Colin P. Williams}
\affiliation{Jet Propulsion Laboratory, California Institute of Technology,Pasadena, CA 91109-8099}
\author{Mark W. Johnson, Murray C. Thom, William G. Macready}
 \affiliation{D-Wave Systems Inc., 100-4401 Still Creek Drive, Burnaby, BC, Canada V5C 6G9}
\author{Kristen L. Pudenz}\thanks{KLP would like to acknowledge support from the Lockheed Martin Corporation under the URI program and the NSF under a graduate research fellowship.}
 \affiliation{Ming Hsieh Department of Electrical Engineering, Center for Quantum Information Science and Technology, and Information Sciences Institute, University of Southern California, Los Angeles, CA 90089 \vspace{0.3in}}

\begin{abstract}
\begin{center}
{\bf Abstract}
\end{center}
\begingroup
    \fontsize{9pt}{12pt}\selectfont
The future of Space Exploration is entwined with the future of artificial intelligence (AI) and machine learning. Autonomous rovers, unmanned spacecraft, and remote space habitats must all make intelligent decisions with little or no human guidance. The decision-making required of such NASA assets stretches machine intelligence to its limits. Currently, AI problems are tackled using a variety of heuristic approaches, and practitioners are constantly trying to find new and better techniques. To achieve a radical breakthrough in AI, radical new approaches are needed. Quantum computing is one such approach.
 
Many of the hard combinatorial problems in space exploration are instances of NP-complete or NP-hard problems. Neither traditional computers nor quantum computers are expected to be able to solve all instances of such problems efficiently. Many heuristic algorithms, such as simulated annealing, support vector machines, and SAT solvers, have been developed to solve or approximate solutions to practical instances of these problems. The efficacy of these approaches is generally determined by running them on benchmark sets of problem instances. Such empirical testing for quantum algorithms requires the availability of quantum hardware.
 
Quantum annealing machines, analog quantum computational devices, are designed to solve discrete combinatorial optimization problems using properties of quantum adiabatic evolution. We are now on the cusp of being able to run small-scale examples of these problems on actual quantum annealing hardware which will enable us to test empirically the performance of quantum annealing on these problems. For example, D-Wave builds quantum annealing machines based on superconducting qubits. While at present noise and decoherence in quantum annealing devices cannot be easily controlled or corrected, these devices have been shown to display multi-spin  tunneling, a distinct quantum phenomenon at the root of the quantum annealing process. In order to attack an optimization problem on these machines, the problem must be formulated in quadratic unconstrained binary optimization form in which the cost function is strictly quadratic in bit assignments (in physics applications this form is often referred to as an Ising model).  The above limitation is not fundamental: all NP-complete problems can be mapped to this form. However, an optimal mapping involving small or no overhead in terms of additional bits is of significant practical interest because of the limited size of early quantum annealing machines.
 
In this article, we discuss a sampling of the hardest artificial intelligence problems in space exploration in the context in which they emerge. We show how to map them onto equivalent Ising models that then can be attacked using quantum annealing. We review existing quantum annealing results on supervised learning algorithms for classification and clustering and discuss their application to planetary feature identification and satellite image analysis. We present quantum annealing algorithms for unsupervised learning for clustering and discuss its application to anomaly detection in space systems.  We introduce quantum annealing algorithms for data fusion and image matching for remote sensing applications. We overview planning problems for space exploration missions applications and introduce algorithms for planning problems using quantum annealing of Ising models. We describe algorithms for diagnostics and recovery as well as their applications to NASA deep space missions and show how a fault tree analysis problem can be mapped onto an Ising model and solved with quantum annealing. We discuss combinatorial optimization algorithms for task assignment in the context of autonomous unmanned exploration that take into account constraints due to physical limitations of the vehicles. We show how these algorithms can be presented in the framework of Ising model optimization with application to quantum annealing. Finally, we discuss ways to circumvent the need to map practical optimization problems onto the Ising model. We demonstrate how this can be done in principle using a \lq\lq blackbox" approach based on ideas from probabilistic computing.
In particular, we provide initial results on Monte Carlo sampling for solving non-Ising problems.
 
In this article, we describe the architecture, duty cycle times and energy consumption of the D-Wave One quantum annealing machine. We report on benchmark scalability studies of D-Wave One run times and compare to state of the art classical algorithms for solving Ising optimization problems on a uniform random ensemble of problems.
Results on problems in the range of up to 96 qubits show improved scaling for median core quantum annealing time compared with simulated annealing and iterative tabu search, though how it will scale as the number of qubits increases remains an open question. We also review existing results of D-Wave One benchmarking studies for solving binary classification problems with a quantum boosting algorithm. The error rates on synthetic data sets show that quantum boosting algorithm consistently outperforms the AdaBoost classical machine learning algorithm. We review quantum algorithms for structured learning for multi-label classification and describe how the problem of finding an optimal labeling can be mapped onto quantum annealing with Ising models, and then introduce a hybrid classical/quantum approach for learning the weights. We review results of D-Wave One benchmarking studies for learning structured labels on four different data sets.
The first data set is Scene, a standard image benchmark set. The second data set, the RCV1 subset of the Reuters corpus of labeled news stories, has a significantly larger number of labels, and more complex relationships between the labels. The other two are synthetic data sets generated using MAX-3 SAT problem instances. On all four data sets, quantum annealing was compared with an independent Support Vector Machine (SVM) approach with linear kernel and exhibited a better performance.
\endgroup
\end{abstract}

\maketitle

\tableofcontents

\section{Introduction}

The future of Space Exploration is entwined with the future of
artificial intelligence and machine learning.
The unique challenges of deep space exploration, such as
speed of light limited communications delays and human susceptibility to
extended radiation exposure, mean NASA missions must become increasingly
reliant on autonomous systems. Moreover, to ensure astronaut safety and
mission success, spacecraft, life-support systems, astronaut habitats,
and the underlying software that controls them must become highly
reliable. It has proven difficult to achieve human-level artificial
intelligence and autonomy, and equally difficult to assure the
reliability of mission software and hardware. Current approaches to
building intelligent systems use a ``systems of systems'' architecture.
Such systems include computationally intensive problem solving software
that attempts to mimic human intelligent behavior \cite{Krishna2004},
and have demonstrated impressive autonomous capabilities
such as Deep Space One \cite{Bernard1999}.
As successful as current techniques have been for
NASA applications and beyond, many challenges remain.
The obstacles to greater autonomy are the core computational problems
that underlie intelligence, design, verification and validation
that typically require the solution of hard combinatorial optimization
problems. Radical transformations in capability
require radically new approaches.

The majority of difficult problems that arise in real-world applications
are NP-complete \cite{Cook1971,Karp1972}. Likewise, the overwhelming majority
of computational bottlenecks faced in Space Exploration
are either NP-complete or NP-hard.
Neither classical (traditional) computers
nor quantum computers are expected to be able to solve all instances
of such problems efficiently.
In engineering applications, the greatest challenge is to find
approximations that reduce optimization tasks
to computationally tractable ones that stay within given time and
memory resources, often at the expense of drastic reductions in the
quality of the optima that can be found.
Many heuristic algorithms, such as
simulated annealing, support vector machines, and SAT solvers, have
been developed to attack practical instances
of these problems.

Quantum computing \cite{NCbook,RPbook,Williams2011} is a potential
approach to developing radically new ways to attack these hard
combinatorial optimization problems.
Servedio and Gortler \cite{Gortler-01} showed, for example, that
there exist classification tasks that can only be performed
accurately and efficiently on a quantum computer.
While theoretical results such as this one are useful,
practical algorithms are generally evaluated
by running them on benchmark sets of problem instances
and testing them on real world examples.
Such empirical testing for quantum algorithms requires the availability
of quantum hardware.

Advances in quantum hardware mean that empirical testing of one
particular family of quantum algorithms, Quantum Annealing algorithms,
may be possible in the near term.
Theoretical studies and classical simulations suggest that Quantum
Annealing \cite{Brooke1999,Santoro2002,Das-book2005,Das-RMP2008} can
provide dramatic improvements, both in the algorithmic runtime and quality
of the solutions, to many instances of hard optimization problems where
state-of-the-art classical approaches fail. With the advent of Quantum
Annealing hardware \cite{Johnson2011} designed to implement Quantum
Annealing for a general type of combinatorial optimization problem,
the Ising model, we can begin empirical testing of this class of
quantum algorithms. As quantum hardware advances, it will be
possible to evaluate these algorithms on ever larger problems.
One commercial company (D-Wave Systems, Inc.) builds computational devices,
based on superconducting circuits, that are designed to implement
Quantum Annealing algorithms for the Ising model.

\begin{figure}[b]
\begin{center}
\includegraphics[width=0.75\textwidth]{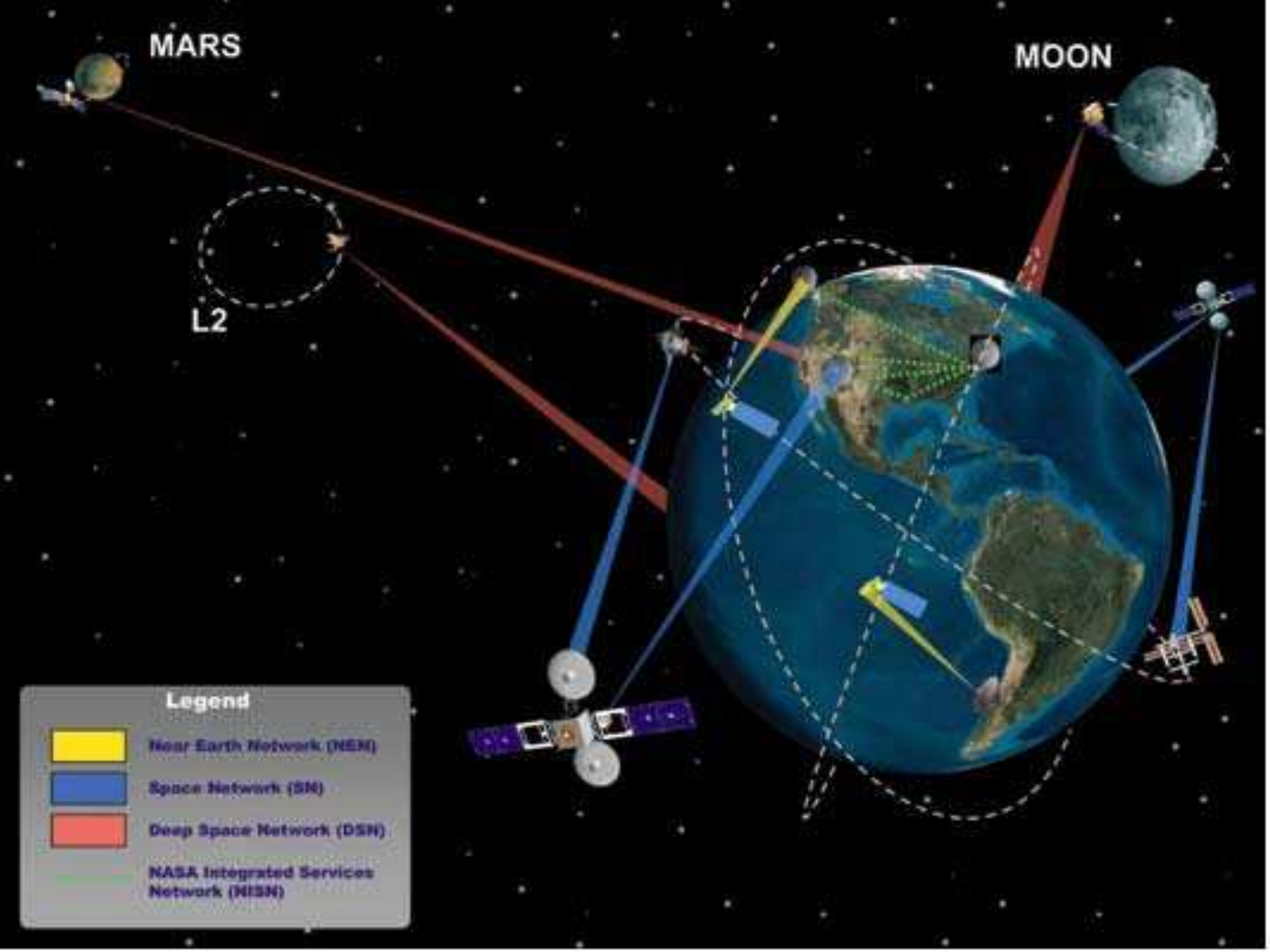}
\end{center}
\caption{
Intelligent data analysis and machine learning techniques are critical
for remote sensing and autonomous space vehicles and systems.
The graphic, from NASA's 2007 Space Communications Plan \cite{SpaceCommPlan},
depicts the three components of the NASA Integrated Services Network,
the Near Earth Network, the Space Network, and the Deep Space Network.
Communication delays and bandwidth are significant issues in all of
these communication networks, particularly the Deep Space Network.
Roundtrip communication between Earth and Mars, for example, takes between
ten and forty minutes. Remote space vehicles and systems must be
able to tackle complex tasks and make quick decisions on their own, with
little or no human guidance.
}
\label{fig:spaceComm}
\end{figure}

In optimization problems, it is routine to construct a cost or
``energy function'' whose global minimum value represents an optimal
solution to the problem. Unfortunately, the energy landscapes induced
by these energy functions are usually replete with local minima
that distract search algorithms away from finding global minima.
The algorithms often become trapped in one of these local minima,
and hence return a sub-optimal solution.
For optimization problems represented by Ising models, the cost (or energy)
is the sum of linear and quadratic functions over binary variables
\begin{equation}
E_{\rm Ising}(s_1,\ldots,s_N)=-\sum_{i=1}^{N}h_i s_i+\sum_{\langle i,j\rangle\in \,\texttt{E}}J_{i,j} s_i s_j,\quad {\rm where} \quad s_i=\pm 1.\label{eq:Ising}
\end{equation}
\noindent
The expression above is written in the customary way for Ising models,
in terms of symmetric binary variables, called Ising spins $s_i\in\{-1,1\}$.
The goal is to find an assignment of Ising spins that provides a
global minimum of the energy. In many conventional formulations, the
cost function (\ref{eq:Ising}) is expressed in terms of the bits
$z_i\in\{0,1\}$, related to Ising spins via $s_i=1-2z_i$.
When written in terms of bits, the cost function (\ref{eq:Ising})
is still the sum of a linear and a quadratic function of its arguments.
The problem of finding a bit assignment that minimizes the cost is usually referred to as the Quadratic Unconstrained Binary Optimization (QUBO) problem, which by the aforementioned transformation  $s_i=1-2z_i$ is equivalent to the Ising  model.

Each Ising spin corresponds to a vertex in a graph, and the connectivity between vertices in (\ref{eq:Ising}) is given by the
matrix $J_{i,j}$. Only when two vertices are connected by an edge
$\langle i,j\rangle$ is the element $J_{i,j}$ nonzero. In
most practical problems, the number of edges scales linearly
with the  number of vertices $N$. In this case, the underlying graph of
interactions is sparse and, in general, non-planar.
The Ising optimization problem is NP-complete \cite{Barahona1982}.
Therefore, any problem from the NP-complete class can be mapped onto an
Ising model defined on the appropriate non-planar graph. This mapping,
or Ising embedding, uses computational resources that scale no faster
than polynomially in the number of Ising spins $N$.
In practical applications, the
difference between linear and nonlinear scaling can be substantial, so
the cost of the mapping itself can be important.
Therefore, computational problems that can be expressed natively
in the Ising model are especially attractive candidates for
Quantum Annealing approaches, as they map directly onto the model
without any overhead.

In this paper, we first give an overview of Quantum Annealing, describe its
implementation on a D-Wave superconducting quantum annealing machine,
and review D-Wave's preliminary benchmarking studies. We follow this
overview with a sampling of hard AI problems from NASA applications.
We show in each case how to phrase the problem as a quadratic binary
optimization problem with cost function $E(z_1, z_2, ... z_N)$ that
can be mapped to an Ising energy function $E_{\rm Ising}(s_1,\ldots,s_N)$
via $s_i = 1-2z_i$ whose lowest energy states can be sought via
Quantum Annealing. These problems and applications are focused around
four major intelligent system domains:
\begin{enumerate}
\item[1.] {\bf Data Analysis and Data Fusion} -- this includes both supervised
and unsupervised machine learning and autonomous analysis to extract information content from data obtained via remote or robotic sensing. These techniques include structural learning, such as multi-label classification of data where there are subtle dependencies between labels; data segmentation and classification; feature identification and matching between different data streams (e.g. images), including data obtained by different physical sensor types (sensor fusion); clustering and pattern recognition.
Such sophisticated data analysis will expand the possibilities for planetary explorations through enhanced precision landing, system navigation, terrain mapping and other related tasks.
\item[2.] {\bf Planning and scheduling} -- this concerns the realization of optimal strategies or action sequences in which various constraints associated with normal operation must be satisfied at all times. Examples in space applications include augmented planning capabilities to support crew autonomy, ISS operations, deep space missions,  autonomous robots and unmanned vehicles.
\item[3.] {\bf Decision making} --  this involves computerized generation of conclusions and decisions, as well as model identification, from available data, using the laws of physics, logical, mathematical, and statistical techniques.
Such autonomous functions are utilized within system health monitoring
hardware and software to detect anomalies and isolate faults, as well
as to develop recovery and avoidance responses. Applications include launch
abort sequences, early warning, and system health management in
complex in-space engineering architectures, such as fuel depots, deep
space missions and habitats.
\item[4.] {\bf Distributed coordination} -- this addresses algorithms for cooperation between unmanned autonomous systems such as unmanned autonomous ground, sea, and aerial vehicles.
We consider a representative example of the type of computational problem that arises in autonomous unmanned exploration: multiple UAVs together determining how to split up tasks and coordinate to achieve a collective mission.
   \end{enumerate}

While not the focus of the present overview, we also note recent work that
exploits the stochastic nature of quantum annealing to develop algorithms
based on sampling probability distributions.
One application of this work is a hybrid classical/quantum approach to problems
for which a mapping onto an Ising model is more difficult.

\section{A High-Level View of Quantum Annealing}
Quantum Annealing (QA) is analogous to a classical algorithm,
Simulated Annealing (SA) \cite{Kirkpatrick:1987}, that attempts to
find a global minimum of an energy function derived from a combinatorial
optimization problem. It mimics the physical process whereby an object is
cooled in a controlled manner so that it freezes just as the object attains
its minimum energy configuration. The choice of cooling schedule is critical:
if the cooling proceeds too fast, it prevents sufficient exploration of
the energy landscape and the search terminates in a sub-optimal energy
configuration corresponding to a local, but not global, energy minimum.
Conversely, if the cooling schedule is too slow, the search process takes
longer than necessary. Intuitively the ``temperature'' controls the
probability the search accepts a move to a higher energy value than
the current configuration. Initially, the temperature is high and the
search hops around accepting moves to configurations that may have higher
or lower energy values than that of the current configuration. As simulated
annealing proceeds and the temperature is lowered, the process tends to
accept only those moves to configurations with lower energy
than that of the current configuration. Eventually, the process
``freezes'' in some configuration of bit values that corresponds to a
local (hopefully global) minimum of the energy function. The key obstacle
faced by Simulated Annealing is that its search is constrained to
the surface of the energy landscape. If the surface happens to possess
tall narrow peaks it can be very difficult for Simulated Annealing to find
even nearby global energy minima.

\begin{figure}[!ht]
\centering
\subfigure[]{
\includegraphics[width=2.8in]{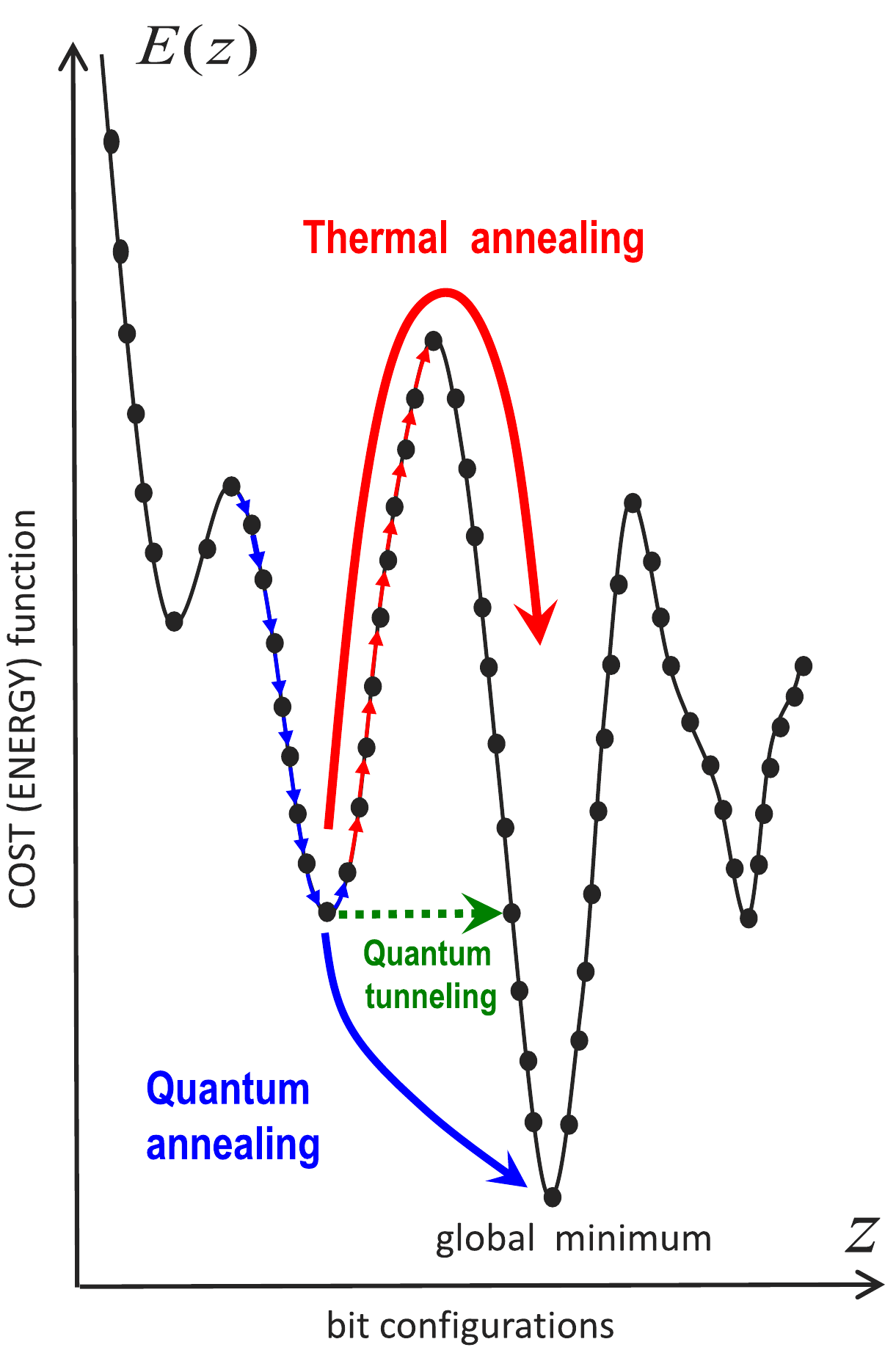}
\label{fig:1a}
}
\subfigure[]{
\includegraphics[width=3.4in]{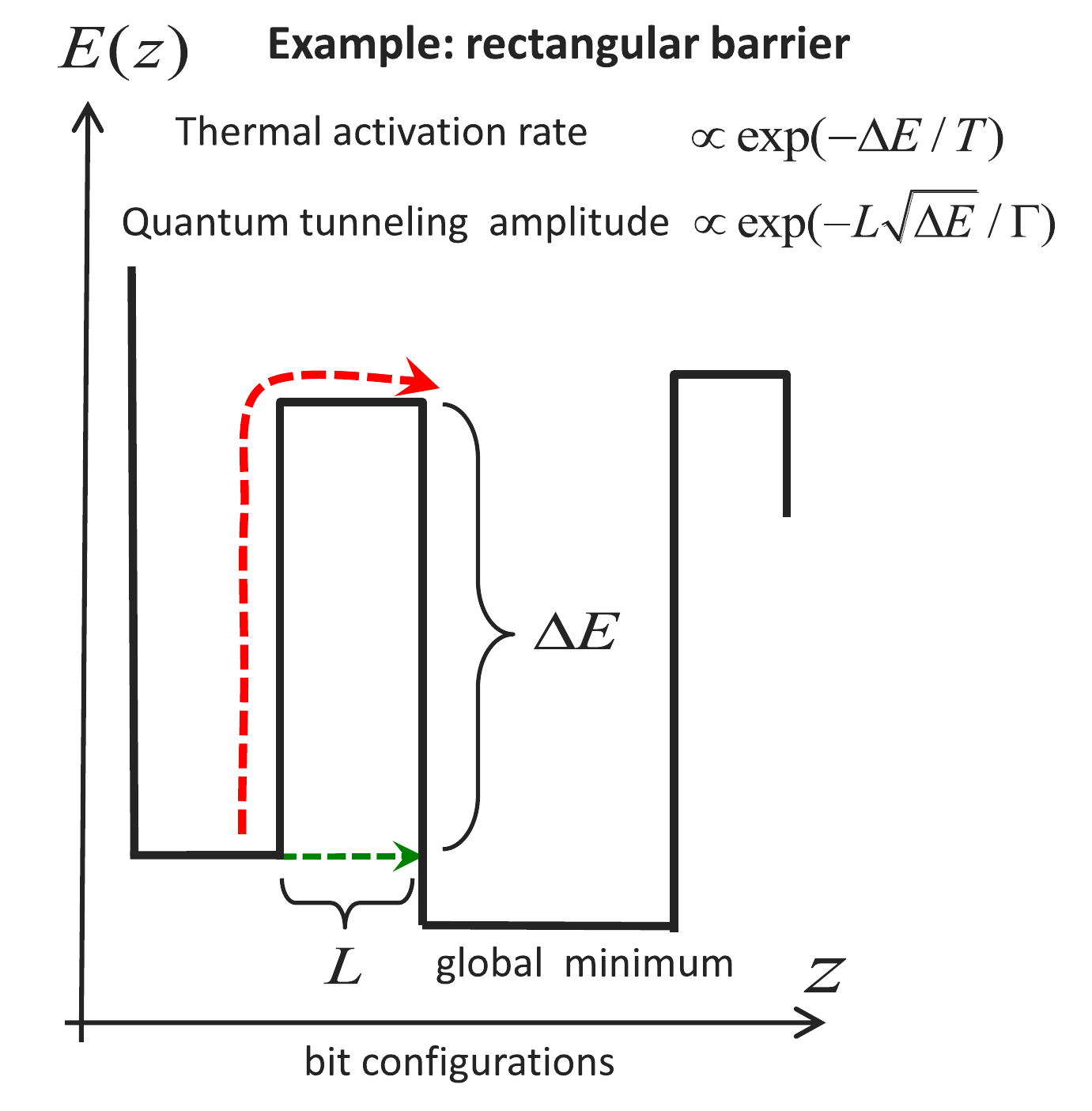}
\label{fig:1b}
}
\label{fig:1}
\caption{ {\footnotesize \subref{fig:1a} Multi-modal energy landscape
of a binary optimization problem. Black points correspond to different
configurations of binary variables and neighboring points correspond
to configurations differing by one bit flip.
Small blue arrows indicate downhill moves followed by greedy optimization
algorithms. In classical stochastic optimization heuristics such as
Simulated  Annealing (SA), in addition to downhill moves, uphill moves
are allowed with probabilities determined by the temperature of the
Gibbs sampler. A quantum computer can implement a Quantum Annealing (QA)
process, illustrated by the thick blue arrow. In this process, the
system can move, not only upward and downward, but also tunnel \emph{through}
barriers, as illustrated by the green dotted line. \subref{fig:1b} An
example of quantum annealing with a rectangular barrier of width $L$ and
height $\Delta E$ \cite{Das2005}. The tunneling rate through the
barrier is $\propto\exp(-L\,\sqrt{\Delta E}/\Gamma)$, where
$\Gamma=\Gamma(t)$ is a QA constant that is gradually reduced to zero
throughout the algorithm. By comparison, the rate of thermal
over-the-barrier activation in SA is $\propto \exp(-\Delta E/T)$,
where $T=T(t)$ is the annealing temperature that is gradually
reduced to zero by the end of SA.}}
\end{figure}

Imagine, instead, that a random walker in an optimization algorithm is now a quantum
object. Unlike a classical particle, a quantum object possesses
wave properties; it can be present in many points of
the search domain simultaneously, and through quantum-mechanical tunneling, can reach
through (not over) energy barriers.
As a result, a quantum object can penetrate
potential barriers of the energy landscape even at low or zero temperature,
as illustrated by the green arrow in Fig. \ref{fig:1a}. It can also explore
multiple transition pathways at the same time (a phenomenon called ``quantum parallelism'').
Edward Farhi and co-workers
\cite{Farhi2001,Farhi2002} first described the ideas of
quantum optimization within the context of quantum adiabatic evolution,
and showed how such an optimization can be implemented, in principle,
on a quantum computer.
This approach to optimization, being a quantum analog of Simulated
Annealing \cite{Farhi2002},
is sometimes called Quantum Annealing \cite{Das-RMP2008}.

Fig.~\ref{fig:1b} illustrates a simple example \cite{Das2005} in
which Quantum Annealing
can perform better than Simulated (Thermal) Annealing.
The tunneling rate through a rectangular potential barrier of height
$\Delta E$ and length $L$ is
$\propto \exp\left(-L \sqrt{\Delta E}/\Gamma\right)$, where $\Gamma$ is
a quantum annealing constant analogous to the temperature in SA.  Unlike
the classical thermal activation rate for SA to cross the barrier,
$\exp(-\Delta E/T)$, quantum tunneling depends not only on the barrier
height but also on its width. We emphasize that the quantum tunneling
rate exponent grows only as $\sqrt{\Delta E}$, whereas the classical
activation rate exponent
displays a faster linear growth with $\Delta E$ . Therefore, QA will have
a tendency to beat SA in landscapes dominated by
\emph{high narrow barriers}, with $L\ll \sqrt{\Delta E}$
for typical values \cite{Farhi2002}.

This result is not surprising and corresponds to a well-known fundamental relation between classical and quantum search efficiency. The spiky energy landscapes are typical for  optimization problems with a lot of noise and little structure. In the extreme, such problems correspond to a search of a completely unsorted domain (finding a needle in a haystack). This task is the most difficult for both classical and quantum algorithms. However, the best possible classical algorithm will search an unstructured data set with $M$ entries in ${\cal O}(M)$ steps, while the best quantum algorithm will do it much faster, in  ${\cal O}(M^{1/2})$ steps \cite{Grover1996}.
Such energy landscapes are not the only scenarios in which QA will outperform classical algorithms. In cases where there is some structure, for example, quantum annealing may be able to exploit this structure better than can classical approaches. An overview of related results is given in Sec.~\ref{sec:mingap}.
More recent ideas that take account of the system noise in QA also show a significant improvement in performance over both SA
and noise-free QA \cite{Amin-noise2011}.

In order to describe quantum annealing more carefully, we first describe
the quantum mechanics that underpins it, notably the quantum adiabatic
theorem, and how cost functions for binary optimization functions can
be be encoded in a Hamiltonian for a quantum system.

\begin{figure}
\centering
\begin{minipage}[t]{0.58\linewidth}
\raisebox{-1cm}{
\includegraphics[width=1\linewidth]{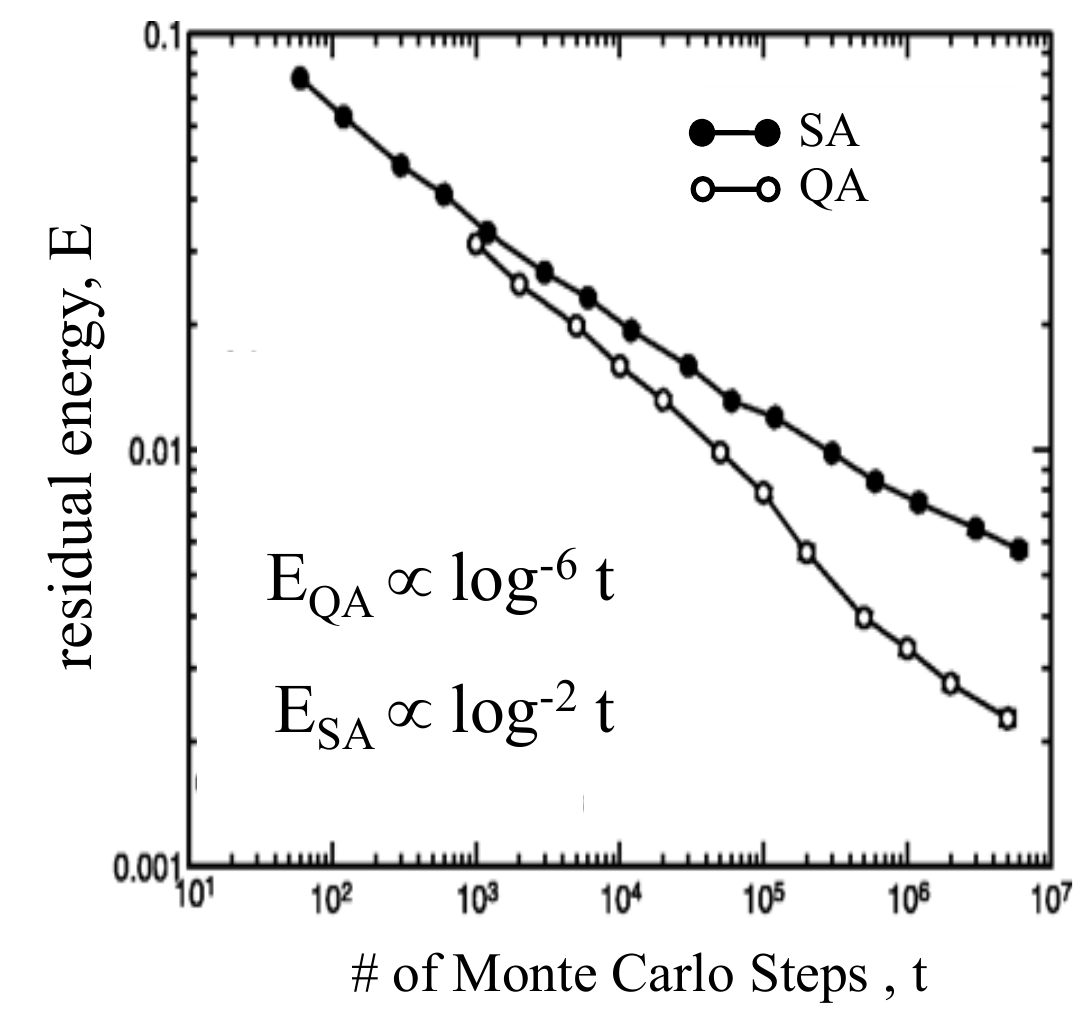}}
\end{minipage}\hfill
\begin{minipage}[b]{0.38\linewidth}
\caption{ \label{fig:2}  {\footnotesize Figure is re-plotted from \cite{Santoro2002}. Empty circles show the energy value per site for an 80x80 disordered 2D Ising model after simulated annealing. Disks show the same for quantum annealing (QA). For fair comparison, the annealing time used in the QA has been rescaled (multiplied by the number of replicas, $P$) so that points corresponding to the same $t$ require the same amount of computer time (number of Monte Carlo steps). For illustration, consider a relative increase of annealing time $t/t_0$ needed to improve the accuracy of a certain annealing, say with $t_0=10^6$, by a factor of 10. In SA, doing so would require $t/t_0$=10$^{13}$. In QA, the same result would be accomplished with $t/t_0$=10$^{2.8}$.}}
\end{minipage}
\end{figure}

\section{Background for Quantum Annealing}

The development of quantum mechanics (QM) in the 1920s and 1930s required
a significant shift in perspective, but the modern presentation of QM is
remarkably concise. The state of an isolated physical system is described
by a vector. If the space of all states is spanned by two vectors, it can
be viewed as a quantum bit or qubit. To connect with traditional computation,
we choose two orthogonal vectors to
represent the values zero and one, such as
\begin{equation}
|0\rangle \equiv \begin{pmatrix}1 \\ 0\end{pmatrix}, \; |1\rangle \equiv \begin{pmatrix}0 \\ 1\end{pmatrix}.\label{eq:01}
\end{equation}
\noindent
The general state $|\psi\rangle$ of a qubit is a superposition of the
two basis states,
$|0\rangle$ and $|1\rangle$:
\begin{equation}
|\psi\rangle=c_{0}|0\rangle+c_{1}|1\rangle \textrm{~~with~~} \quad |c_0|^2+|c_1|^2=1.\label{eq:qubit}
\end{equation}
\noindent
The coefficients $c_0,c_1$ are, in general, complex numbers. Measurement
of a qubit (or \textit{readout})  destroys  its quantum superposition
state $|\psi\rangle$. The state found as a result of the measurement will
be either $|0\rangle$ or $|1\rangle$. The quantities  $|c_0|^2$ and $|c_1|^2$
give the probabilities  of the measurement outcome, and so must
sum to $1$. This explains the normalization property in (\ref{eq:qubit}).
The coefficients $c_0$ and $c_1$
in the quantum superposition are called amplitudes. In general, each qubit
exists in a quantum superposition of its two possible bit assignments (0 and 1).

In quantum mechanics, column and row vectors are represented with
Dirac's ``bra-ket'' notation. A state vector is a column vector of complex
numbers denoted as ket $|\psi\rangle$. Conjugate to this state vector is a
row vector denoted as bra $\langle \psi|$:
\begin{equation}
\langle \psi|=c_{0}^{*}\langle0|+c_{1}^{*}\langle1|.\label{eq:cc-qubit}
\end{equation}
\noindent
The norm or length of a state vector is given by the dot product, sometimes
referred to as a bracket, of bra and ket vectors $\langle\psi|\psi\rangle=1$,
which explains the origin of Dirac's ``bra-ket'' terminology.
It is instructive to note that
\begin{equation}
c_0=\langle 0|\psi\rangle, \qquad c_1=\langle 1|\psi\rangle,\qquad
c_{0}^{*}=\langle\psi|0\rangle, \qquad c_1^{*}=\langle \psi|1\rangle.
\label{eq:c}
\end{equation}

In general, quantum systems have more then two basis states, in fact, \emph{many} more than two basis states! Consider, for example,  a quantum register  with $N$ qubits. Each qubit $i$ is characterized by  a pair of basis states $|0\rangle_{i}$ and $|1\rangle_{i}$ corresponding to the  two possible bit assignments. There are in total $2^N$ basis states of the register, each corresponding to a specific assignment of $N$ bits
\begin{equation}
{\bf z}=\{z_1,z_2,\ldots,z_N\}. \label{eq:z}
\end{equation}
\noindent
Each of the $2^N$ basis states of the register  $|{\bf z}\rangle$ is an outer product  of basis states of individual qubits
\begin{equation}
 |{\bf z}\rangle=|z_1\rangle_{1}\otimes|z_{2}\rangle_{2}\otimes\ldots\otimes|z_{N}\rangle_{N}.\label{eq:register}
 \end{equation}
 \noindent
where a convenient short-hand notation is $|{\bf z}\rangle=|z_1,z_2,\cdots,z_N\rangle$.  In an analogy with (\ref{eq:01}), each state $|{\bf z}\rangle$
corresponds to a column vector of length $2^N$, with all elements
equal to 0 except for one element that is equal to 1.
This unit element occupies the position $z$ in the column  vector, where
the binary encoding of the integer $z$ is given by a bit assignment (\ref{eq:z})
 \begin{equation}
 |{\bf z}\rangle=\begin{pmatrix}0\\0\\\cdot\\1\\\vdots\\0\\0\end{pmatrix} \longrightarrow \,\,{\rm single \,\, unit\,\,element\,\,occurs\,\,at\,\,} z^{\rm th}\,\,{\rm position\,\, in\,\, a \,\,column}.\label{eq:xcol}
 \end{equation}

 The general quantum state of the $N$-qubit register is a linear superposition
\begin{equation}
|\psi\rangle=\sum_{z_1=0,1}\cdots\sum_{z_N=0,1}C_{z_1,\ldots,z_N}|z_1,\ldots,z_N\rangle\equiv \sum_{{\bf z}\, \in \{0, 1\}^N}C_{\bf z}|{\bf z}\rangle\label{eq:psi}
\end{equation}\noindent
Here and in what follows, we shall use the shorthand notation
$C_{z_1,z_2,\ldots,z_N}\equiv C_{\bf z}$. The summation above is
understood to be over the
$2^N$ bit configurations ${\bf z}$ (\ref{eq:z}). Upon readout,
the quantum register will be found in one of its basis states
$|z_1,\ldots,z_N\rangle$ with probability $|C_{\bf z}|^2$,
\begin{equation}
\sum_{{\bf z}\, \in \{0, 1\}^N} |C_{\bf z}|^2=1.
\end{equation}
\noindent
A particular example of the quantum state is $|\psi\rangle=|{\bf z}\rangle$. In this case, all coefficients in the  superposition but one equal zero. Clearly,
upon readout of this state, the system will be found \emph{with certainty  }
in the state $|{\bf z}\rangle$.

\subsection{Evolution of quantum systems}

State vectors $|\psi \rangle$ evolve in time according the
Schr\"{o}dinger equation,
\begin{equation}
i \hbar \frac{d}{dt} |\psi\rangle = H |\psi\rangle, \label{eq:SchroEq}
\end{equation}
where, for an $N$ qubit system, the Hamiltonian $H(t)$ is a
$2^N\times 2^N$ Hermitian matrix.
Here, the notation $H |\psi\rangle$ conventionally denotes a
matrix-vector product. It is said that the Hamiltonian ``acts'' on the system state vector $|\psi\rangle$.
For an $N$-qubit register, the Hamiltonian is a $2^N\times 2^N$
Hermitian matrix that acts on  basis states (\ref{eq:xcol}) and
their superpositions (\ref{eq:psi}). In the basis consisting of states
of the form of Eq.~\ref{eq:xcol}, the $z', z$ entry of the Hamiltonian matrix
can be written as
 \begin{equation}
 H_{{\bf z^\prime},{\bf z}}=\langle {\bf z^\prime}|H|{\bf z}\rangle.\label{eq:Hzz}
 \end{equation}
It is common to compactly represent matrices in quantum mechanics using
a set of matrices each with only one nonzero unit element located in
row  ${\bf z}$ and column ${\bf z^\prime}$.
A short-hand notation for such a matrix, in ket and bra notation,
is the following:
\begin{equation}
|{\bf z}\rangle\langle{\bf z^\prime}|\label{eq:zz}
\end{equation}
\noindent
An arbitrary  Hamiltonian matrix $H$ can be expressed via its matrix
elements as follows
\begin{equation}
H=\sum_{{\bf z},{\bf z^\prime}\, \in \{0, 1\}^N} H_{{\bf z^\prime},{\bf z}} |{\bf z}\rangle\langle{\bf z^\prime}|.\label{eq:Hbracket}
\end{equation}

\subsection{Stationary states and system energies}
In general, a system Hamiltonian $H=H(t)$ is time-dependent. When it is
not, the quantum system is called stationary. The Schr\"{o}dinger equation
(\ref{eq:SchroEq}) in this case corresponds to a set  of $2^N$ linear ordinary differential equations with constant coefficients. Using ket notation for
state vectors, its solution can be written in a well-recognized form
\begin{equation}
|\psi(t)\rangle=\sum_{n=0}^{2^N-1}c_{n}|\Phi_{n}\rangle \exp\left(-\frac{i}{\hbar}\lambda_n t\right).\label{eq:Schro-stat-sol}
\end{equation}
\noindent
Here,  the $\lambda_n$ are eigenvalues of the Hermitian Hamiltonian matrix $H$
are called system energies, and  $|\Phi_{n}\rangle$ are the corresponding
eigenstates, called the system's stationary states. They are solutions to
the matrix-eigenvalue problem
\begin{equation}
H|\Phi_n\rangle=\lambda_n|\Phi_n\rangle,\quad n=0,\ldots,2^N-1.\label{eq:Schro-stat}
\end{equation}
\noindent
The eigenstate $|\Phi_0\rangle$ corresponding to the lowest energy level $\lambda_0$  is called a {\em ground state}.

The coefficients $c_{n}$ in (\ref{eq:Schro-stat-sol}) determine the initial system state as a superposition over its stationary states
\begin{equation}
|\psi(0)\rangle=\sum_{n=0}^{2^N-1}c_n|\Phi_n\rangle.\label{eq:Schro-stat-sol-init}
\end{equation}
\noindent
A  remarkable property of the solution (\ref{eq:Schro-stat-sol}) is that if the system's initial state at $t=0$
 is an exact eigenstate of the Hamiltonian  then at any time its state is the same eigenstate times the phase factor
 \begin{equation}
{\rm if}\quad |\psi(0)\rangle=|\Phi_{n}\rangle \quad {\rm then}\quad  |\psi(t)\rangle=e^{-\frac{i}{\hbar}\lambda_n t} |\Phi_{n}\rangle
 \end{equation}

\subsection[Pair of superconducting flux qubits]{\label{subsec:I} Example Hamiltonian for a pair of superconducting flux qubits}
\begin{figure}[!ht]
\begin{center}
\includegraphics[width=6.0in]{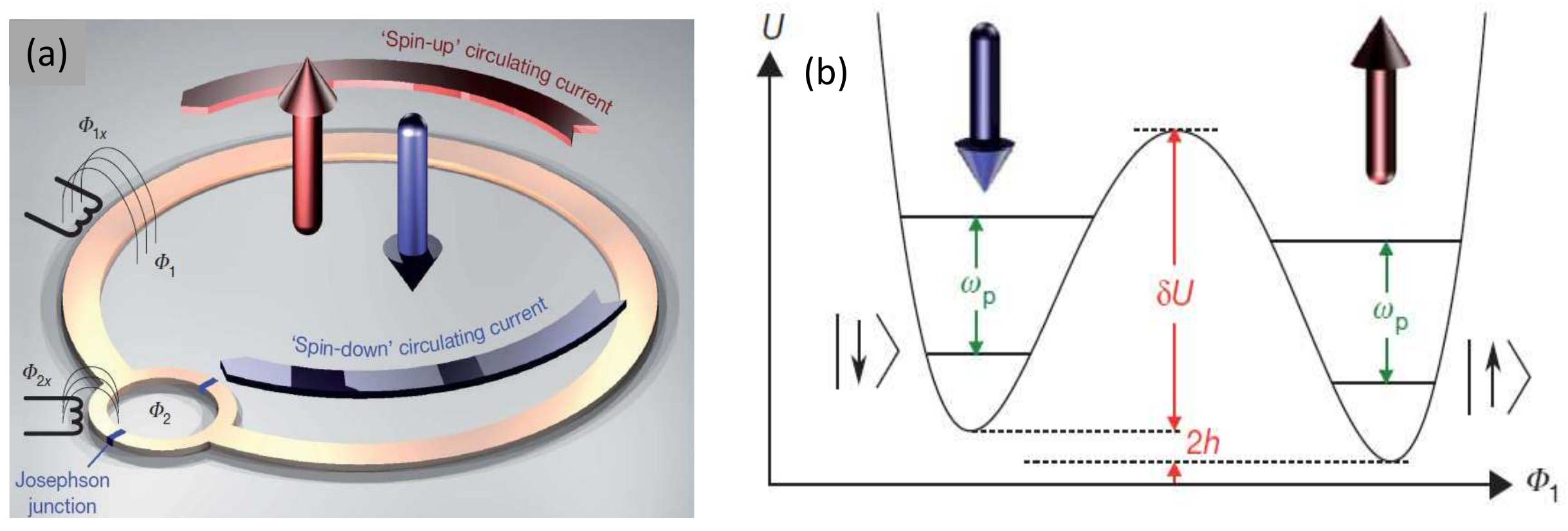}
\end{center}
  \caption{Superconducting qubit. \label{fig:4} }
\end{figure}

The superconducting flux qubit is implemented by a simple wire ring interrupted by a small dc SQUID as illustrated in Fig.~\ref{fig:4}(a).  Flux within a superconducting loop is quantized, as is current circulating around it.  For appropriate choice of external flux biases $\Phi_{1x}$  and  $\Phi_{2x}$, the smallest amount of flux that can be in the loop corresponds to either  $+\Phi_{0}/2$ and $-\Phi_{0}/2$, where
$\Phi_{0}=\hbar/2e$ is the magnetic flux quantum.  These lowest two states represent the two possible values  of a classical bit,  ``1'' and ``0'',  or  the two basis states of a qubit $|0\rangle $ and $|1\rangle$. These two possible states  may be thought of as current circulating either clockwise or counterclockwise in the loop.

In this configuration, the potential energy of the qubit has the double-well shape shown in the Fig.~\ref{fig:4}(b).  Classically the ``1'' and ``0'' states correspond to each of the two potential energy minima. The barrier between them, tunable with bias $\Phi_{2x}$, is low enough that the system can tunnel back and forth between $|0\rangle $ and $|1\rangle$.  In fact, it can occupy a state that is a superposition of $|0\rangle $ and $|1\rangle$, meaning that current can circulate both clockwise and counterclockwise at the same time.
A single qubit Hamiltonian representing this system can be written in the form
\begin{equation}
H_{1}=-\Delta \hat X-h \hat Z\label{eq:H1}
\end{equation}
\noindent
where $\hat X$ and $\hat Z$ are Pauli matrices,
specific $2\times 2$ matrices that act on single qubit states (\ref{eq:01})
\begin{equation}
 \hat X=\begin{pmatrix}0&1 \\ 1&0\end{pmatrix},
\quad  \hat Z=\begin{pmatrix}1&0 \\ 0&-1\end{pmatrix}
\label{eq:XZ}
\end{equation}
\noindent
In the Hamiltonian (\ref{eq:H1})
the bias parameter $h$ describes the energy separation between
the bottom of the potential wells in Fig.~\ref{fig:4}(b), and $-\Delta$
is the tunneling amplitude between the wells.

Instead of a bit variable  $z=0,1$, one could use the symmetric binary
variable $s=1-2z$ (Ising spin) to signify qubit states. This representation
will prove  especially useful below because of the Pauli matrix property
\begin{equation}
\hat Z |z\rangle= s|z\rangle,\quad s=1-2z,\quad z=0,1.\label{eq:s}
\end{equation}
\noindent
When  $\Delta$=$0$, quantum tunneling is suppressed and the qubit
Hamiltonian $H_1$=$-h\hat Z$. The qubit states corresponding to
$s=\pm 1$ (magnetic fluxes  $\pm \Phi_0$) are stationary eigenstates of
$H_1$ with eigenvalues equal to $-s h=\mp h$. In Figure \ref{fig:4}(b)
these eigenstates  are denoted as $\us$ and $\ds$ and correspond to the
system being localized either in the left or right well, respectively.
For positive bias $h>0$, the right well is the ground state. When tunneling
is finite, the state is a superposition of the right well $\us$ and left
well $\ds$.
For example, at zero bias $h=0$, corresponding to the symmetric
double-well potential in Figure \ref{fig:4}(b), the qubit Hamiltonian
is of a pure tunneling nature. Its two eigenstates are symmetric
and anti-symmetric superpositions of the right and left states
\begin{equation}
H_{1}\frac{\us \pm \ds}{\sqrt{2}}=\mp \Delta\,\frac{\us \pm \ds}{\sqrt{2}},\quad\quad H_1=-\Delta \hat X,\label{eq:symm}
\end{equation}
\noindent with the symmetric superposition being a ground state.

\begin{figure}[!ht]
\begin{center}
\includegraphics[width=6.0in]{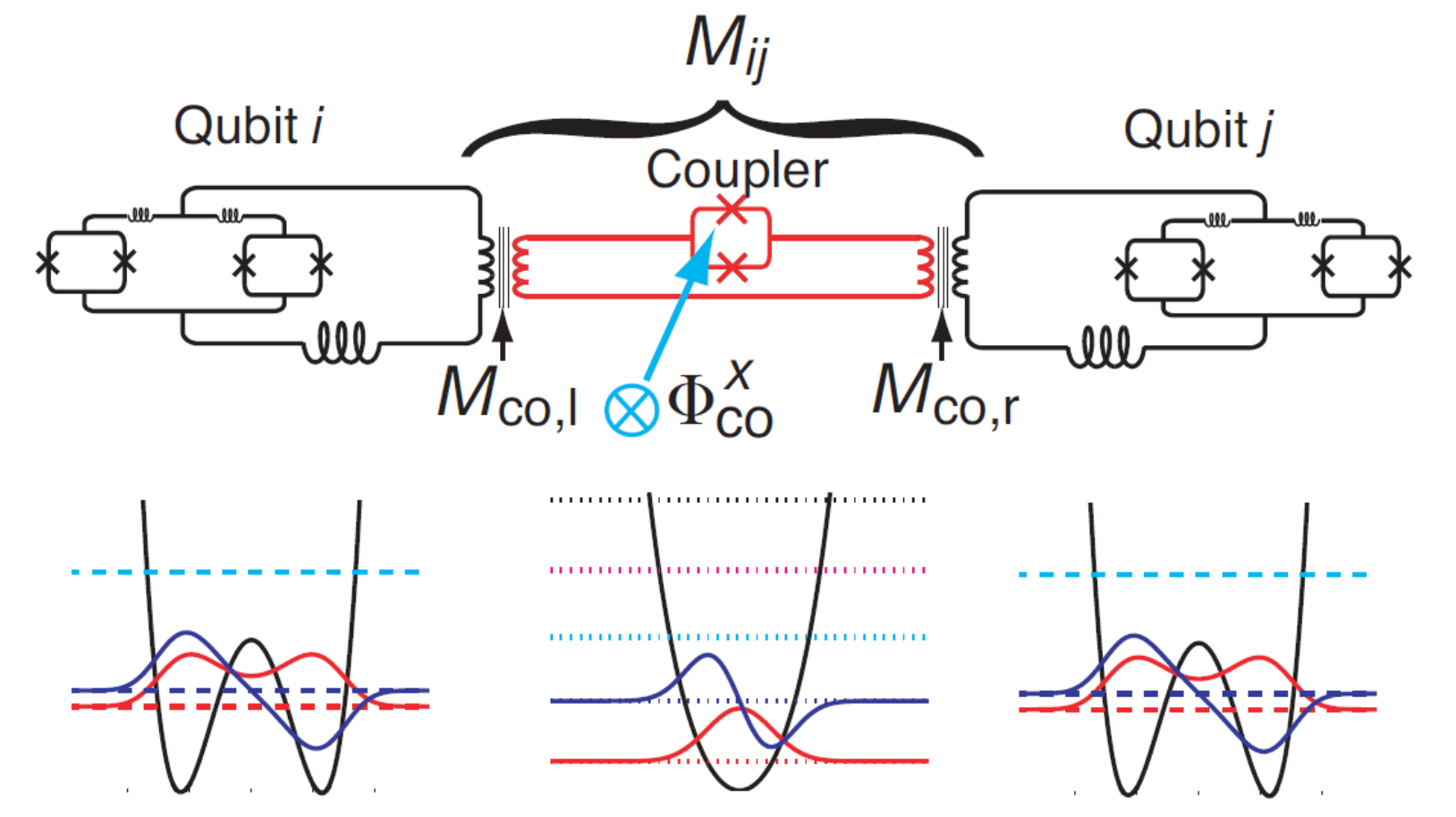}
\end{center}
  \caption{Tunable magnetic coupling between two superconducting flux qubits . \label{fig:5} }
\end{figure}

Tunable magnetic coupling is achieved between qubits with a two-junction rf-SQUID as shown in the Fig.~\ref{fig:5}.   The aggregate magnetic coupling between the two qubits is
\begin{equation}
M_{12}=\chi_i\,M_{\rm co,l}\cdot M_{\rm co,r}
\end{equation}
\noindent
where $\chi_i$ is the susceptibility of the intermediate rf-SQUID coupler,
tunable through the external flux bias $\Phi_{\rm co}^{x}$. The coupler's
susceptibility can be of either sign, so the coupler can be tuned either
so that it is energetically favorable for the qubits to be in the same
state as each other (both ``1'' or both ``0''), or so that it is
energetically favorable for the qubits to be in opposite states
(either ``1,0'' or ``0,1'').

In the language of Ising model (\ref{eq:Ising}), this inter-qubit coupling
is described by parameter $J_{12}$. The case $J_{12}< 0$ means the qubits
want to be in the same state (ferromagnetic coupling), and $J_{12}> 0$
means the qubits want to be in different states (anti-ferromagnetic coupling).
We now can extend the Hamiltonian above to two qubits:
\begin{equation}
H_{2}=-\Delta_{1}\hat X_{1}-\Delta_{2}\hat X_{2}-h_1 \hat Z_{1}-h_2\hat Z_{2}+J_{12} \hat Z_{1}\hat Z_{2}
\end{equation}
\noindent
Each of the operators in $H_{2}$ is a product of two operators
one acting on the states of  qubit 1 and the other acting on the states
of qubit 2. In total, the Hamiltonian $H_2$ refers to the composite
two-qubit system. In the standard basis
$\{|00\rangle, |01\rangle, |10\rangle,  |11\rangle \}$:
\begin{equation*}
|00\rangle = \begin{pmatrix} 1 \\ 0 \\ 0 \\0 \end{pmatrix}, \; |01\rangle = \begin{pmatrix} 0 \\ 1 \\ 0 \\0 \end{pmatrix}, \; |10\rangle = \begin{pmatrix} 0 \\ 0 \\ 1 \\0 \end{pmatrix}, \; |11\rangle = \begin{pmatrix} 0 \\ 0 \\ 0 \\ 1 \end{pmatrix}.
\end{equation*}
\noindent
$H_2$ is the $4\times4$ matrix
\begin{equation}
\left(
  \begin{array}{cccc}
    -h_1-h_2+J_{12} & -\Delta_2 & -\Delta_1 & 0 \\
    -\Delta_{2} & -h_1+h_2-J_{12} & 0 & -\Delta_{1} \\
    -\Delta_{1} & 0 & h_1-h_2-J_{12} & -\Delta_{2} \\
    0 & -\Delta_{1} & -\Delta_{2} & h_1+h_2+J_{12} \\
  \end{array}
\right)
\end{equation}
This scheme can be extended to include multiple controllable pairwise
coupling between qubits as well as single qubit terms. The total
Hamiltonian of $N$ superconducting qubits has the form
\begin{equation}
-\sum_{i=1}^{N}\Delta_{i}\hat X_{i}-\sum_{i=1}^{N}h_i \hat Z_{i}+\sum_{\langle i j\rangle\in \texttt{E}} J_{i j} \hat Z_{i}\hat Z_{j}.\label{eq:IsingN}
\end{equation}
\noindent
Here, $\texttt{E}$ is a set of pairs of coupled qubits that is completely analogous to the set of spin couplings in the Ising model  (\ref{eq:Ising}).

We now give an example of a Hamiltonian that encodes a binary
optimization problem. If, for each qubit, the potential barrier between
the wells is high enough, the quantum mechanical tunneling is suppressed.
Then, the parameters $\Delta_i$ can be set to zero and the
Hamiltonian (\ref{eq:IsingN}) takes the form
\begin{equation}
H_{P}=-\sum_{i=1}^{N}h_i \hat Z_{i}+\sum_{\langle i j\rangle\in \texttt{E}} J_{i j} \hat Z_{i}\hat Z_{j}.\label{eq:HP-Ising}
\end{equation}
\noindent
It follows from (\ref{eq:s}) that $H_P$ will be diagonal in the standard
basis of $N$-qubit states, and one can  immediately check that the eigenvalue
for the eigenstate $|z_1,z_2\ldots z_N\rangle$ of $H_{P}$ equals the
energy of the Ising model (\ref{eq:Ising}) for
the appropriate  assignment of binary variables $s_i$:
\begin{equation}
H_{P}|z_1,z_2\ldots z_N\rangle=E_{\rm Ising}(s_1,\ldots,s_N)|z_1,z_2\ldots z_N\rangle,\quad {\rm where}\quad  s_i=1-2z_i,\quad s_i=\pm 1.\label{eq:Ising-encoding}
\end{equation}
\noindent
Therefore, the problem $H_P$ Hamiltonian (\ref{eq:HP-Ising}) encodes the  optimization problem given by the Ising  model $E_{\rm Ising}$: there is one-to-one correspondences between the Ising spins $s_i$  in $E_{\rm Ising}$ and
Pauli matrices  $\hat Z_i$ in  $H_P$.  Of course, in order to make
use of this encoding we need a quantum mechanism that can realize
the ground state of this Hamiltonian with high probability.

\subsection[Hamiltonian for binary optimization]{\label{sec:P}Hamiltonian for a binary optimization problem}
Consider a classical binary optimization problem with cost (energy)
function $E({\bf z})$ defined over the space of bit configurations
${\bf z}=\{z_1 \ldots z_N\}$. A solution of the problem is a bit
configuration ${\bf z^*}$ such that
  \begin{equation}
 {\bf z^*}=\underset{\bf z}{\operatorname{argmin}}\, E({\bf z})\quad {\rm or}\quad  E({\bf z})\geq E({\bf z^*})\quad {\rm for\,\, all}\quad {\bf z}\in\{0,1\}^N.\label{eq:opt}
 \end{equation}
\noindent
A particular example of a quantum Hamiltonian that will be central to
our discussion is the one whose eigenstates are the standard basis states
of a quantum register $|{\bf z}\rangle$ with eigenvalues equal to the
classical cost function values $E({\bf z})$ for the corresponding bit
assignments  ${\bf z}$. Thus, for any classical binary optimization
problem, we can define a \emph{problem Hamiltonian} $H_P$
  \begin{equation}
  H_P|{\bf z}\rangle=E({\bf z}) |{\bf z}\rangle,\quad {\bf z}=\{z_1,\ldots,z_N\},\quad z_i=0,1.\label{eq:HP}
  \end{equation}
Matrix elements  of the problem Hamiltonian $(H_P)_{{\bf z^\prime},{\bf z}}$ are nonzero only if they are located on diagonal and $(H_P)_{{\bf z},{\bf z}}=E({\bf z})$. Such a  Hamiltonian matrix encodes the classical optimization problem on a quantum register. The compact notation for this Hamiltonian following is
\begin{equation}
H_P=\sum_{{\bf z}\,\in \{0, 1\}^N}E({\bf z})|{\bf z}\rangle\langle{\bf z}|.\label{eq:P}
\end{equation}
 \noindent
The ground state of this Hamiltonian is the eigenstate with lowest energy,
the basis state corresponding to the bit string
${\bf z^*} = z_1^*z_2^*\dots z_N^*$
that is a solution of the classical optimization problem
 \begin{equation}
 |\Phi_0\rangle=|{\bf z^*}\rangle.\label{eq:ground-P}
 \end{equation}
 \noindent

\subsection[Fully-symmetric ground state]{\label{sec:D} Hamiltonian with fully-symmetric ground state}
Consider now another example of a Hamiltonian matrix $H_D$ called a driver Hamiltonian. It will be constructed so that it causes transitions between the
eigenstates $|{\bf z}\rangle$ of $H_P$ corresponding to energies $E({\bf z})$ of the classical optimization model. We will assume that diagonal elements of $H_D$ in the basis of the quantum register states $|{\bf z}\rangle$ are all zero. The only nonzero matrix elements  $(H_D)_{{\bf z},{\bf z^\prime}}$ are those corresponding  to pairs of bit configurations ${\bf z}$ and ${\bf z^\prime}$ that differ by a single bit flip. We shall assume that  all these matrix elements have the same value, say $-\Delta$.  Then a row  of $H_D$ corresponding to a bit string ${\bf z}$ has exactly $N$ elements equal to $-\Delta$  and the rest of the elements equal to 0. The elements equal to $-\Delta$ are located in columns corresponding to the bit assignments ${\bf z^\prime}$ with one  bit flipped compared to  ${\bf z}$. Using the compact matrix representation, one can write
\begin{equation}
  H_D =  -\Delta\sum_{k
=1}^{N}  \sum_{\boldsymbol{z} \in \{0, 1\}^N} |z_1 \ldots z_k \ldots z_N \rangle
  \langle z_1 \ldots \bar{z}_k \ldots x_N |, \quad \bar z_k=1-z_k. \label{eq:D}
\end{equation}
\noindent
This Hamiltonian causes transitions between the
eigenstates $|{\bf z}\rangle$ of $H_P$ in an ``unbiased'' fashion, attaching to each of them an equal weight $-\Delta$.  The ground state of the Hamiltonian is the fully symmetric superposition of  $|{\bf z}\rangle$ and the corresponding eigenvalue equals to $-N\Delta$. We shall denote this fully-symmetric state as $|\Phi_S\rangle$
\begin{equation}
H_D |\Phi_S\rangle=-N\Delta |\Phi_S\rangle\quad {\rm where}\quad |\Phi_S\rangle=2^{-N/2}\sum_{{\bf z}\in\{0,1\}^{N}}|{\bf z}\rangle.\label{eq:Phi0}
\end{equation}
\noindent
This state can be represented by a column vector of the length $2^N$ with
all elements equal to $2^{-N/2}$.

Let us provide a specific example in the context of superconducting qubits (Fig.~\ref{fig:4}). Unlike Sec.~\ref{sec:P}), we assume here that all magnetic coupling constants $J_{ij}$ and single qubit bias fields $h_i$ are set to zero. We will also set individual qubit tunneling splittings to the same value. Then the Hamiltonian for $N$ superconducting qubits is
\begin{equation}
H_{D}=-\Delta\sum_{i=1}^{N}\hat X_{i}.\label{eq:D-Ising}
\end{equation}
\noindent
Comparing this with Eq.~(\ref{eq:symm}) one sees that for each qubit a ground state is a symmetric superposition $\frac{|0\rangle_i+|1\rangle_i}{\sqrt{2}}$. Because the  Hamiltonian is additive with respect to individual qubits,
its ground state is the $N$-qubit product state
\begin{equation}
|\Phi_{S}\rangle=\frac{1}{\sqrt{2}}(|0\rangle_1+|1\rangle_1)\otimes\ldots \otimes \frac{1}{\sqrt{2}}(|0\rangle_N+|1\rangle_N)\label{eq:PhiS}
\end{equation}
\noindent
with eigenvalue equal to $-N \Delta$. After multiplying out the left hand
side, one can immediately see that this ground state is nothing but the fully-symmetric state $|\Phi_{S}\rangle$ (\ref{eq:Phi0}). In fact the Hamiltonians in Eqs.~(\ref{eq:D}) and (\ref{eq:D-Ising}) are identical.

The practical importance of the state $|\Phi_S\rangle$  is that it can be  be prepared easily. Indeed, the  driver Hamiltonian  $H_D$ (\ref{eq:D-Ising})  can be implemented easily by setting biases and inter-qubit couplings to zero. The energy gap separating the ground state energy of $H_D$ from the first excited state
equals $\Delta$. Assuming that  $\Delta$ is made sufficiently large compared to
the temperature of the environment, the system will reach the state $|\Phi_{S}\rangle$ by itself in the course of thermal relaxation.  This property will be used later.

\subsection{Basics of Adiabatic Theorem} Consider the case when the matrix elements of the system Hamiltonian $H=H(t)$ are  varying in time but very slowly (the exact criterion on speed will be specified below).
In this case, many properties of a quantum system at an instant $t$ will be determined by the instantaneous values of the matrix elements in $H(t)$. Intuitively, we therefore introduce the instantaneous energies and the instantaneous eigenstates of the Hamiltonian $H(t)$ in analogy with (\ref{eq:stat})
 \begin{equation}
H(t)|\Phi_n(t)\rangle=\lambda_n(t)|\Phi_n(t)\rangle,\quad n=0,\ldots,2^N-1.\label{eq:stat}
\end{equation}
\noindent
According to Adiabatic Theorem of quantum mechanics, if at $t=0$ the system state is  prepared as an  eigenstate of the Hamiltonian $H(0)$, then at all future times its state will remain very close to the {\em  instantaneous} eigenstate of the Hamiltonian $H(t)$ with the same quantum number $n$, up to a phase factor.
As an example, if the system is initially prepared in the ground state of $H(0)$, i.e., if $\quad |\psi(0)\rangle=|\Phi_{0}(0)\rangle$, then
\begin{equation}
 |\psi(t)\rangle\simeq \exp\left[ -\frac{i}{\hbar}\int_{0}^{t}d\tau \lambda_0(\tau)\right] |\Phi_{0}(t)\rangle.\label{eq:adiabatic}
 \end{equation}
\noindent
The slow quantum evolution where the state vector closely follows the instantaneous eigenstates of $H(t)$ having the same quantum number at all times is called {\em adiabatic}.

We now revisit the criterion on the slowness of $H(t)$ which is a condition on the magnitude of the matrix elements of the  matrix of time derivatives $\dot H(t)$. The criterion for the ground state adiabatic evolution reads
\begin{equation}
|\langle \Phi_{0}(t)|\dot H(t)|\Phi_{1}(t)\rangle|\ll \frac{|\lambda_{1}(t)-\lambda_0(t)|^2}{\hbar},\label{eq:slow}
\end{equation}
\noindent
where $|\Phi_1\rangle$ and $\lambda_{1}$
are the instantaneous first excited state of the system Hamiltonian $H(t)$ and the corresponding eigenvalue. The criterion for adiabatic evolution is local in time: the instantaneous rate of change of $H(t)$ must be much less than the square of the gap separating the energies of the ground state and the first excited state.
It is said that adiabatic evolution is \emph{asymptotically} exact. The smaller the Hamiltonian derivative matrix $\dot H(t)$ the closer the system  state will be to the instantaneous eigenstate of $H(t)$ as in Eq.~(\ref{eq:adiabatic}).

\section[ Quantum Annealing]{\label{sec:QA}  Quantum Annealing Concept}

The Adiabatic Theorem suggests how classical optimization problems can be solved with the help of quantum mechanics using the so called Quantum Annealing (QA) process.  Quantum Annealing can be implemented by varying the \emph{control }Hamiltonian $H(t)$ slowly in time in a manner that interpolates between a driver Hamiltonian at the beginning, $H(0)=H_D$ (\ref{eq:D-Ising}), and a problem Hamiltonian at the end, $H(T)=H_P$ (\ref{eq:HP-Ising}), where $T$ is the duration of the quantum evolution. For example, a simple linear interpolation is
\begin{equation}
H(t)=\left(1-\frac{t}{T}\right) H_D+\frac{t}{T} H_P\label{eq:H-int}
\end{equation}
 \noindent
where
\begin{equation}
H_{D}=-\Delta\sum_{i=1}^{N}\hat X_{i},\qquad H_{P}=-\sum_{i=1}^{N}h_i \hat Z_{i}+\sum_{\langle i j\rangle\in \texttt{E}} J_{i j} \hat Z_{i}\hat Z_{j}.\label{eq:DP}
\end{equation}
\noindent
At the beginning, the state of the quantum register is prepared to be a ground state of the initial Hamiltonian $H(0)=H_D$. This state is a fully symmetric product state (\ref{eq:PhiS})
\begin{equation}
|\psi(0)\rangle=|\Phi_{S}\rangle.\label{eq:psi0-PhiS}
\end{equation}
Assuming that time variation of the control Hamiltonian (\ref{eq:H-int}), (\ref{eq:DP}) is slow enough, the quantum evolution is adiabatic. The criterion for slowness is given in the next Section.  According to (\ref{eq:adiabatic}), the state of a quantum register $|\psi(t)\rangle$  at all times will follow closely the instantaneous ground state of the control Hamiltonian $H(t)$ given in (\ref{eq:H-int}), (\ref{eq:DP}).  Therefore, at the end of the adiabatic evolution, the state of the quantum register will approach the ground state of the problem Hamiltonian $H_P$ (up to the phase factor --- see Eq.~(\ref{eq:adiabatic}))
 \begin{equation}
 {\rm if}\quad |\psi(0)\rangle=|\Phi_{S}\rangle \quad {\rm then}\quad  |\psi(T)\rangle\simeq \exp\left[ -\frac{i}{\hbar}\int_{0}^{T}d\tau \lambda_0(\tau)\right] |{\bf z^*}\rangle.\label{eq:QA}
 \end{equation}
 \noindent
This final state corresponds to a bit assignment ${\bf z^*}$ that solves the optimization problem at hand. By performing a readout at $t=T$, one can recover this bit assignment. The state and energy mapping produced by this QA process is shown in Figs.~\ref{fig:QA-conv}.

\subsection{Quantum measurements at intermediate times}
\begin{figure}[!ht]
\centering
\subfigure[]{
\includegraphics[height=2.7in]{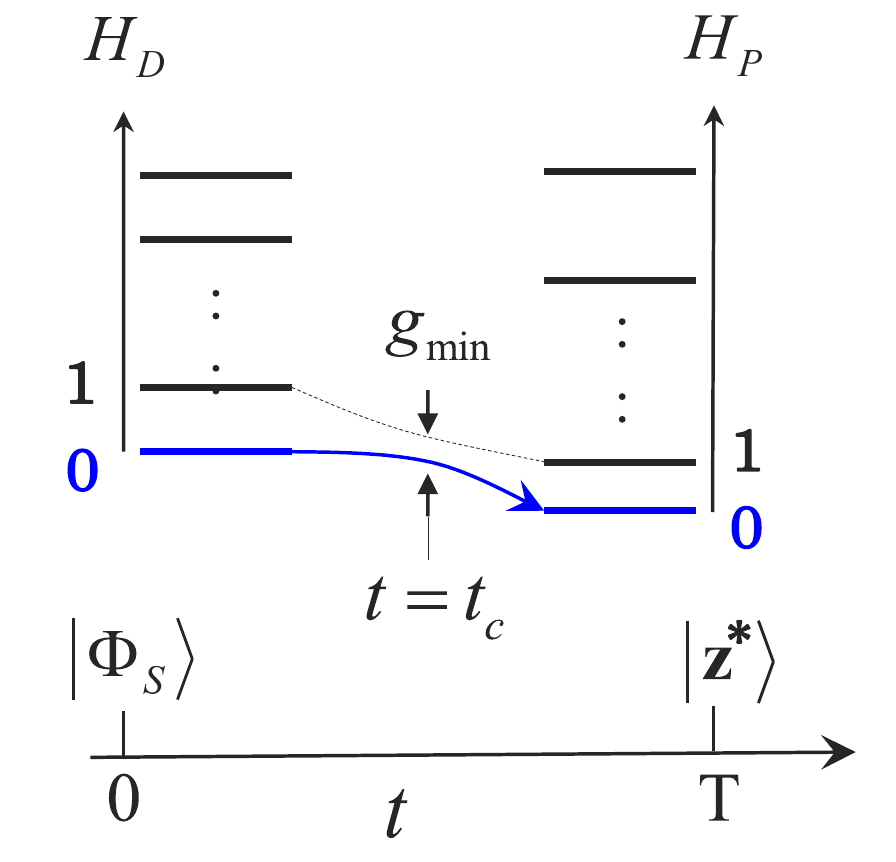}
\label{fig:3}
}\hspace{0.1in}
\subfigure[]{\includegraphics[width=2.7in]{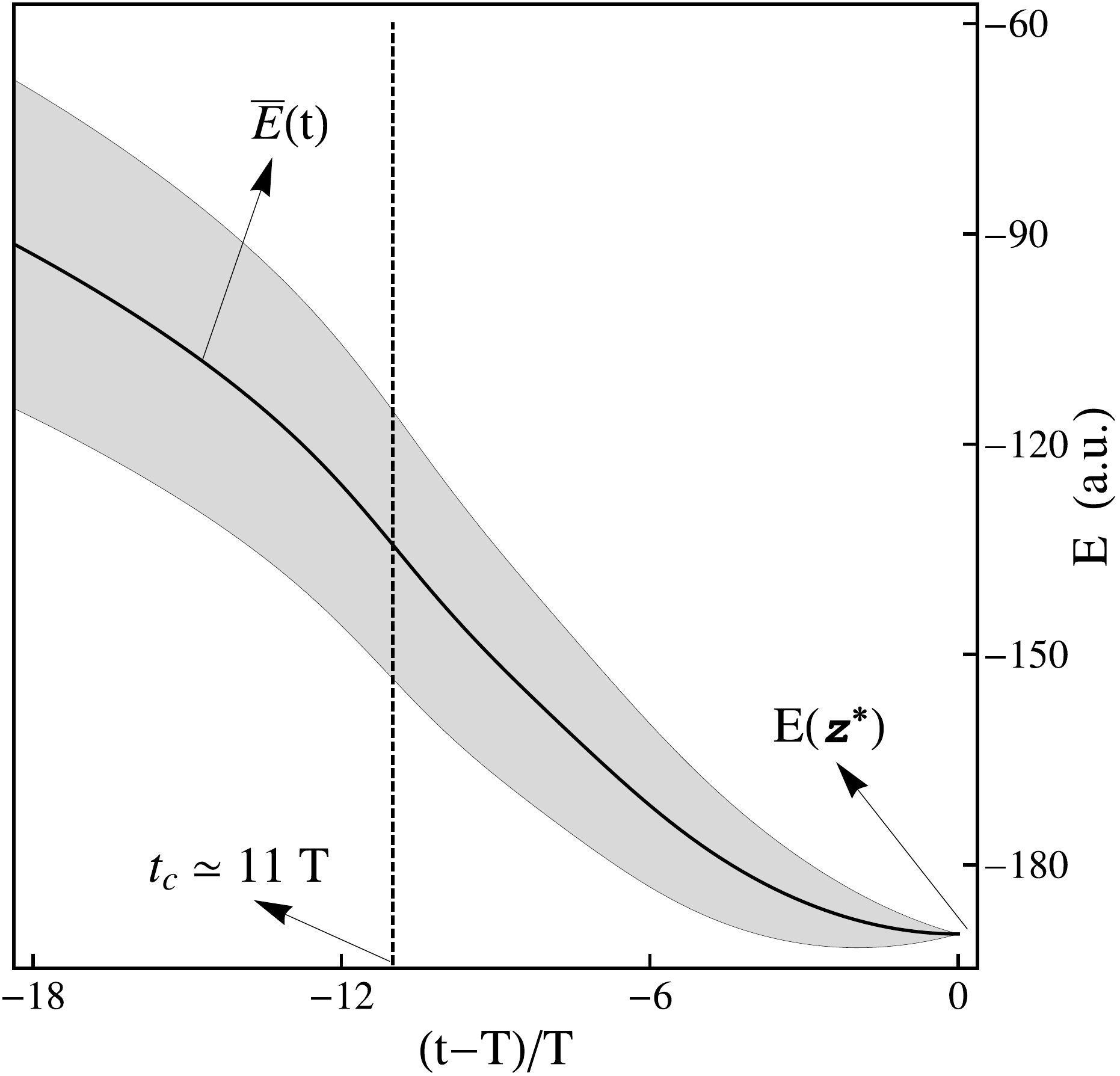}
\label{fig:Emean}
}
\label{fig:QA-conv}
\caption{{\footnotesize \subref{fig:3} Illustration of  the mapping in QA between the two lowest eigenvalues and the ground states of the initial  ($H_D$) and final  ($H_P$)  Hamiltonians  as time varying from 0 to $T$. Pair of vertical arrows mark the instant $t=t_c$ where the minimum energy gap is reached between the two eigenvalues.  This region is the QA bottleneck (cf. Sec.~\ref{sec:mingap}). 
If QA is sufficiently slow and the quantum evolution is adiabatic at all times than at the end ($t=T$)  the ground-state wavefunction corresponds to a string ${\bf z^*}$ giving the solution of optimization problem.
\subref{fig:Emean}  Plots of the mean energy $\overline{E}(t)$ and its variance  vs interpolation variable $(t-T)/T$ during QA convergence process. At the end of QA ($t=T$) variance shrinks to zero and mean energy approaches the ground state of the Hamiltonian $H_P$ (\ref{eq:DP}) of the Ising problem giving the solution to the latter. Note that the region near the minimum gap corresponds to the steepest variation of the mean register energy.} }
\end{figure}

It will be useful for control purposes to investigate the outcome of QA when terminating the quantum evolution prematurely at some \emph{intermediate} time $t<T$. At that instant, the  state of the quantum register $|\psi(t)\rangle$ will be neither the initial state nor the final state of the QA process in (\ref{eq:QA}). Instead, it would be given by the solution of the Schr\"{o}dinger equation (\ref{eq:SchroEq}).  Performing a readout on such a state would cause it to collapse into  one of  the many possible  outcomes $|{\bf z}\rangle$. The probability of obtaining state $|{\bf z}\rangle$ would be given by $|C_{\bf z}(t)|^2=\langle {\bf z}|\psi(t)\rangle|^2$. The samples from this distribution can be obtained by repeating the QA process many times and terminating it with a readout at the same time $t$.  For each of the outcomes the energy (cost) function value $E({\bf z})$ can be calculated.  One will obtain a distribution of energy values having mean and variance given by
\begin{equation}
\overline{ E}(t)=\sum_{{\bf z}\in{0,1}^{N}}|C_{\bf z}(t)|^2\, E({\bf z}),\quad \delta^{2}E(t)= \sum_{{\bf z}\in\{0,1\}^N}|C_{\bf z}(t)|^2(
E({\bf z})-\overline{E}(t))^2.
\label{eq:Emean}
\end{equation}
\noindent
At the end of the QA process, as  $t\rightarrow T$, the probability
distribution over the register states is increasingly peaked around
the solution state
$|C_{\bf z}(t)|^2\rightarrow \delta_{{\bf z},{\bf z^*}}$ and the  sums in  (\ref{eq:Emean}) are  increasingly dominated by one term  corresponding to  ${\bf z^*}$. As a result, the mean cost value approaches the minimum of the energy function $E_{\rm min}=E({\bf z^*})$ and the  variance shrinks to zero as shown in Fig.~\ref{fig:Emean}).

\subsection[Complexity]{\label{sec:mingap} Complexity of Quantum Annealing}
In its most straightforward form, the complexity of any optimization algorithm (including Quantum Annealing) is defined by a functional relation describing how the algorithm runtime $T$ necessary to find a global minimum of the energy function  scales with the number of bits $N$ in the \emph{asymptotic limit} of large $N$. Clearly, the complexity is not defined for a particular problem instance (whose number of bits $N$ is  fixed) but rather for a statistical ensemble of instances of a given $N$-bit optimization problem. For different random  instances the parameters of the cost function are taken at random while the structure of the cost function (defined by the optimization problem) is preserved. One example is the uninform ensemble of instances of the satisfiability problem with 3 bits in a clause  and a fixed clause-to-bit ratio. For each $N$ the runtime can be computed for the hardest instance   (worst case complexity), or for a typical  instance. In many cases the latter ``typical case'' complexity is of more practical significance than the (easier to compute) ``worst-case'' complexity beloved of theoretical computer scientists.
\begin{figure}[b]
\includegraphics[width=5.0in]{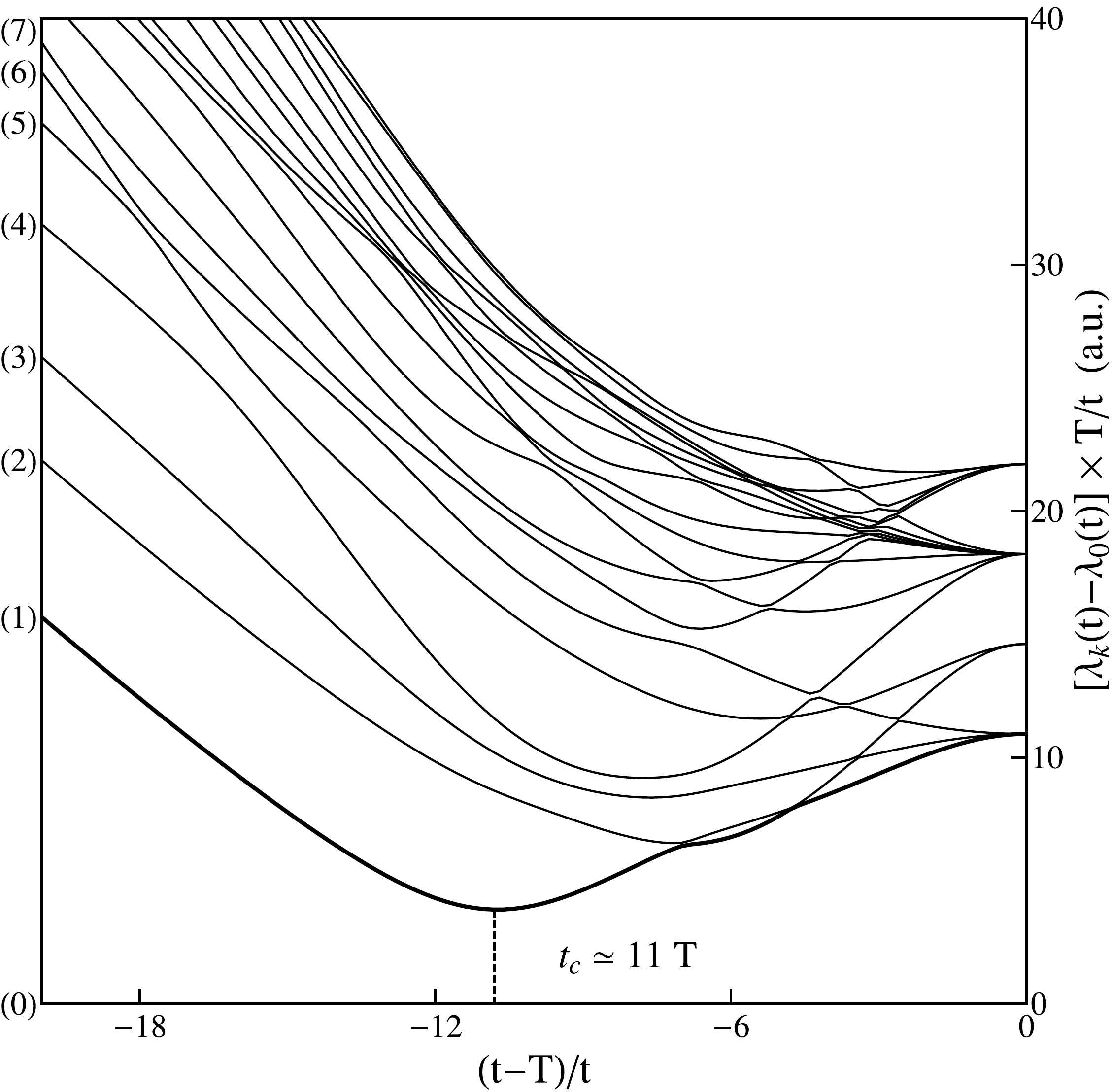}
  \caption{
  Plots of the scaled  differences $[\lambda_{k}(t)-\lambda_{0}(t)]\times \frac{T}{t}$ between the energies  of the excited states and that of the ground state of the control Hamiltonian $H(t)$  (\ref{eq:H-int}), (\ref{eq:DP}).  This Hamiltonian corresponds to  Quantum Annealing of  16-quit  Ising problem whose connectivity graph is shown in Fig.~\ref{fig:16Q}. Horizontal axis corresponds to scaled interpolating variable $(t-T)/t$ that changes from $-\infty$ to 0 during the Quantum Annealing.
  Labels for individual  energies  $(k)$ with $k=0,1,2,\ldots$ are shown at the left. Energy difference for the first excited state ($k=1$) is shown with  bold line. Dashed line corresponds to the instant $t_c=11 T$ when the  gap $\lambda_{1}(t)-\lambda_{0}(t)$ goes through its minimum.
 \label{fig:eigen}}
\end{figure}
To ensure that at the end of Quantum Annealing  the system  state approaches   the ground state of the problem Hamiltonian (\ref{eq:QA})  the adiabatic criterion (\ref{eq:slow}) must be satisfied at \emph{all} intermediate times $t\in(0,T)$. For the case of the control Hamiltonian that interpolates linearly from the initial to the problem Hamiltonian (\ref{eq:H-int}) the adiabatic criterion reads
 \begin{equation}
 T\gg \tau(t)=\hbar \frac{|\langle \Phi_{0}(t)|H_D|\Phi_1\rangle(t)|}{|\lambda_{1}(t)-\lambda_0(t)|^2},\quad {\rm for \,all }\quad t\in(0,T).\label{eq:tau}
 \end{equation}
\noindent
Then the condition for the adiabatic evolution can be re-written by maximizing  the value of the inverse-rate-limiting parameter  $\tau(t)$ over the interval $(0,T)$
\begin{equation}
T\gg \tau_{\rm max}=\frac{\hbar V}{g_{\rm min}^{2}},\quad g_{\rm min}=\min_{t\in(0,T)}|\lambda_1(t)-\lambda_0(t)|,\quad V=\max_{t\in(0,T)}|\langle \Phi_{0}|H_D|\Phi_1\rangle|.\label{eq:T}
\end{equation}
\noindent
Here $g_{\rm min}$ is the minimum energy gap over the entire QA process  as shown in Fig.\ref{fig:3}. We note that the required duration $T$ of the QA process is inversely proportional to the inverse square of the minimum energy gap.    The factor $V$ in the numerator typically scales linearly  with $N$. Therefore it is the minimum gap scaling with $N$ that really determines the complexity of QA.

\begin{figure}
\centering
     \begin{minipage}[t]{0.58\linewidth}
 \raisebox{-3cm}{
     \includegraphics[width=1\linewidth]{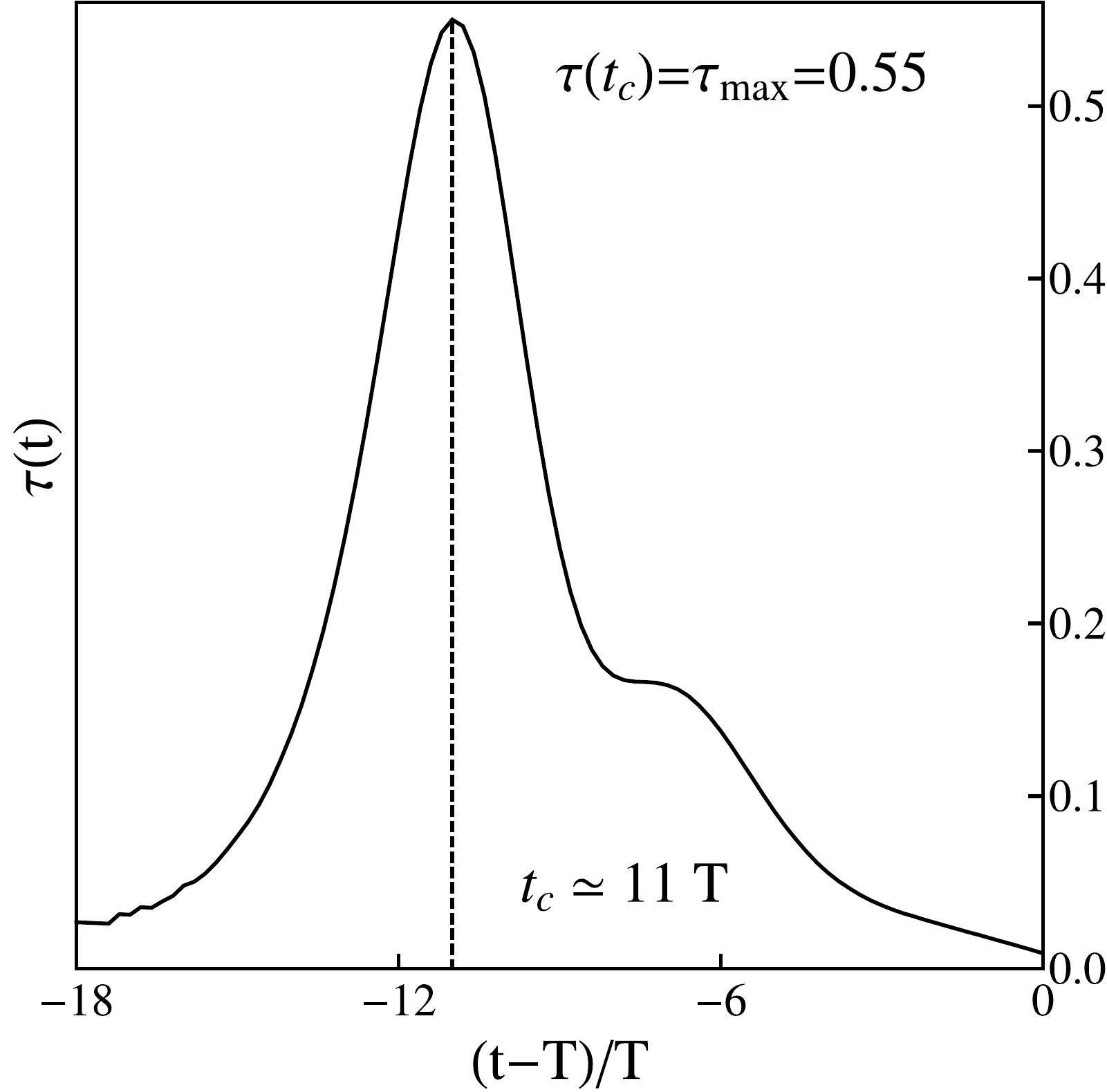}}
    \end{minipage}\hfill
    \begin{minipage}[b]{0.38\linewidth}
      \caption{\label{fig:tau} {\footnotesize Inverse rate-limiting parameter $\tau(t)$ vs interpolation variable $(t-T)/T$. Vertical dashed lines in both figures  correspond to the maximum of $\tau(t)\sim 0.55$ archived very near the point of minimum gap $t_c\simeq 11 T$. At that instant  the first excitation gap $\lambda_{1}(t)-\lambda_{0}(t)$ in Fig.~\ref{fig:eigen}) goes through its minimum.}}
    \end{minipage}
\end{figure}

It follows from the above that in the ideal case of a Quantum Annealing process completely uncoupled from noise and dissipation its performance is determined by the instantaneous energy spectrum of the control Hamiltonian $H(t)$ (\ref{eq:H-int}) especially near the point  of the minimum energy gap as illustrated in Fig.~\ref{fig:3}. We simulated the time-dependent energy spectrum of $H(t)$ for a 16-qubit  problem (\ref{eq:H-int}), (\ref{eq:DP})
defined for the following  Ising model (\ref{eq:Ising}). Individual qubits correspond to the numbered nodes of two $4\times 4$ arrays as shown in Fig.~(\ref{fig:16Q}). The Ising  coefficient $h_i$ corresponds to $i^{\rm th}$ node.  An edge connecting nodes $i$, $j$ corresponds to a nonzero coefficient $J_{ij}$.  Each array displays a bi-partite connectivity, i.e., each qubit in the group numbered 1-4 is connected, respectively, to each qubit in the group numbered 5-8 and likewise for the second array. There are 4 nonzero connections between the qubits across the two arrays each connecting a distinct pair of qubits as shown in Fig.~\ref{fig:16Q}. All the $h_i$, and $J_{ij}$ are varied in sign, and we set $\Delta=1$ in (\ref{eq:DP}).

Fig.~(\ref{fig:eigen}) shows the variation in the  (scaled) differences  in eigenvalues $[\lambda_{k}(t)-\lambda_{0}(t)]\,t/T$ as a function of the interpolating variable  $(t-T)/T$. The Quantum Annealing process corresponds to the interpolating variable  varying from $-\infty$ to 0. The minimum gap is achieved at the point $t_c=11 T$.  The condition for the duration $T$ of the QA process can be obtained by calculating the inverse-rate-limiting parameter  $\tau(t)$. This parameter is plotted in Fig.~\ref{fig:tau}), and its maximum value $\tau_{\rm max}\simeq 0.55$. Therefore in dimensionless units we find that the duration of the QA process needed to find a solution to the above Ising  problem is $T \gg 1$.

For the overwhelming majority of hard optimization problems  this answer is impossible to obtain reliably via theoretical  analysis because  the underlying quantum problems  are too complex (see, e.g., \cite{Santoro2002,Suzuki:2011}). Determination of $g_{\rm min}$ via direct numerical digitization of the control Hamiltonian can only be done for the very small sizes  ($N \lesssim $ 25-30) \cite{Farhi2} where the asymptotical behavior of $T$ with $N$ does not yet manifest itself. Quantum Monte Carlo (QMC) methods allow to investigate the minimum gap for the problems of a much large sizes (with $N$ in access of 100) \cite{YKS:2010}. In recent years a lot of attention was given to Quantum Monte Carlo (QMC) studies of the complexity of QA for the random satisfiability problem. However due to its inner workings the QMC method can only be applied to a very limited and artificially prepared subset of instances of Satisfiability, all having a unique satisfying assignment \cite{YKS:2010,YKS:2011,Young:2011,Raedt:2011}. Problems having more than one satisfying assignment has a very different degeneracy structure to their eigenspectrum. Preliminary evidence suggests that QA can be much more efficient than classical solvers on hard satisfiability problems having many satisfying assignments \cite{KS:2010}. Additionally, recent results show that by tailoring the sweep rate of the interpolation between the initial Hamiltonian $H(0)$ and the problem Hamiltonian nonlinearly one can dramatically  increase the efficiency  of QA  even for classically hard optimization problems \cite{Farhi:2009,Dickson:2011}. Finally, the  complexity expression for the minimum required duration of QA based on the minimum energy gap is only relevant in the limit of very low temperatures (much smaller then the minimum  gap) and vanishingly small dissipation rates. A theoretical analysis of computational problems in the limit of large $N$ that includes finite noise and non-negligible dissipation has not been done to date. Overall,  the studies of the complexity of QA in the last decade have shown that it would be extremely difficult to  obtain reliable answers without actually  executing QA on real quantum hardware. With development of the D-Wave quantum annealing processors this has  now become possible for the first time. Recent studies, as described in the next section, based on the D-Wave quantum hardware indicate very promising results for QA  even in realistic regimes where noise and dissipation are significant.

\begin{figure}[!ht]
\begin{center}
\includegraphics[width=2.9in]{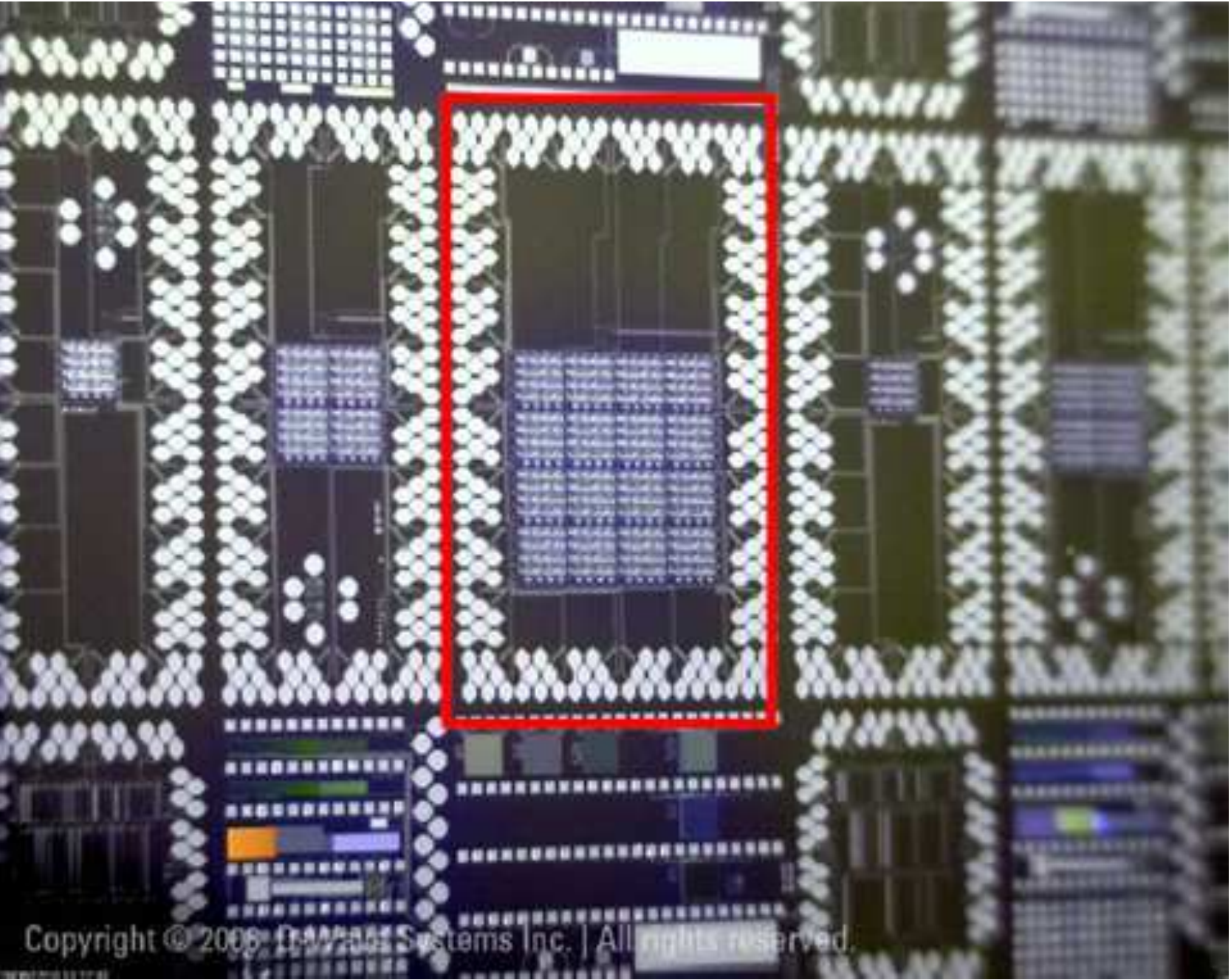}
\hspace{2cm}
\includegraphics[width=2.4in]{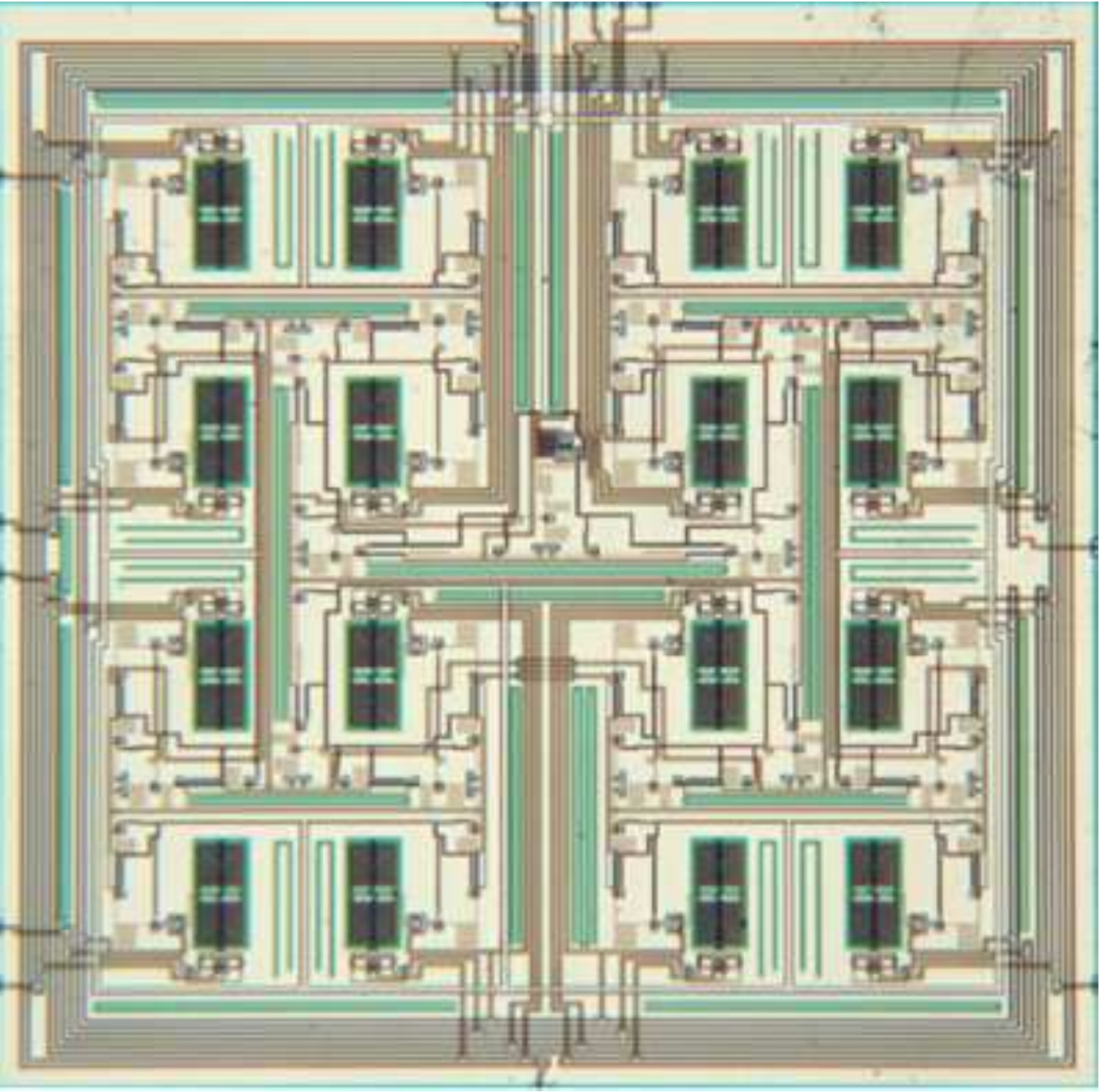}
\caption{Details of the hardware circuitry.
(\emph{Left figure.}). 128-qubit Processor:
This device contains 1,024 SFQ-based 2-stage digital-to-analog converters for controlling ultra-low noise analog circuitry comprising 128 qubits, and 256 couplers. It contains 24,000 Josephson junctions and 32,000 integrated resistors. This device is fabricated to commercial standards in a state-of-the-art semiconductor fabrication facility. (\emph{Right figure.}) Single Flux Quantum Logic: The figure shows a prototype 4-level single flux quantum logical binary de-multiplexing tree for efficient addressing. This device successfully routed 100 million flux quanta to precise qubit inductor locations throughout a test circuit without error.\label{fig:chip}
}
\end{center}
\end{figure}

\section{\label{sec:QA-super}  Superconducting Quantum Annealing Machine}

Our discussion of the practical implementation of  Quantum Annealing will be  substantially  different from conventional  quantum computer research. The latter is focused  on perfecting single qubits and elementary quantum gates. Instead, we will focus on the architecture currently advanced by  D-Wave Systems Inc.   that is designed ``top-down''. This mindset is centered on building a functional computer rather than single qubits or gates. This shift in thinking has proven to be powerful for it has forced focus on the end goal --- the performance of some computation --- rather than on the
fundamental condensed matter physics itself.

 The choice of superconducting chips and Josephson junction substrate to implement Quantum Annealing was made because commercial integrated circuit foundries can be used to construct high-quality chips, and the superconducting electronics manifesting the quantum behavior at low temperatures can easily be coupled to standard electronics.

\begin{figure}[!ht]
\centering
\subfigure[]{
\includegraphics[width=3in]{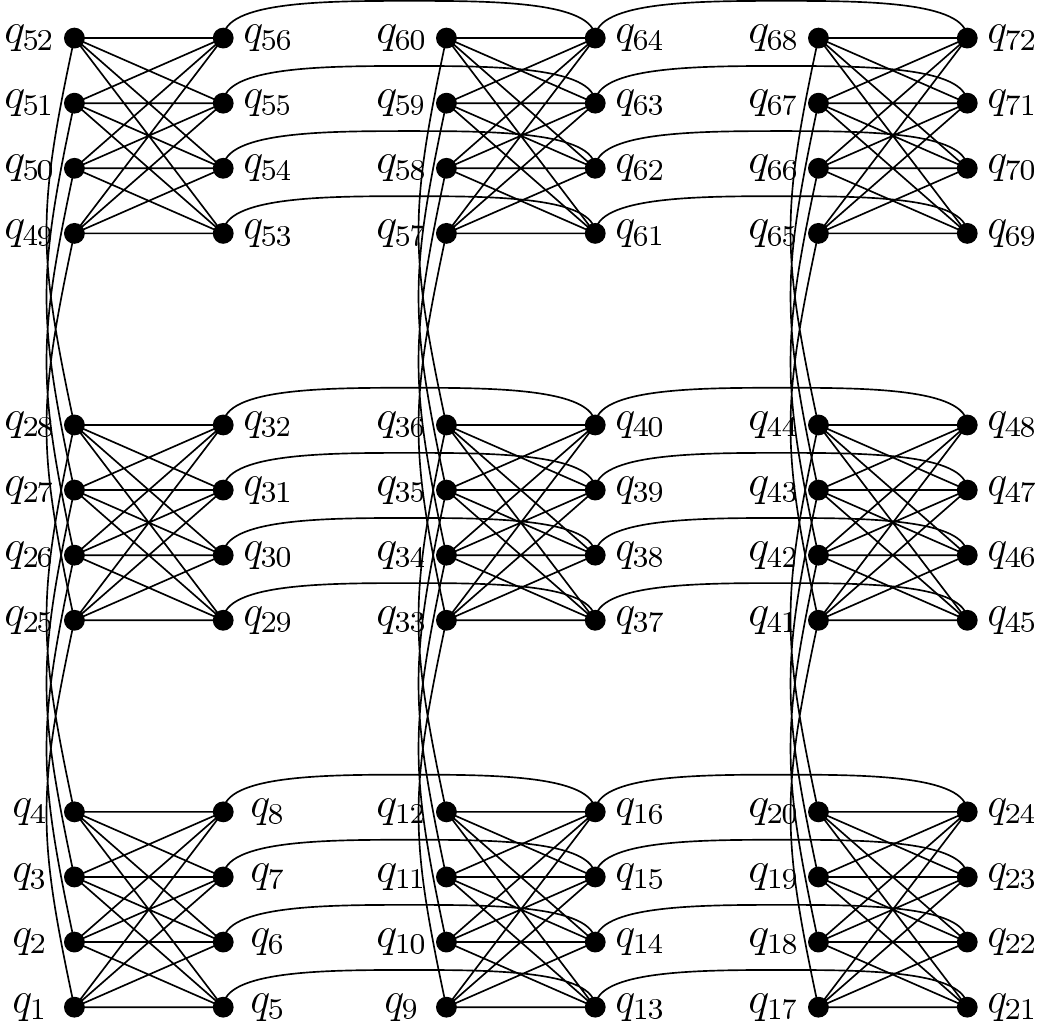}
\label{fig:architecture}
}\hspace{0.2in}
\subfigure[]{
\includegraphics[height=3in]{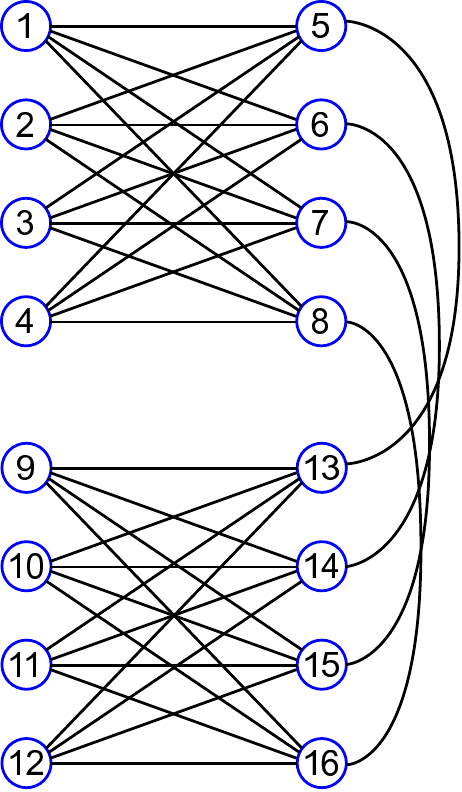}
\label{fig:16Q}
}
\label{fig:ARCHITECTURE}
\caption{  \subref{fig:architecture} The D-Wave One processor (Rainier) implements a fixed non-planar connection graph topology on 128 qubits.  Different problems are encoded by changing the coupling strengths $J_{ij}$ between qubits and the biases $h_i$ of each qubit.\subref{fig:16Q} Graph for 16-qubit Ising problem which is a sub-graph of the 128 node Rainier graph topology shown in the left figure. Instantaneous eigenvalues of control Hamiltonian $H(t)$ (\ref{eq:H-int}),(\ref{eq:DP}) for  this problem are simulated in Fig(\ref{fig:eigen}).}
\end{figure}

 The D-Wave One (Rainier) machine contains a processor with 128 qubits.  The red-boxed region of Fig \ref{fig:chip}(left) shows a chip consisting of a $4\times 4$ array of units cells with each unit cell consisting of 8 qubits. Individual qubits from the same cell and from different cells are coupled to each other according to a hardware graph shown in Fig.~\ref{fig:architecture}.   In operation, the chip is cooled in a dilution refrigerator to 20mK where quantum effects arise. The chip is programmed with room-temperature electronics which route currents to individually addressed qubits and couplers inducing small magnetic fluxes through the Josephson junctions. Signals are multiplexed on the chip to minimize external noise sources (see Fig. \ref{fig:chip}(right)). Further details describing the qubits and the couplers between qubits given in Hamiltonian (\ref{eq:Ising}) can  be found in \cite{Harris2010a} and \cite {Harris2009} respectively. The processor implements a quantum annealing algorithm ~\cite{Johnson2011}  as will be described below.
D-Wave's next generation processor (``Vesuvius'') to be released in Q4 2012 will have  4 times more units cells compared to Rainier (512 qubits). As discussed in Sec.~\ref{sec:energy-benchmarking} its  circuitry is re-designed to  effectively eliminate a cooling overhead.

 The D-Wave quantum annealing processor  solves discrete optimization problems of the  Ising (QUBO) form given in (\ref{eq:Ising}) where   Ising spins  $s_i=\pm 1$ are related to bits $z_i=0,1$ as $s_i=1-2z_i$.  The problem specification is given by a vector with elements $\{h_i\}_{i=1}^{N}$  and sparsely populated matrix of numbers, $\{J_{ij}\}_{i,j=1}^{N}$. Only those $J_{ij}$ allowed that belong to hardware graph shown in Fig.~\ref{fig:architecture}. The problem  is encoded in the Hamiltonian (\ref{eq:HP-Ising}). The solution state is uncovered as a result of QA process that is conceptually described in Sec.~\ref{sec:QA}. The specifics of D-Wave quantum annealing machine duty cycle is discussed below.

\subsection{\label{sec:duty-cycle} Duty Cycle}
\noindent
{\bf Step 1. Input stage\\} At the beginning of the processor's duty cycle there is an input stage when an Ising optimization problem (\ref{eq:Ising}), (\ref{eq:HP-Ising})  is loaded onto the quantum processor. The problem is  defined by the local biases, $\{h_i\}$, and the coupling biases, $\{J_{ij}\}$, which define a vector of numbers that are recorded in on-chip magnetic memory. During this input phase signals are sent down $\sim$ 100$+$ serial signal lines and then demultiplexed into the processor circuitry. The demultiplexing circuitry incorporates some shunting resistance which produces a small, but measurable, amount of heating in the processor circuitry ($\sim$ 1nW).

\hfill

\noindent
{\bf Step 2.  Cooling\\}
In the second phase of the processor's duty cycle a wait period of programmable length $\sim$ 1s is imposed that allows heat to dissipate from the processor circuitry into the cooling elements of the fridge.
 This  cooling wait period   is required every time new optimization problem parameters are  loaded into the magnetic memory.  After cooling  the problem can be solved and read out (Steps 3\&4) repeatedly (1000X typ.) while the processor circuitry remains in thermal equilibrium with the fridge. In D-Wave's next generation 512 qubit processor design (``Vesuvius''), the demultiplexing circuitry has been re-designed \textit{without} resistors, effectively eliminating cooling overhead from the processor duty cycle and the need for the wait time. For details  on residual energy dissipation see Sec.~\ref{sec:energy-benchmarking}.

\hfill

\noindent
{\bf Step 3. Processor state is initialized\\}
The tunnel barrier, shown in Fig.~\ref{fig:4},  is completely suppressed in each qubit. In this case the system's initial Hamiltonian is equal to $H_D$  (\ref{eq:D-Ising}) and each qubit independently relaxes into its ground state given by the symmetric superposition  $\frac{|0\rangle_i+|1\rangle_i}{\sqrt{2}}$. The  state of the quantum register is a fully symmetric product state  (\ref{eq:PhiS}) that includes all possible $2^N$ bit-assignment basis states $|{\bf z}\rangle$ (\ref{eq:Phi0}) and there is no barrier to tunnel between any of these states

 \hfill

 \noindent
{\bf Step 4. Quantum annealing\\}
Annealing occurs as  the  tunnel barrier of each qubit is gradually increased (tunneling matrix element $-\frac{T-t}{T}\Delta$ is gradually decreased), while the components of the Hamiltonian that encode the problem (those involving    $\frac{t}{T}h_i $ and $\frac{t}{T}J_{ij}$)  are increased in magnitude. Two important things happen simultaneously during annealing: (i) because the relative weight of $H_P$ in the control Hamiltonian (\ref{eq:H-int}) increases the states representing the lowest energy solutions to the optimization problem are progressively preferred,  and (ii) because the relative weight of $H_D$ in the control Hamiltonian (\ref{eq:H-int}) decreases the ease of tunneling between states is gradually diminished and eventually terminated.
    As the tunneling amplitudes approach zero the  state of the quantum register is dominated by the bit assignment that provides the solution of optimization problem.

\hfill

\begin{figure}[!ht]
\includegraphics[width=5.0in]{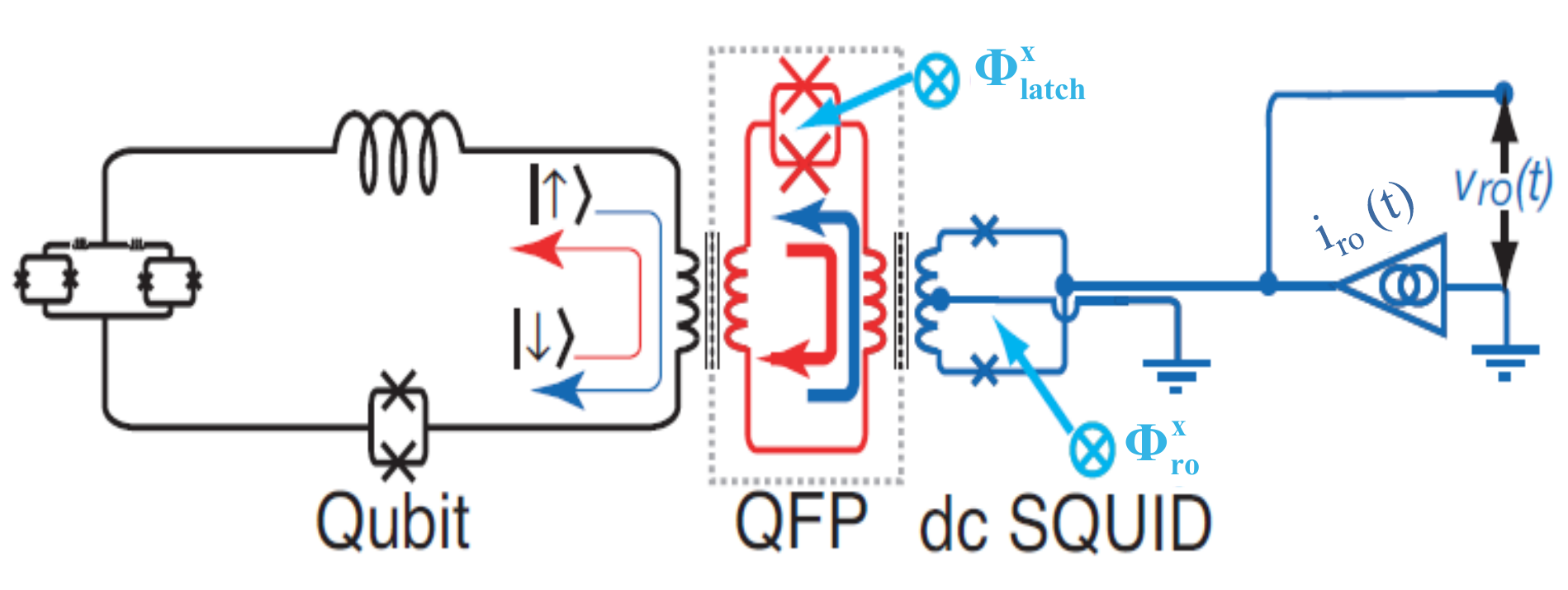}
  \caption{Schematic of readout circuitry. \label{fig:READOUT} }
\end{figure}

 \noindent
 {\bf Step 5. Readout\\} It follows from above  that  at the end  of the quantum annealing  every qubit is in  a classical state corresponding to its bit value  (0 or 1)  in the solution string $\{z_{1}^{*},z^{*}_{2},\ldots,z^{*}_{N}\}$.  Physically this bit value is represented by the direction of current in the qubit loop, either clockwise or counter-clockwise. This state is recorded by the Quantum Flux Parametron (QFP) when triggered by an external signal as illustrated in Fig.~\ref{fig:READOUT}. The QFP is subsequently readout with very high fidelity (in excess of 99.99\% in practice)  by a sensitive magnetometer -- a simple latching dc-SQUID. For full details of the readout process see \cite{A.J.Berkley2010}.
\newcolumntype{C}[1]{>{\centering\let\newline\\\arraybackslash\hspace{0pt}}m{#1}}
\renewcommand{\arraystretch}{1.7}
\begin{table}[!ht]
\begin{tabular}{| c |@{\hspace{0.15in}} C{4.5cm} @{\hspace{0.15in}} |@{\hspace{0.15in}} C{6.5cm} @{\hspace{0.15in}}|}
\hline
Duty Cycle Stage  &  Time for 128 qubit `Rainier'  processor [ms]   &  Projected Time for 512 qubit  `Vesuvius' processor (next gen) [ms] \\ \hline\hline
Input             & 143    & 33 \\ \hline
Cooling           & 1,000  & 1  \\ \hline
Annealing (x 1000)& 5      & 5  \\  \hline
Output (x 1000)   & 2,300  & 51  \\ \hline
Total (approx.)   & 3,450  & 90  \\ \hline
\end{tabular}
\caption{Duty cycle times for Rainier and projected timings for Vesuvius D-Wave quantum processors\label{tab:duty-cycle}}
\end{table}
\noindent
 Table ~\ref{tab:duty-cycle} shows  duty cycle times for  D-Wave's Rainier processor and projected timings for the next processor generation, code-named 'Vesuvius'. In  'Vesuvius', the processor circuitry has been scaled up by 4 times and redesigned to eliminate heating and accelerate the timing stages that surround the core quantum annealing step.

 Cooling times are chosen to be sufficient to return the processor circuitry to equilibrium with base operating temperature. They are calibrated using a direct measure of electron temperature in the qubit circuitry fit to a closed form expression. Full details are available in Sec.~II.D of the supplementary materials  \cite{Johnson2011a}.

\section{Benchmarking Studies of the D-Wave Processor}

\subsection{\label{sec:energy-benchmarking} Energy dissipation}
\begin{figure}[t]
 \includegraphics[width=6in]{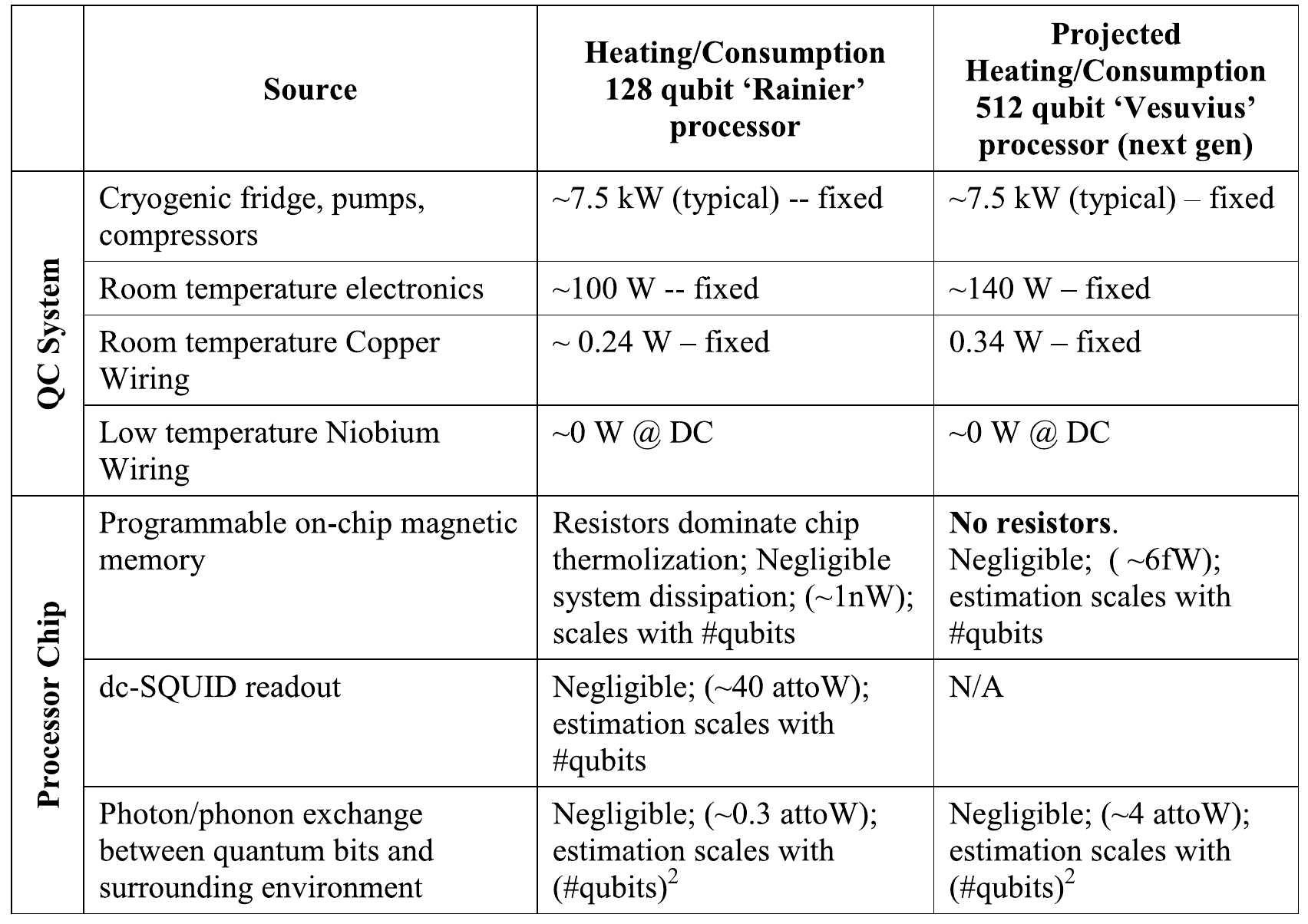}
  \caption{Summary of energy consumption and dissipation sources in D-Wave Quantum Computers. Where measured values were unavailable, estimates of energy dissipation were averaged over problem-solving duty cycles for an estimate of dissipated power. \label{fig:heat-specs} }
\end{figure}

The D-Wave quantum processor circuitry is fabricated out of superconducting metals which form the basis of the quantum bits. A well  known advantage to superconducting circuitry is its complete lack of electrical resistance when operated below its critical temperature. This advantage has been maximized with careful engineering so that the bottom half of the system wiring, the cryogenic filtering, chip packaging, and wirebonds are all superconducting. The use of superconductors in the wiring and filters closest to the chip helps us to keep the chip cold in spite of the more than 100 signal lines physically contacting it, and allows the creation of a quiet environment in which it is possible  to harness the quantum mechanical resources necessary for the quantum annealing algorithm. The stratification of heat generating sources on the chip is  described in Fig.~\ref{fig:heat-specs}
\noindent
With many of the modes of electrical dissipation operating near fundamental physical limits the Quantum Computer's energy consumption is dominated by the cryogenic refrigeration. That is to say that the wall power required is, for practical purposes, flat with respect to processor growth.
  As mentioned above in Sec.~\ref{sec:QA-super}, in the  current 128 qubit D-Wave One (``Rainier'') processor design there is some heating that occurs on-chip. This small heating of $\sim$ 1nW (see 6$^{\rm th}$ row of the Fig.~\ref{fig:heat-specs}) is produced by resistors in the demultiplexing circuitry during the processor's input programming. The resulting cooling dominates the timing of the processor duty cycle ($\sim $ 1s). In D-Wave's next generation processor (``Vesuvius'') with 512 qubits, the demultiplexing circuitry has been re-designed without resistors, effectively eliminating cooling overhead from the processor duty cycle down to about 6fW with cooling time of about 1$ms$.

\subsection[Processor speed]{\label{sec:speed}Processor Speed}

\begin{figure}[!ht]
\includegraphics[width=6.0in]{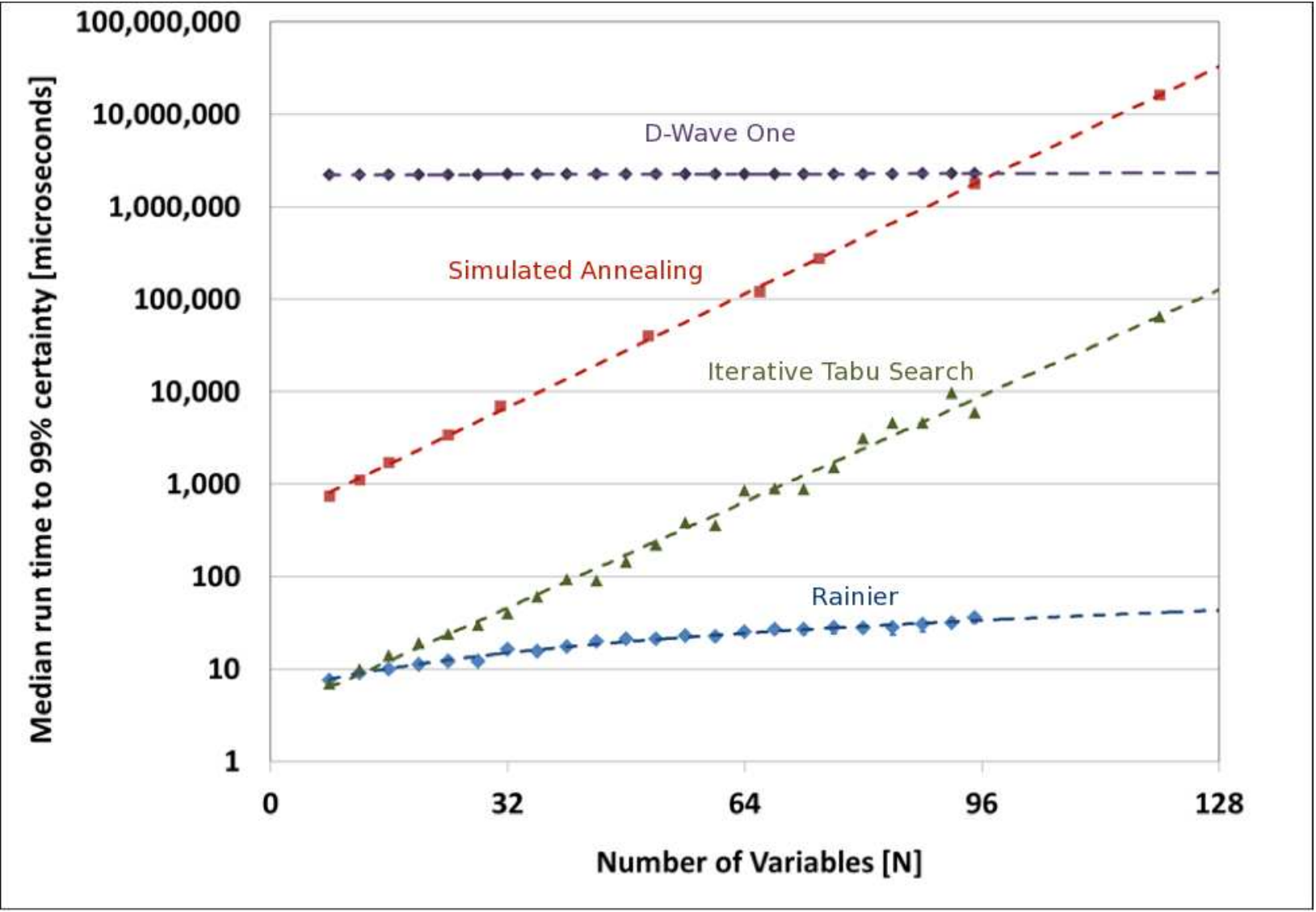}
\caption{Benchmark performance as time to fixed accuracy versus the number of variables in the candidate problems. Core computing time reported as the accumulated time spent in repeated quantum annealing steps (blue), classical heuristic approaches Iterative Tabu Search (ITS - green) and Simulated Annealing (SA - red) and total wall-clock time for the Quantum Computing system with all overhead (purple).
The runtime of D-Wave One is dominated by very long cooling wait time.
Runtimes of simulated annealing increase exponentially with $N$, in proportion to $\exp(0.088 N)$ and $\exp(0.083 N)$ respectively.
Computing time of the D-Wave One (Rainer) machine is consistent with the linear fit $5+0.29 N$ shown, or with an exponential scaling of the runtime exponent $\sim 8 \exp(0.017 N)$. How it scales as the number of qubits increases remains an open question.
\label{fig:performance-bench} }
\end{figure}

To benchmark the performance scaling capabilities of the D-Wave One processor, or any other heuristic solver capable of discrete optimization, a large set of randomly generated optimization problems (\ref{eq:Ising}) is generated. For each problem values of $\{h_i\}$ and  $\{J_{i,j}\}$ in (\ref{eq:Ising}) are generated randomly with uniform distribution from the set of numbers $[-1,-2/3, -1/3,0,1/3,2/3, 1]$. Problems were generated for various numbers of discrete variables: $N= \{8,12,16,\ldots,96\}$  and then optimized repeatedly. The problem sizes are still small enough so that
for each problem an exact solution was found using commercial Integer Programming solver CPLEX. The metric for performance was the median number of optimization iterations required to achieve a 99$\%$ accuracy for a given number of variables, $N$.
\begin{figure}[!ht]
\includegraphics[width=5.0in]{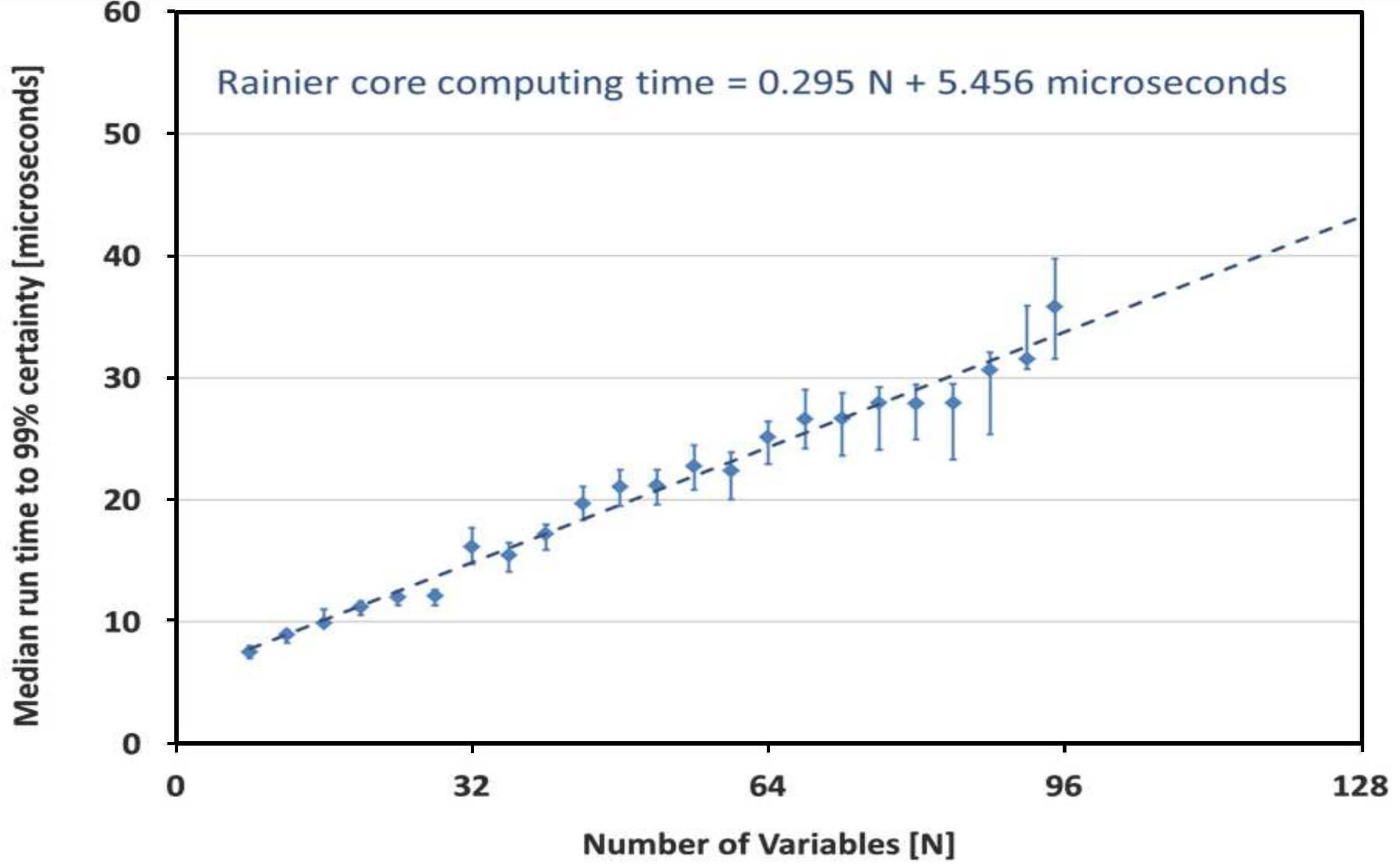}
\caption{Figure shows the D-Wave one (Rainier) core computing time from the benchmark study with a linear fit (cf. also with Fig.~13). Lower  and upper \lq\lq error bars" show the 40$^{\rm th}$ and 60$^{\rm th}$  percentiles respectively.
 \label{fig:performance-bench-core} }
\end{figure}
\noindent
The results of the processor were compared to a workhorse heuristic public algorithm (Iterative Tabu Search-ITS) for optimizing this type of discrete optimization problem. The results are given in Fig.~\ref{fig:performance-bench}.
They  show that the median time to fixed accuracy scaled exponentially with problem size for the classical solvers, ITS and Simulated Annealing (SA). For the Rainier quantum processor, the dark blue curve in  Fig.~\ref{fig:performance-bench} corresponds to the median time dependence on the problem size.
This time refers to the \lq\lq wall-clock time" of the \emph{full} QA duty cycle, including all its stages described in 
Sec.~\ref{sec:duty-cycle}.
The light blue curve  at the bottom of the plot shows results for the core quantum annealing stage of the duty cycle with $N$. This time is much faster because it does not involve a long cooling wait time following the parameter input stage. As we discussed in Secs.~\ref{sec:duty-cycle} and \ref{sec:energy-benchmarking} this cooling time is being dramatically shortened by changing the circuitry in the the next-generation "Vesuvius" 512 qubit processor (see also Table~\ref{tab:duty-cycle}). Therefore the dependence  of core annealing time on the problem size is of the  most interest. It is given in Fig.~\ref{fig:performance-bench-core}.
For additional performance results comparing runs on the D-Wave
processor to classical algorithms, see \cite{Karimi}.

We now turn to a series of combinatorial optimization problems, grounded
in NASA applications, for which quantum annealing approaches can be
constructed.
The strategy in all cases is to phrase the problem as a Quadratic
Unconstrained Binary Optimization (QUBO) problem which can then
be translated into a problem  Hamiltonian $H_P$ (\ref{eq:HP-Ising})
that can be implemented on quantum annealing hardware for Ising models.

\section[Classification for
 Feature Identification]{\label{sec:classification} Classification for Planetary
 Feature Identification}

\begin{figure}
\centering
\includegraphics[width=0.5\textwidth]{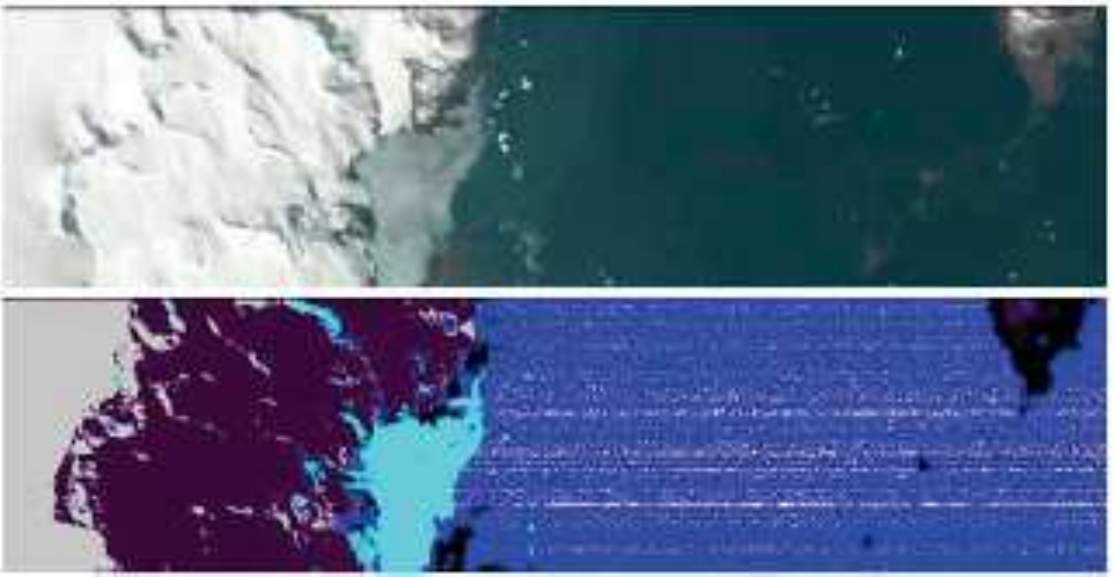}
\caption{Automatic classification can classify pixels in images such
as the upper image. The classification can then be visualized
by coloring the pixels according to their classification: water (blue),
land (black), ice (cyan), snow (purple), cloud (gray), unknown (white).}
\label{fig:iceClass}
\end{figure}

This section describes a quantum classification algorithm useful
for planetary feature identification.
Classification supports identification and characterization of a
myriad types of remotely sensed objects, from rocks
and craters on Mars to pulsars and other fast transients.
Figure \ref{fig:iceClass} shows classification of different regions
of an image as water, land, ice, snow, cloud, or unknown.
Terrain classification is critical to
hazard detection and autonomous landing \cite{Brady,Epp,Epp2,Johnson,Rohr}.
Classification of both directly and remotely sensed objects could
be improved from its current state. As one example, currently
identification of rocks and craters is done by human catalogers,
in spite of efforts to develop automated solutions
\cite{Stepinski,Thompson,Urbach}.

\begin{figure}
\centering
\includegraphics[width=0.5\textwidth]{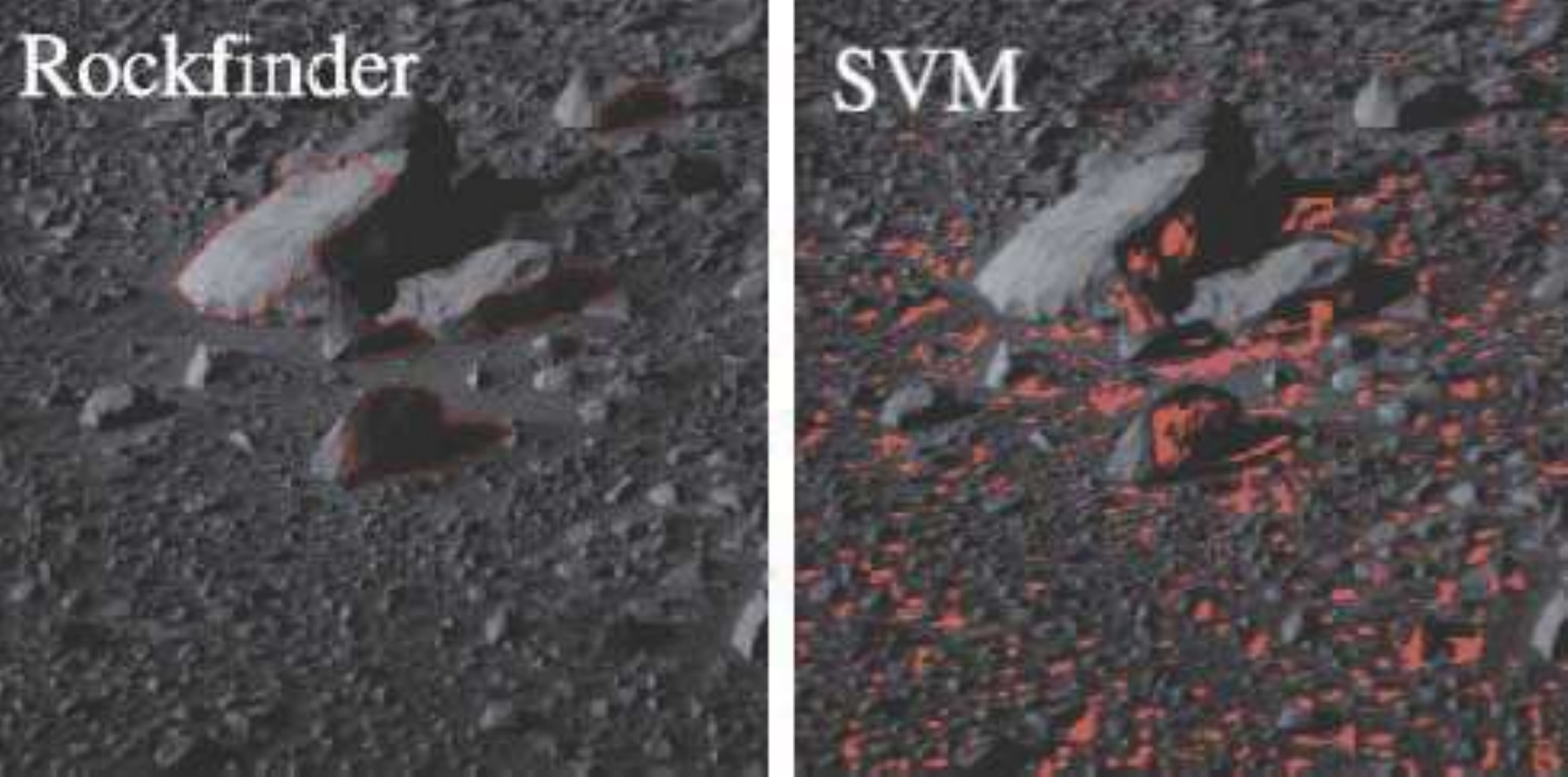}
\caption{Results of two rock finding algorithms on a Mars Exploration Rover
surface image, from \cite{Thompson}. The two algorithms find different rocks,
with the Rockfinder algorithm finding all large rocks and the
support vector machine (SVM) algorithm finding
the many small rocks in the image.}
\label{fig:rocks}
\end{figure}

Boosting provides a way to combine a diverse set of weak classifiers
to obtain a strong classifier.
Existing classification algorithms for rock identification, for example,
have different strengths and weaknesses;
some perform better on larger rocks, for example, while others
are excellent for distinguishing smaller rocks (Figure \ref{fig:rocks})
\cite{Thompson,Shang}. We first describe traditional
boosting, and then QBoost \cite{Neven09}, a quantum approach to boosting.
Boosting, and therefore QBoost, can also be applied in conjunction
with other machine learning techniques, but here we focus attention
on classification.

\subsection[Boosting]{\label{sec:boost} Boosting for Binary Classification}

\begin{figure}
\centering
\includegraphics[width=2.0in]{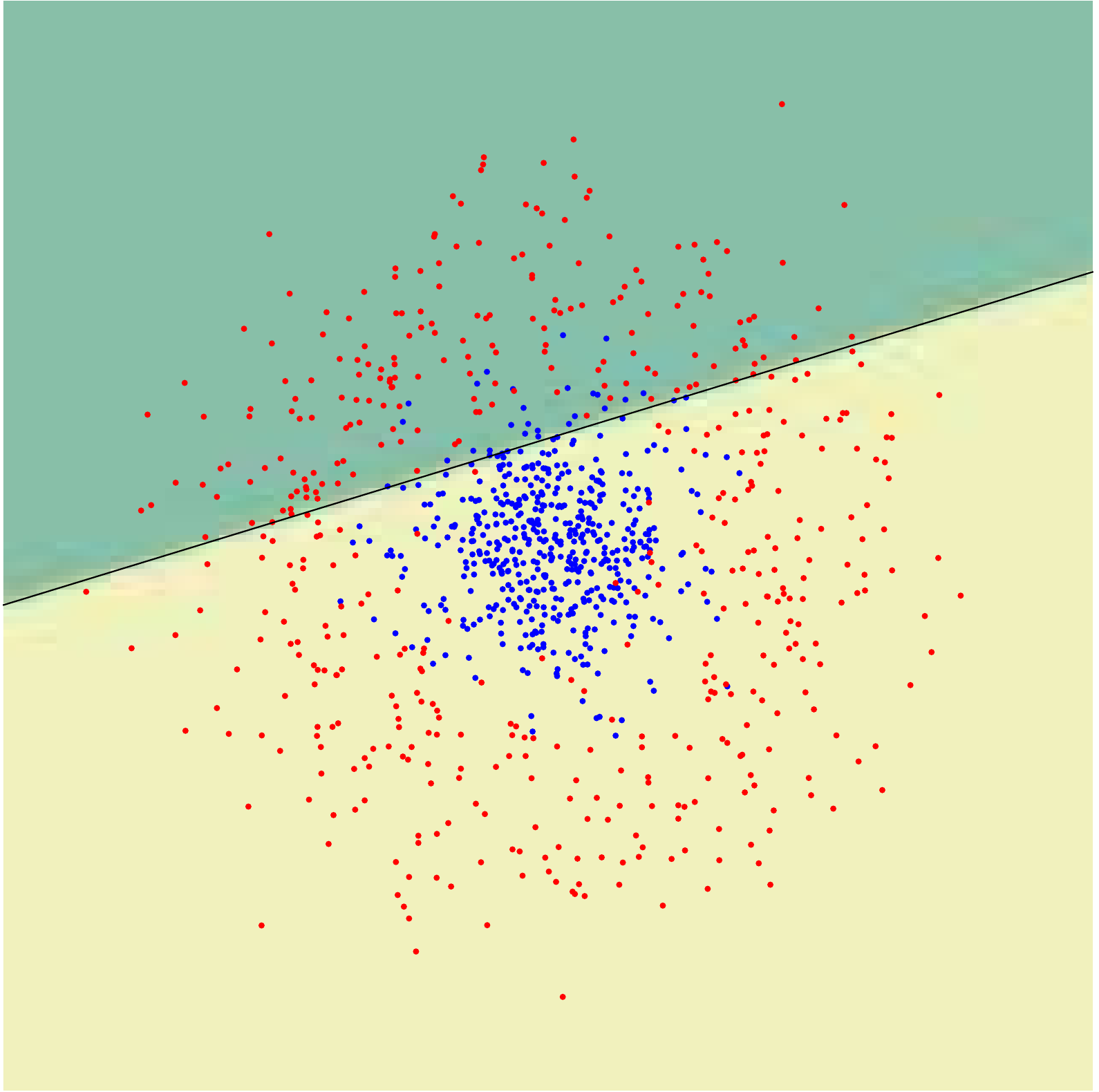}
\includegraphics[width=2.0in]{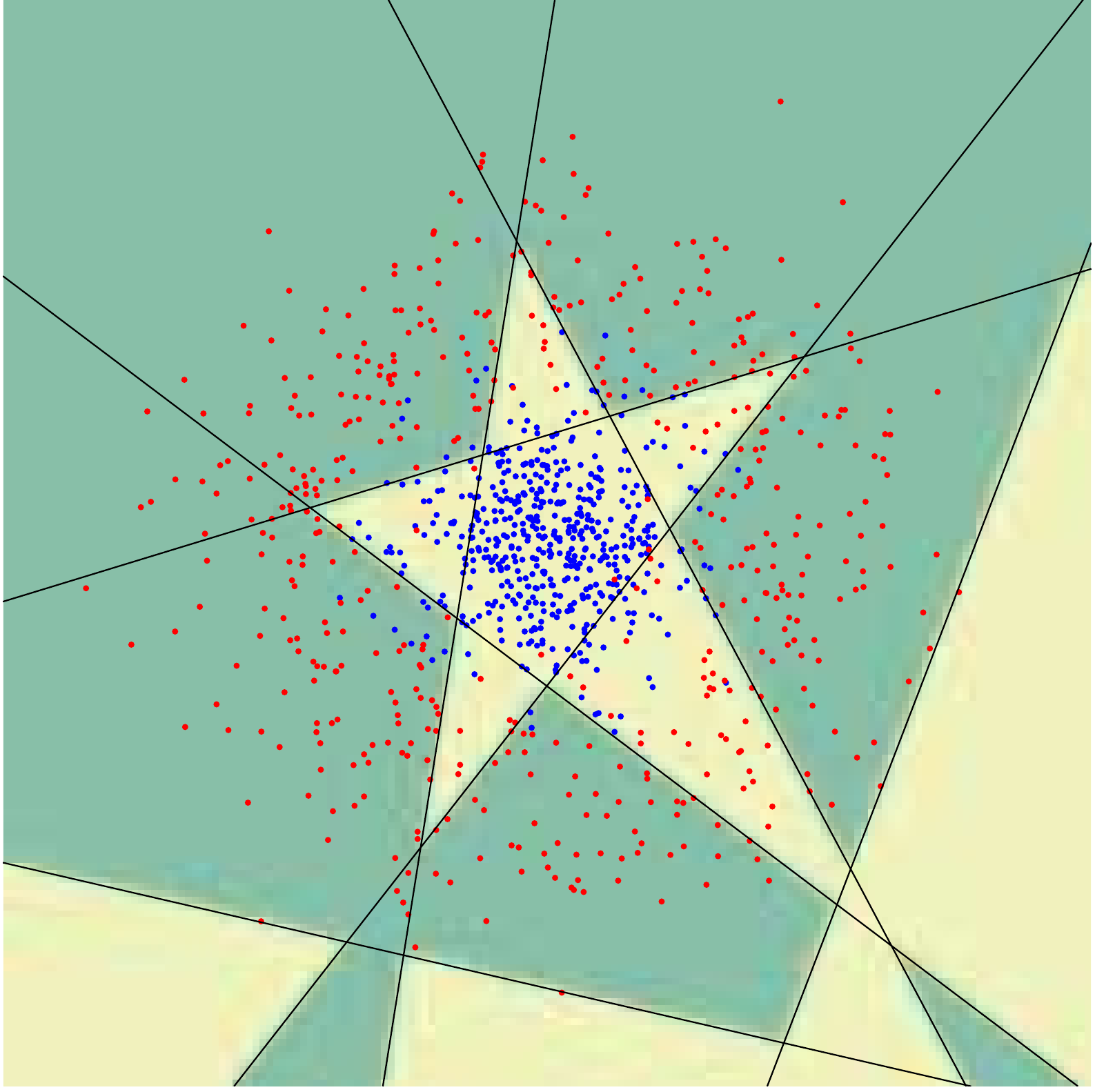}
\includegraphics[width=2.0in]{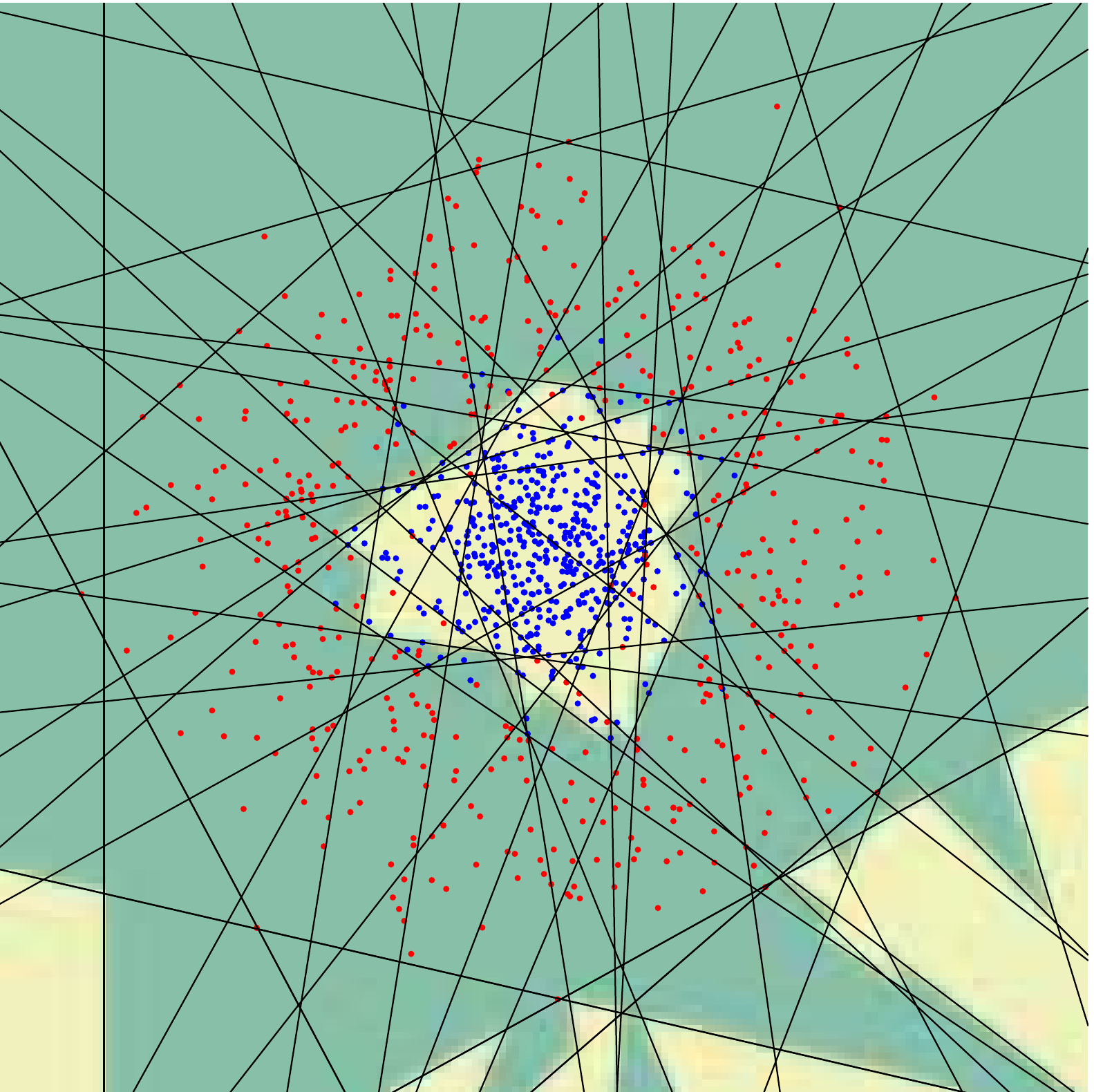}
\caption{How a collections of weak classifiers can form a strong classifier.
The image on the left shows a single weak classifier. It does better than
chance in classifying the points, but does not do a good job. The middle
figure shows a small number of weak classifiers, together, do a better job.
The figure on the right show how a large number of weak classifiers combine
to give a strong classifier with high accuracy. Figures are from Jan
Sochman's AdaBoost talk \cite{Sochman}. }
\label{fig:adaboost}
\end{figure}

The motivation behind boosting is that it is relatively easy to
generate automatic methods that can do better than chance when asked
to classify an object such as a rock, but that it is hard to design
automatic methods that have high accuracy. Starting with a large
number of ``weak classifiers,'' that work moderately better than chance, boosting creates a ``strong classifier'' that has higher accuracy
(see Figure \ref{fig:adaboost}) \cite{Meir,Schapire}.
One way to create a ``strong classifier'' is to poll all of the weak
classifiers and weigh the answers in some way to make a classification
decision. Boosting is an iterative algorithm that, starting from labeled
training data, determines a good weighting for the weak classifiers.
There are many different boosting algorithms. Here, we consider a
particularly simple case of boosting in which
the algorithm makes a binary YES/NO decision as to whether to poll
a classifier at all, and then simply takes the majority vote of the
polled classifiers as the final classification.
To further simplify the exposition, we assume that the classifiers
themselves are binary classifiers, labeling each point with one of two
labels, such as ``is a rock'' or ``is not a rock.'' The same
approach to boosting, choosing a set of weak classifiers and then
taking a majority vote, can also be applied to non-binary,
multiple labels cases in a similar way.

We consider binary classification problems such as ``Is this object a rock?
YES or NO?''
A binary classifier is a function that assigns a label of $1$ or $-1$
to an input data vector $\vec{x}$ depending on its membership in some set $A$.
Boosting, like all supervised machine learning algorithms,
requires labeled training data in order to learn.
In our case that means that the algorithm has access to a set of data
consisting of data vectors, $\vec{x}$, corresponding to objects,
together with a label $1$ or $-1$ indicating whether, YES, it is a rock,
or NO it is not.

Weak classifiers are generally simple and very fast to compute.
As a result, the strong classifier, which simply polls a subset
of the classifiers and takes a majority vote,
can compute the classification of new data points efficiently,
as long as the subset of chosen classifiers is not too large.
How the polled weak classifiers are chosen varies from boosting
algorithm to boosting algorithm. We turn now to QBoost,
a quantum approach to finding an optimal choice of
weak classifiers to include.

\subsection[QBoost]{\label{sec:Qboost} QBoost: Mapping Boosting onto Quantum Annealing}

Neven et al.'s QBoost algorithm generates a binary classifier through
boosting, that is, combining the output of many ``weak classifiers''
each of which exhibit lower than desired accuracy into a ``strong classifier''
which is more accurate than any of its parts alone.
The goal is to output a subset
of ``weak classifiers'' that, when polled, and a majority vote is taken,
give the correct label with high probability.
Our description of QBoost follows that of Pudenz and Lidar \cite{Pudenz}.

The input set of weak clasifiers are of the form
\begin{equation}
\label{bin_classifier}
h(\vec{x})=\begin{cases}
+1 & \vec{x}\in A\\
-1 & \vec{x}\notin A.
\end{cases}
\end{equation}
QBoost determines a weight of $0$ or $1$ to give to each of the $N$
weak classifiers in its ``dictionary.'' Each weak classifier is simply
a binary classifier with its weight normalized to the size of the
dictionary, $h_i\in\{-1/N,1/N\}$. QBoost outputs a strong classifier
that classifies input by taking a majority vote over the weak
classifiers assigned weight $1$ by the algorithm.
Let $\vec{z} = (z_1, \dots, z_N) \in \{0,1\}^N$ be the set of
weights. The strong classifier is then
\begin{equation}
Q_{\vec{z}}(\vec{x})
=\mathrm{sign}\left(\sum_{i=1}^{N}z_{i}h_{i}(\vec{x})\right)
\in [-1,1].
\end{equation}

The weights will be optimized using quantum annealing. We wish to
find weights that optimize the performance of the resulting strong
classifier on the training data $\vec{x_s}\in \cal{S}$
and, at the same time, encourage few
positive weights in order to avoid overfitting and to keep the computation
of the strong classifier very fast.
Let $\vec{y}\in\{-1,1\}^S$  be a $S$-dimensional vector consisting
of the correct label for each of the training data points
$\vec{x_s}\in \cal{S}$, where $S$ is the size of the training set.
Let $\vec{R}_{\vec{z}}$ be the dimension $S$ vector consisting of
the labels estimated by a strong classifier with weight vector
$\vec{z}$, where $z_j$ is the weight given to classifier $h_j$
on each of the training data points $x_s$.
We wish to minimize the distance between $\vec{y}$ and $\vec{R}_{\vec{z}}$:
\begin{align}
\delta(z)  &  =\| \vec{y}-\vec{R}_{\vec{z}}\| ^{2}=\sum
_{s=1}^{S}\left\vert y_{s}-\sum_{i=1}^{N}z_{i}h_{i}(x_{s})\right\vert
^{2}\nonumber\\
&  =\| \vec{y}\| ^{2}+\sum_{i,j=1}^{N}C_{ij}^{\prime}%
z_{i}z_{j}-2\sum_{i=1}^{N}C_{iy}^{\prime}z_{i},
\end{align}
where%
\begin{align}
C_{ij}^{\prime}&=\vec{h}_{i}\cdot\vec{h}_{j}=\sum_{s=1}^{S}h_{i}(x_{s}%
)h_{j}(x_{s}),\\
\quad C_{iy}^{\prime}&=\vec{h}_{i}\cdot\vec{y}=\sum_{s=1}%
^{S}h_{i}(x_{s})y_{s}.%
\end{align}
We also add a ``sparsity term'' $2\sum_{i=1}^{N}\lambda z_i$
where the parameter $\lambda$ enables a choice as to how to balance the
desire for accuracy, by minimizing the distance, with the
desire for sparsity, by minimizing the sparsity term \cite{Blumer}.
We wish to find the set of weights binary weights $z_i$ that minimize the
the total cost function
\begin{align}
E(\mathbf{z})  =   \sum_{i,j=1}^{N}C_{ij}^{\prime}z_{i}z_{j}%
+2\sum_{i=1}^{N}(\lambda-C_{iy}^{\prime})z_{i}.   \label{QUBO}%
\end{align}
\noindent
Eq.~\ref{QUBO} is a quadratic binary optimization (QUBO) problem
that can be related to the Ising model \ref{eq:Ising} in a straightforward way.
To do so we need to work with symmetric binary variables (Ising spins)
with range $s\in \{-1,1\}$, not $\{0,1\}$. Let
$s_{i}=2(z_{i}-1/2)\in\{-1,1\}$. In terms of Ising spins, the
cost function  is
\begin{align}
E_{\rm Ising}(s_1,s_2,\ldots, s_N) = \sum_{i,j=1}^{N}C_{ij}s_{i}s_{j}+\sum_{i=1}%
^{N}(\lambda-C_{iy})s_{i}  , \label{q_opt}%
\end{align}
where
\begin{equation}
C_{ij}=\frac{1}{4}C_{ij}^{\prime},\quad C_{iy}=C_{iy}^{\prime}-\frac{1}{2}%
\sum_{j=1}^{N}C_{ij}^{\prime}. \label{Cij}%
\end{equation}
The problem Hamiltonian encoding the quantum weight-learning problem is
\begin{equation}
H_{P}=\sum_{i,j=1}^{N}C_{ij}\hat Z_{i}\hat Z_{j}+\sum_{i=1}^{N}(\lambda-C_{iy})\hat Z_{i},
\label{Hf}
\end{equation}
where $\hat Z_{i}$ is the Pauli spin-matrix  acting on the $i$th qubit.
This Hamiltonian is a particular instance of the Hamiltonian given by
Eq.~(\ref{eq:HP-Ising}) of Sec.~\ref{sec:P}. Its ground state can be sought
using the QA procedure described in Secs.~\ref{sec:QA}.
When the non-zero terms of this Hamiltonian match the edge connections
in the D-Wave Ising hardware graph, the ground state can be sought using
the current hardware as described in Sec.~\ref{sec:QA-super}.

\begin{figure}
\centering
\includegraphics[width=0.7\textwidth]{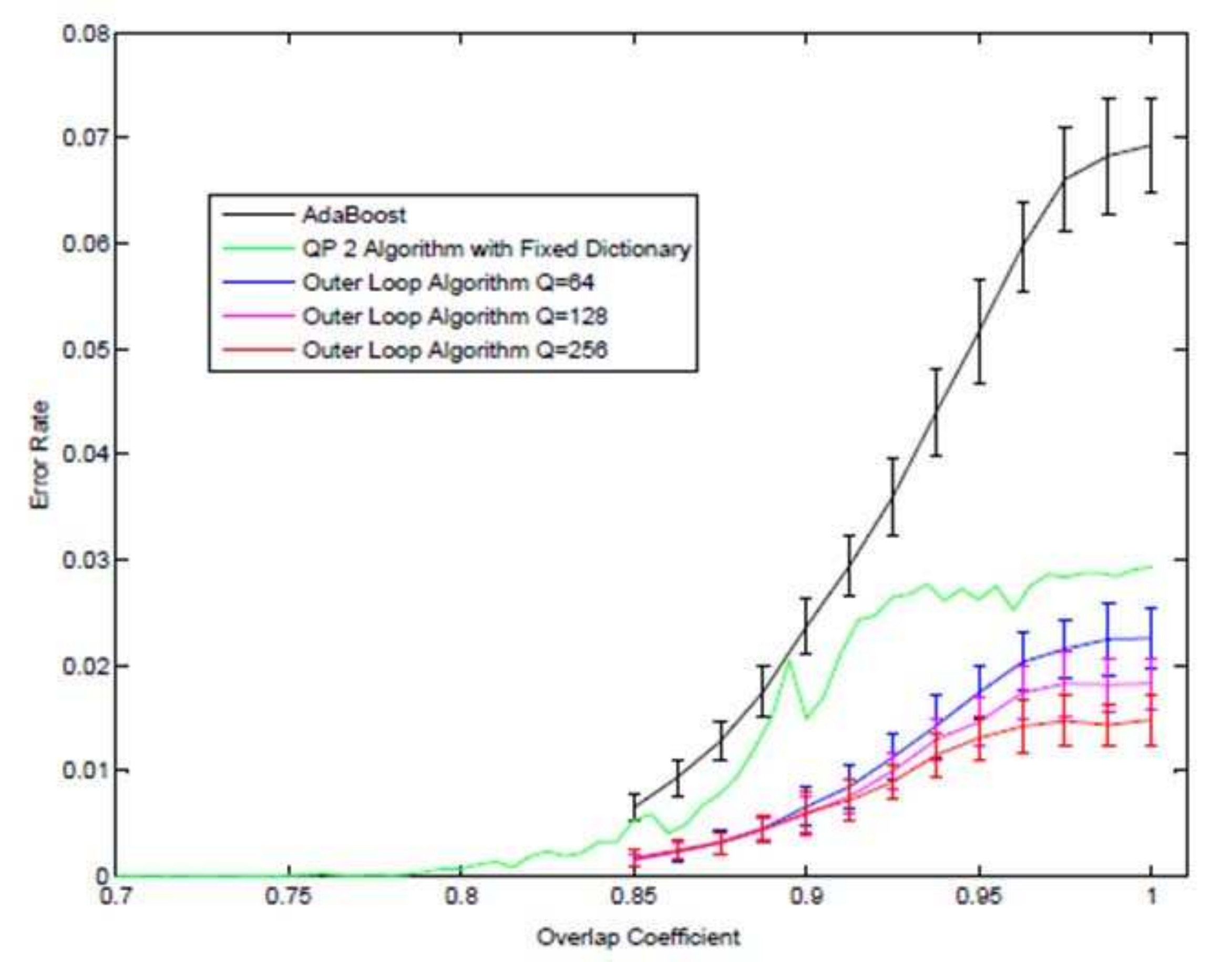}
\caption{Results of QBoost on a synthetic dataset with increasing overlap \cite{Neven09}. All variations of QBoost attempted consistently outperformed the popular AdaBoost classical machine learning algorithm in accuracy.}
\label{fig:QBoost}
\end{figure}

QBoost has outperformed in accuracy classical boosting methods for 
classification in numerical studies \cite{Neven09} (Figure \ref{fig:QBoost})
and for detection of cars in an image \cite{Neven09demo}.
Care should be taken in interpreting these results. While the accuracy is
better than that of Adaboost, the runtime is not. It takes longer just
to set up the problem prior to the quantum annealing step than it does
to run the Adaboost algorithm; 
computing the coefficients $C_{ij}$ takes $O(SN^2)$ where $N$ is
the number of weak classifiers and $S$ is the number of test samples, whereas
the entire Adaboost training runs in $O(TSN)$ where $T$ is the number of
iterations or, equivalently, the number of weak classifiers included
in the strong classifier, which means $T << N$. This issue, and the 
difficulty of analyzing the complexity of quantum annealing as discussed
in Sec.~\ref{sec:mingap}, means that it remains a tantalizing 
research question as to how to determine the reasons for the improved 
accuracy, with possible answers ranging from essentially classical 
differences between the algorithms to purely quantum properties of 
the QBoost algorithm.

\section{Clustering for Pattern Recognition}

\begin{figure}
\begin{center}
\includegraphics[width=3.0in]{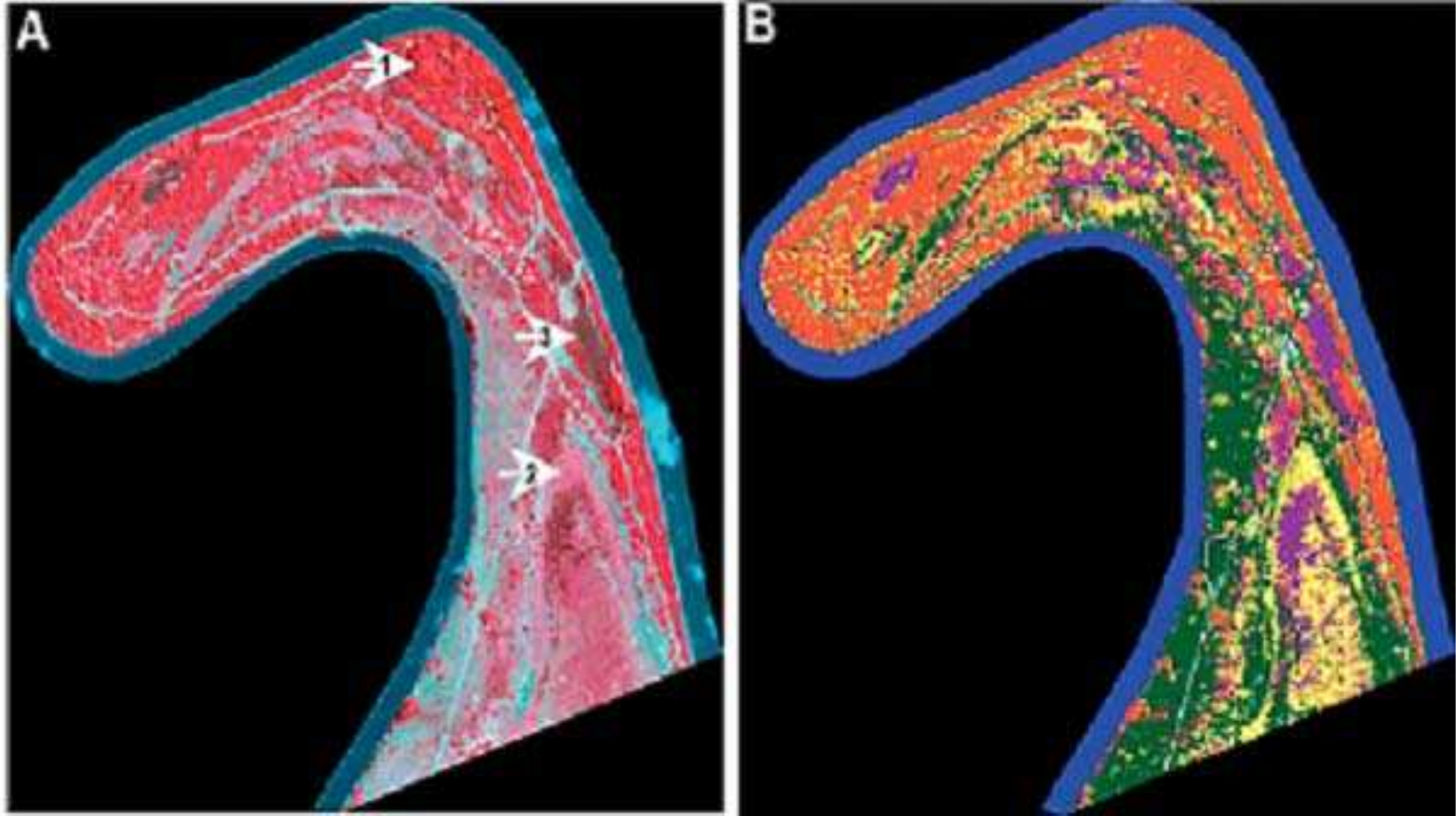}
\end{center}
\caption{Two false color segmentations of a satellite image
\cite{Barra09}.
The one on the left classifies pixels into three classes:
riparian woodland, green herbaceous vegetation, and spiny aster.
On the right, clustering was used to segment the image, giving a
more detailed view. The six clusters correspond to the three classes
used riparian woodland, green herbaceous vegetation, and spiny aster,
plus three new classes corresponding to
stressed herbaceous vegetation, sparsely vegetated/bare soil, and water.}
\label{fig:unsupervisedSeg}
\end{figure}

Much of the data NASA receives, whether from astronomical instruments,
autonomous vehicles, or space station sensors, comes in unstructured
form, without any labels, or even known classes.
Supervised machine learning techniques, such as the classification techniques
discussed in Sec.~\ref{sec:classification} and Sec.~\ref{sec:strLearning}
cannot be applied when there is
no labeled training data. Instead, this type of problem must be
attacked by unsupervised machine learning techniques that learn without having
access to labeled data. In unsupervised machine learning,
the aim is to learn something about the structure of the data.
Instead of classifying the data by assigning labels, unsupervised
algorithms finds clusters, for example, within the data. New data will be
associated with the closest cluster.
Clustering is often a first, critical step in
extracting scientific or engineering information from
unstructured and unlabeled data.
Clustering partitions the data space into usable
blocks or chunks of similar data.
Representative samples from each cluster can then be analyzed by
human experts or fed to an automated analysis tool.
Even when classes are known, clustering can provide additional
information beyond classification.
Figure \ref{fig:unsupervisedSeg} shows two false
color segmentations of a satellite image. The one obtained
using clustering provides a more detailed picture, with six classes
instead of the three obtained using classification.

Clustering has a long history of use at NASA. As early as 1973,
NASA had developed an unsupervised clustering program to automatically
extract features from remotely sensed data \cite{Jayroe73},
find patterns in ``In-Close Approach Change'' (ICAC) reports
\cite{Statler03}\cite{ICACslides},
to cluster trajectories for airspace monitoring \cite{Gariel10},
to identify recurring anomalies \cite{ReADS} or lessons learned \cite{NEN},
and to segment a variety of images with unknown features, including
hyperspectral images \cite{HiiHat}
and extreme ultraviolet images \cite{Barra09}.

\subsection[MAXCUT clustering algorithm]{The MAXCUT clustering algorithm}

There are numerous clustering methods, with different measures
of what makes a good clustering.
For many measures, finding the optimal cluster is an NP-hard problem.
For this reason, there are many different clustering algorithms with
different tradeoffs with respect to efficiency, approximation, and
generalizability, some aimed at specific applications.
In this section, we give an example using one measure of a good
cluster, and show how finding the optimal clustering according to
this measure can be phrased as binary optimization problem
that can be translated into a Hamiltonian so that quantum annealing
can be applied. One area for future work is to investigate
quantum analogs of other clustering algorithms, or to invent quantum
clustering algorithms with no traditional analog.

We start by examining how best to partition a data set into two
clusters. One measure of a good partitioning is how far
apart the data points grouped together in one cluster are from each other.
One approach to clustering is ``maxcut'' partitioning of the training
data, viewed as points in generally high dimensional space.
In other words, we wish to minimize the intracluster distances:
we want to find clusters $C_0$ and $C_1$ that minimize
\begin{equation}
\sum_{c,c'\in C_0}d(c, c') + \sum_{c_1,c_1'\in C_1}d(c_1, c_1'),
\end{equation}
where the first sum is over pairs of data points $c$ and $c'$ in $C_0$,
the second sum is over pairs of data points $c_1$ and $c_1'$ in $C_1$,
and $d(c,c')$ is the distance between the points $c$ and $c'$.
This minimization is equivalent to the MAXCUT problem which asks for
the clusters $C_0$ and $C_1$ that maximize the intercluster distances
\begin{equation}
\sum_{c_0\in C_0,\,c_1\in C_1}d(c_0, c_1),
\end{equation}
where, in this case, the sum is over pairs of points $c_0$ and $c_1$
which are in $C_0$ and $C_1$ respectively.
The MAXCUT problem is known to be NP-complete, so practical instances are
attacked through a variety of heuristic and approximation techniques.
An interesting open question is whether quantum algorithms can improve
on existing classical techniques.

\subsection[QCut algorithm]{\label{sec:Qcut} QCut: Mapping Clustering onto Quantum Annealing}

One quantum approach is to write MAXCUT as a binary
optimization problem that can then be translated to the quantum setting
where it can be attacked by quantum annealing.
Let $\{x_i\}$ be the set of all training points.
The MAXCUT problem can be written as a binary optimization problem with
respect to binary membership variables $z_i$, which indicate which
cluster the point $x_i$ is placed in:
\begin{equation}
z_i = \left\{\begin{array}{c}
1 \mbox{ if } x_i\in C_1 \\
0 \mbox{ if } x_i\in C_0
\end{array}.\right.
\end{equation}
The following cost function in terms of the binary variables
$z_i$ is  a QUBO problem
\begin{equation}
E({\bf z}) = -\sum_{i,j} C_{ij} z_i(1-z_j),
\end{equation}
where each coefficient $C_{ij}$ is the distance between the
training points $x_i$ and $x_j$, $C_{ij} = d(x_i, x_j)$.
As we did for the cost funcion in Sec.~\ref{sec:Qboost}, this cost function
can be mapped directly onto an equation of the form Eq.~\ref{eq:Ising}
with Ising spins $s_i$, that take on values $\{-1, 1\}$ as binary variables,
by taking $s_i = 1 - 2z_i$.
The resulting Ising energy function can then be translated into a quantum problem
Hamiltonian $H_P$ of the form of Eq.~\ref{eq:HP-Ising} by replacing
the Ising spin variables with Pauli matrices:
\begin{equation}
H_P = \sum_{i,j}C_{ij}(I - \hat Z_i \hat Z_j)/2
\end{equation}
It can be used to attack the MAXCUT problem using quantum annealing
as described in Sec.~\ref{sec:QA}. When the non-zero terms of the Hamiltonian
match the edge structure of D-Wave's architecture, the problem can
be run on D-Wave's current hardware
as discussed in Sec.~\ref{sec:QA-super}

One of the main classical approaches for attacking MAXCUT has been
simulated annealing. Given its success on the problem, and the
advantages that quantum annealing has over simulated annealing,
this problem is a natural one to attack with quantum annealing.
In fact, the quantum annealing algorithm for this Hamiltonian
has been run on an NMR quantum computer to solve a toy version
of the MAXCUT problem with three data points \cite{Steffen03}.
Advances in quantum computational hardware mean that it is now
possible to evaluate the algorithm on significantly larger data sets.

Many applications require partitioning the data sets into multiple clusters,
not just two.
One way to obtain multiple clusters is to run the binary clustering
algorithm repeatedly. Here, we describe an alternative which determines
$K$ clusters $C_k$ all at once.

Let $z_i^{(k)}$ be a binary variable indicating whether or not
data point $x_i$ is placed in cluster $C_k$:
\begin{equation}
z_i^{(k)} = \left\{\begin{array}{c}
1 \mbox{ if } x_i\in C_k \\
0 \mbox{ if } x_i\notin C_k
\end{array}.\right.
\end{equation}
We phrase the clustering problem as a binary optimization problem with
the following cost function
\begin{equation}
E({\bf z}) = \sum_k \sum_{i,j} C_{ij} z_i^{(k)} z_j^{(k)}
+ A \sum_i \left( \sum_k z_i^{(k)} -1 \right)^2
\end{equation}
where, as before, the coefficients $C_{ij}$ are the distances between
training points, $C_{ij} = d(x_i, x_j)$. The role of the second term
(where $A>0$ is a large constant) is to ensure that every point belongs
to one and only one cluster.
In the same way as before, this quadratic unconstrained binary
optimization problem (QUBO) can be directly translated into
the Ising model (\ref{eq:Ising}) by substituting spin variables
$s_i = 1 - 2z_i$ for the binary variables $z_i$.
The problem Hamiltonian $H_P$ (\ref{eq:HP-Ising}) that can be employed in the
QA procedure is obtained by replacing the
spin variables with Pauli matrices.
When the non-zero terms in the Hamiltonian match that of the
current edge architecture of the D-Wave machine from Sec.~\ref{sec:QA-super},
the problem can be attacked on currently available hardware.

Note that for $K=2$ this encoding of the problem is somewhat different
from the one we
we used before. In this encoding, the two cluster case looks like
\begin{equation}
E({\bf z}) = \sum_{i\neq j} C_{ij} z_i^{(0)} z_j^{(1)} +
A \sum_{i} \left( z_i^{(0)} + z_i^{(1)} -1 \right)^2 .
\end{equation}

\section[Anomaly Detection]{Anomaly Detection for  Space Systems}

Anomaly detection aims to determine if systems are running in
a normal mode of operation or if they have entered an anomalous
state. Detected anomalies alert operators to ongoing or potential
problems. The anomalies can be analyzed to help discover faults
before they become catastrophic, and to provide insight
that could help keep faults from occurring in the first place.
In any case, the anomalies first must be detected.
In some situations, training sets containing data for both anomalous and
normal modes of operation may be available, in which case the QBoost
algorithm of Sec.~\ref{sec:classification}, or
other classification methods, could be applied. More frequently, however,
little if any anomalous data is available, let alone labeled as such.
The lack of data on anomalies poses a serious challenge to anomaly
detection. A number of methods exist, one of the most successful being
NASA's Inductive Monitoring System (IMS).

\begin{figure}
\begin{center}
\includegraphics[width=0.75\textwidth]{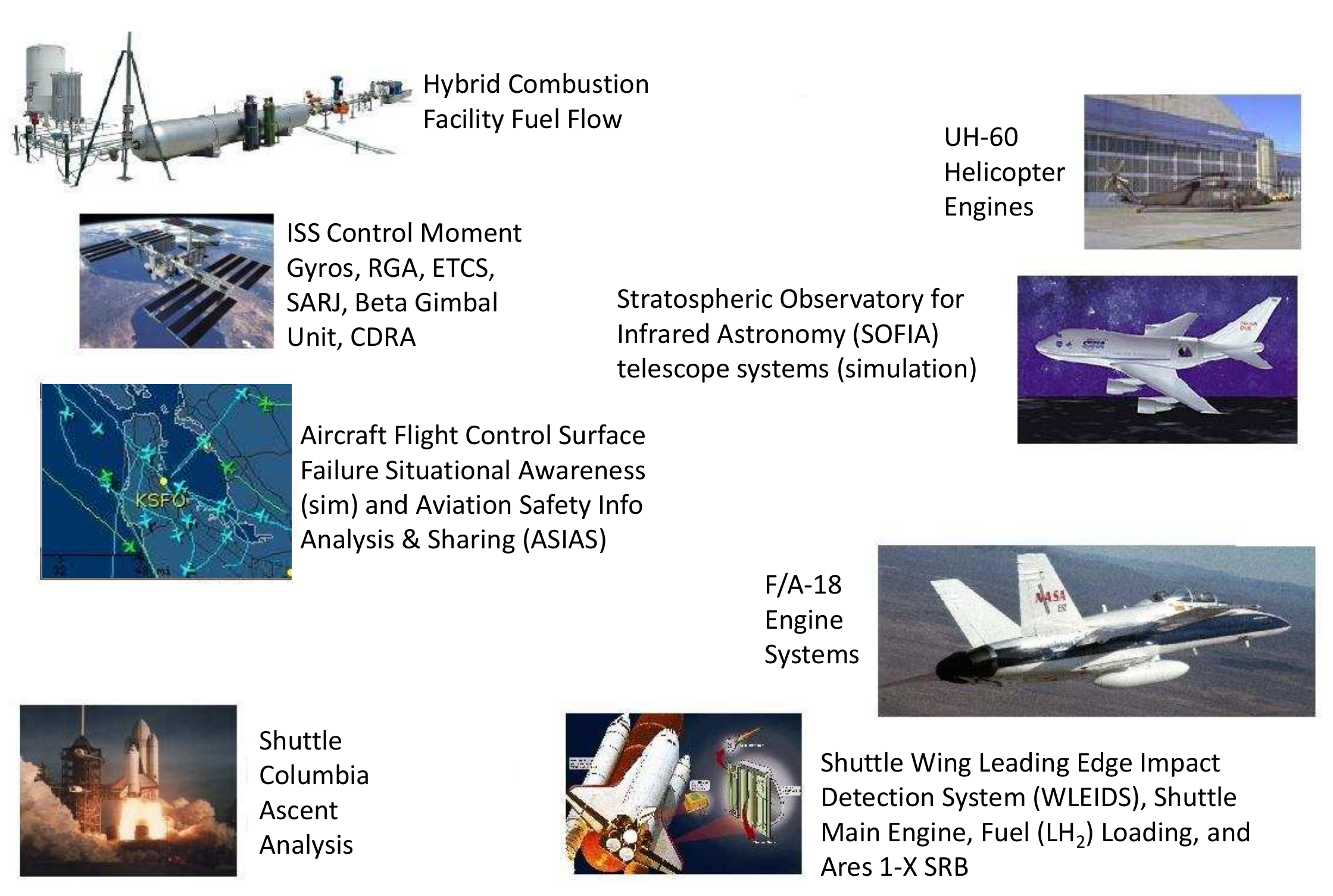}
\end{center}
\caption{
Diverse NASA projects benefit from machine learning.
Pictured are a sampling of NASA projects which have used the
machine learning based Inductive Monitoring System (IMS) to detect
anomalies.}
\label{fig:IMSapps}
\end{figure}

Two main alternatives to automated approaches exist: 1) analysis
by human experts and 2) simulating the system. For a complex
system, simulation may not be possible, or
may be too slow. Even if the system itself can be effectively and
efficiently simulated, the environment in which it operates may not be.
A system with $N$ simple sensors, each outputting a simple binary YES/NO
response, has $2^N$ possible input states. As soon as $N$ is in the hundreds,
let alone thousands, full simulation across all input states is computational
intractable. Similar considerations limit the effectiveness of even
large teams of human experts. Furthermore, for many space applications,
the communication time alone rules out human experts as the primary
approach; roundtrip communication between Earth and Mars takes between
ten and forty minutes. For complex systems, both human expertise
and system modeling can supplement machine learning based
data-driven approaches, but are not sufficient on their own.

\subsection[IMS]{\label{sec:ims} The Inductive Monitoring System (IMS)}

\begin{figure}
\begin{center}
\includegraphics[width=0.75\textwidth]{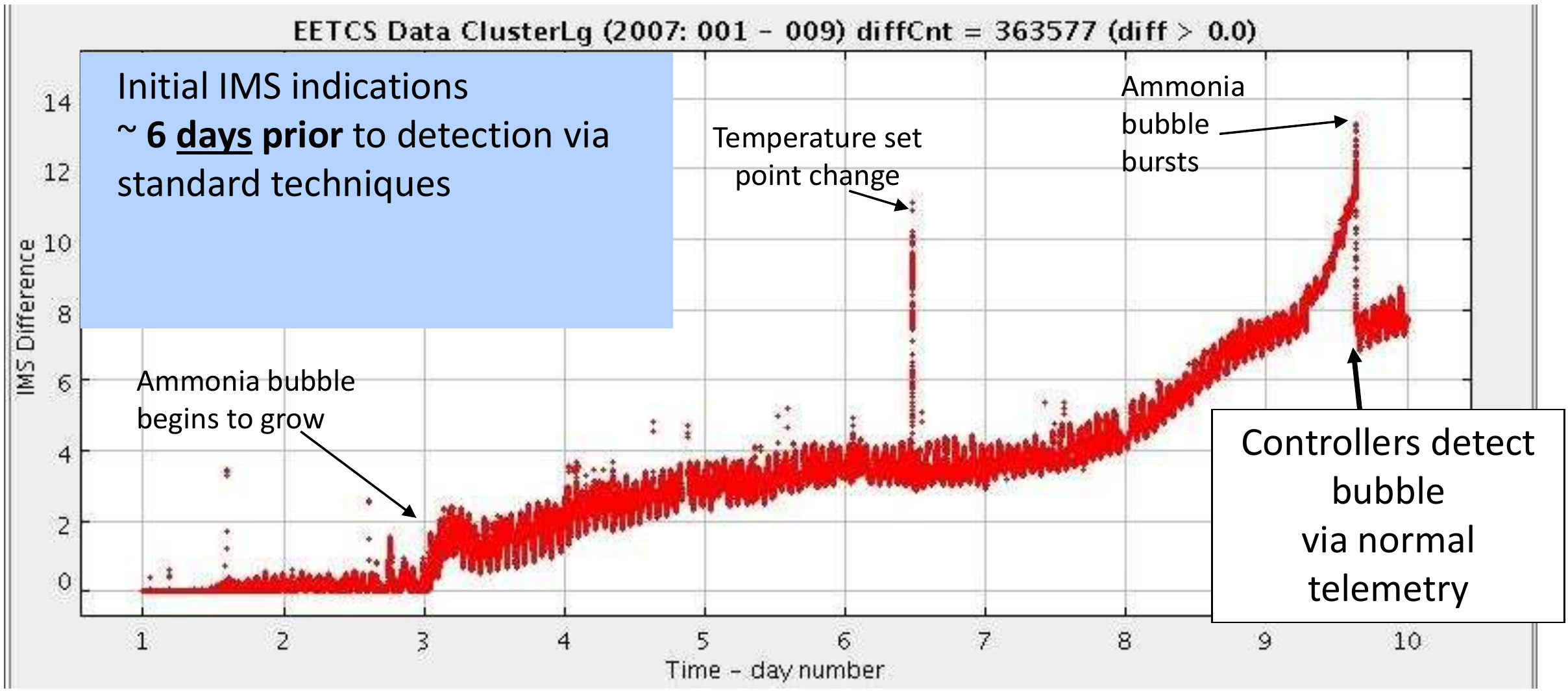}
\end{center}
\caption{IMS trained on 185 days of data that includes 23 parameters
including data from various pressure, temperature, and pump speed sensors
from the International Space Station's thermal control system
detects the anomalous readings, due to a growing ammonia bubble,
six days earlier than standard techniques. \cite{Iverson08}}
\label{fig:IMSgraphs}
\end{figure}

The Inductive Monitoring System (IMS) has been applied to a
large number of NASA systems, including
control systems on the International Space Stations,
sensor systems on space shuttle wings, and rocket and helicopter engines
\cite{Iverson04}.
For example, it has been trained on data from the
ISS Early External Thermal Control System (EETCS), and succeeded
in detecting an anomaly six days before standard techniques
would have detected it (Figure \ref{fig:IMSgraphs}) \cite{Iverson08}.
Martin \cite{Martin09} and Schwabacher et al. \cite{Schwabacher09}
compared IMS with other data-driven anomaly detection approaches
on NASA mission data sets. IMS outperformed the other techniques, but
while all of the techniques found some anomalies, no one technique
found them all.
They conclude that it is useful be able to ``run multiple anomaly
detection algorithms on a data set.''
There is clearly room for improved algorithms, or new and different
algorithms that detect different anomalies.
As Schwabacher point out, it is hard to know how much room for improvement
there is because it is unclear ``how many more anomalies exist in the data''
or ``what fraction of the anomalies'' are detected by current algorithms.
Quantum computational approaches to anomaly detection could
provide new techniques.

IMS \cite{Iverson04} uses unsupervised clustering to
obtain a representation of normal operation that can be used to quickly
check whether a system is operating in the normal range or should be
flagged as anomalous.
The data for normal operation - the training set -
consists of data from onboard sensors. Each data point can be represented
as a set of sensor readings, which can be viewed as a point in a high
dimensional space.
In this case, we are trying to understand something about
the structure of this data - its boundary.

\begin{figure}
\begin{center}
\includegraphics[width=3.0in]{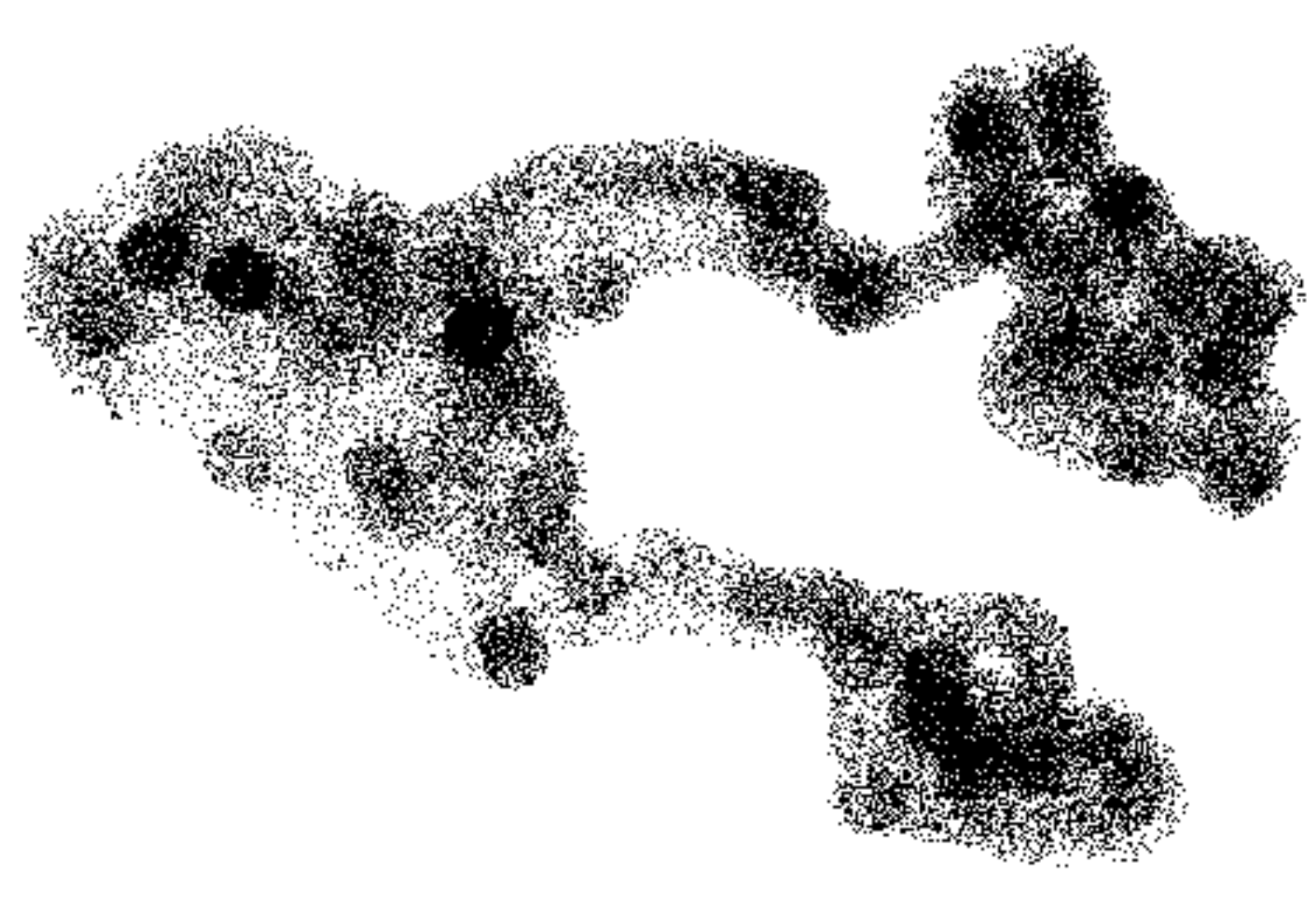}
\includegraphics[width=3.0in]{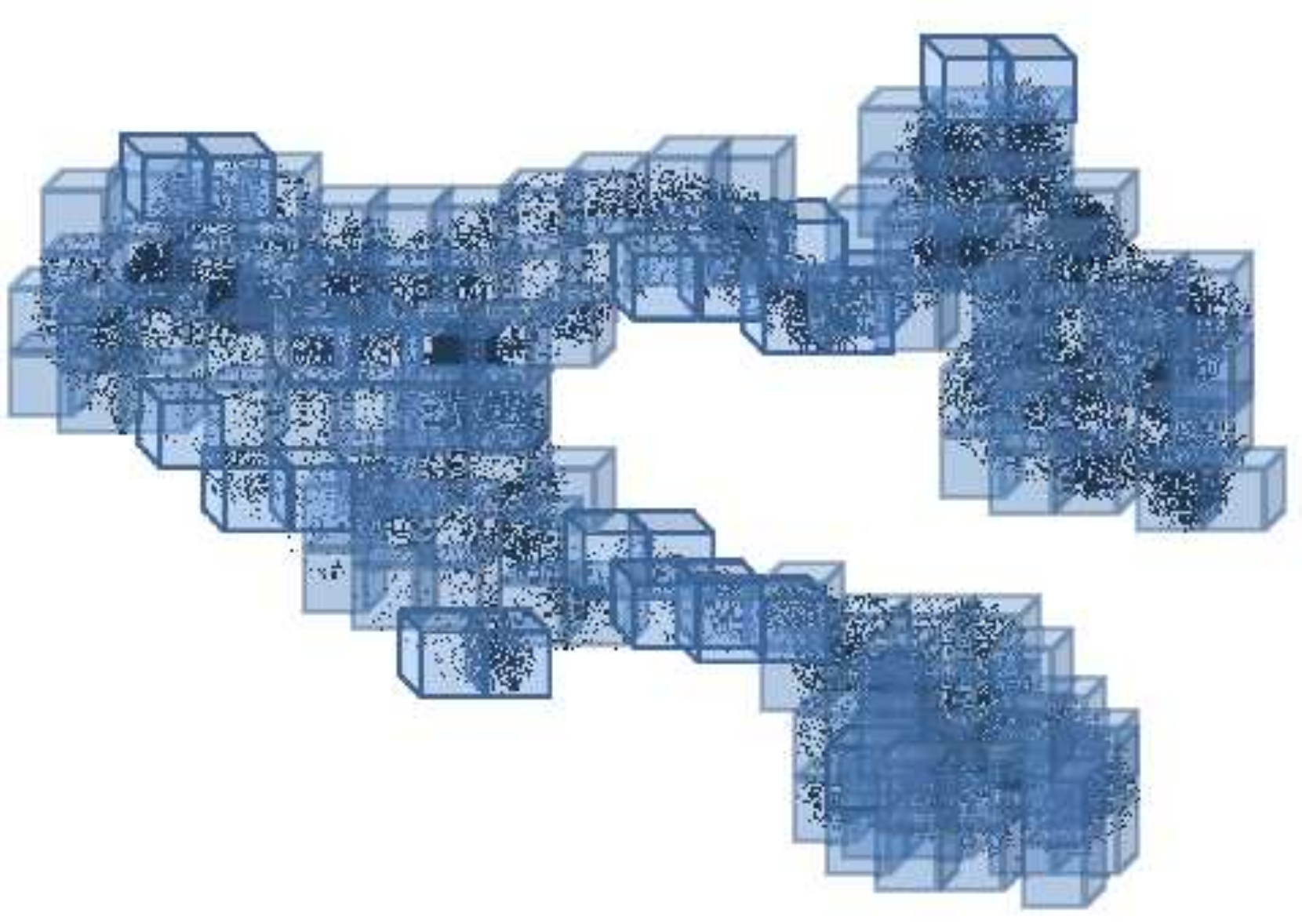}
\end{center}
\caption{The figure on the left shows data points representing normal
behavior. In real applications, these data points are usually high
dimensional, and the pattern would be much more complicated. The
figure on the right depicts output from the IMS algorithm, a set of
axis aligned boxes that tightly cover the region of normal behavior.}
\label{fig:IMSalg}
\end{figure}

More specifically, IMS is a type of ``online unit clustering'' algorithm,
where ``online'' means that it considers each training point one at a time, and
``unit'' means that the clustering is into fixed units, most often spheres
of a fixed radius or boxes of fixed side length \cite{Epstein08}.
It is considered a clustering problem because the centers of the spheres
or boxes are allowed to move during the course of the algorithm.
Problems in which the centers are fixed are called ``covering'' problems.
The problem of finding a minimal unit clustering for a set of points is
an NP-complete problem \cite{Chan09}.
The IMS algorithm returns a set of boxes of side length $2\epsilon$
that tightly cover the space of known behavior (Figure \ref{fig:IMSalg}).
The parameter $\epsilon$ determines how tightly.
More precisely, the IMS algorithm
returns the center points of these boxes.
The boxes are ``axis aligned'' in that each box can be represented by a set
of inequalities on each sensor value separately.
The center points are obtained as weighted sums of training points.
The algorithm is designed to generate boxes
that cover all known points of normal operation.
IMS includes a mode that allows for more complicated geometry than
boxes, but in many instances the axis aligned boxes are preferred,
because whether or not a new point falls within an axis aligned box
is very quick to compute.

We illustrate the quantum approach with QIMS, a family of novel quantum
anomaly detection approaches related to IMS. Other approaches, including
more radically different approaches to anomaly detection, are possible.

\subsection[QIMS]{\label{sec:qims} QIMS: Mapping IMS onto Quantum Annealing}

We describe a parameterized family of unsupervised clustering techniques
that bound the training data with a set of boxes similar to those
returned by IMS, i.e., axis-aligned boxes that can be evaluated quickly
on new data.
The quantum algorithms return the centers of boxes of side length $2\epsilon$
that either fully or approximately cover the training data -- the
extent of the coverage depending on the parameters.
Our quantum algorithm solves a related problem to the minimal unit clustering
problem, the minimal unit covering problem: given a set of
points $P$, and a set $S$ of disks covering these points, find the minimal
subset $S_{min}$ of $S$ that covers $P$. This problem is also known to be
NP-complete \cite{Narayanappa06,Carmi07}.

For each point $x_i$ in the training data, define the $\epsilon$-box $B_i$
to be the set of points whose coordinates are all within $\epsilon$ of
$x_i$:
\begin{equation}
B_i = \{x| \linfnorm{x_i - x} \leq \epsilon\}.
\end{equation}
Let ${\cal B}$ be the set of such boxes.
We now show how to phrase the search for a minimal subset of ${\cal B}$
that covers the training set as
a binary optimization problem that can then be translated into
a Hamiltonian that can be used in a quantum annealing algorithm.
Let $|B_i|$ be the number of training points in $B_i$. We define the
binary optimization problem as finding the binary variables $z_i$,
where $z_i$ indicates whether or not the box $B_i$ should be included in
the final set, that minimize the following QUBO cost function:
\begin{equation}
E({\bf z}) = \sum_{i,j} C_{ij} z_iz_j
- \mu\sum_i C_iz_i + \lambda\sum_i z_i ,
\end{equation}
where $C_{ij} = |B_i\cap B_j|$ and $C_i = |B_i|$.
The first term penalizes overlap
among the boxes, the second encourages all points to be included in
at least one box, and the optional third term encourages there
to be fewer boxes.
Once again, this QUBO cost function can be directly mapped onto the
Ising model  \ref{eq:Ising} via $s_i = 1 - 2z_i$
and encoded in a problem Hamiltonian $H_P$ (\ref{eq:HP-Ising})
by replacing the spin variables $s_i$ with Pauli matrices.
Problems in which non-zero terms in the problem Hamiltonian match the edge
architecture of the D-Wave QA machine from  Sec.~\ref{sec:QA-super}
can be run on the current hardware.

Depending on how the weights $\mu$ and $\lambda$ are set, outliers may
or may not be included. For example, if there is a training point
that is at least $\epsilon$ away from all of the other training points,
when $\mu > \lambda$, the $\epsilon$ box around this point would be
included in the final model, but when $\mu < \lambda$, this box would not
be included, and thus the training point $x_i$ would not be covered
by the model. Thus, $\mu < \lambda$
returns a model more similar to that returned by IMS in that it covers all
points, but our setup allows for other strategies that remove outliers.
Because IMS is a greedy algorithm and QIMS is not, it is reasonable to 
expect that the accuracy of QIMS will be better than that of IMS. On 
the other hand, just as the set up time was expensive for QBoost, the 
computation needed simply to set up the QIMS problem
prior to the quantum annealing step is greater than running the entire
IMS algorithm. Whether this overhead is worth it depends on how significant
the gains in accuracy would be.

Like IMS, our algorithm returns points that are the centers of the
boxes that make up the model. In IMS, the centers are weighted
sums of training points, so are not generally themselves training points.
In our model, all centers are training points, so the two models differ
in this respect.
On the other hand, IMS is usually applied only in cases in which the
training data is dense, since only in that case does the IMS clustering
substantially reduce the time it takes to evaluate a new data point over
an approach like Orca that compares a new data point with all training
data points. For dense data, there will generally be a training point
very near each IMS box center.

For a select set of parameters, this approach can be viewed as a one-class
special case of the supervised learning method of
Sec.~\ref{sec:classification}.
As the set of weak classifiers, consider the set of functions, one for
each training point $x_i$, defined as
\begin{equation}
h_i = \left\{\begin{array}{c}
1 \mbox{ if } \linfnorm{x_i - x} \leq \epsilon\} \\
0 \mbox{      else. }
\end{array} \right.
\end{equation}
Since all of the training data are examples of normalcy,
$\vec{y} = (1,1,\dots,1)$.
In this case, the coefficient
\begin{equation}
C_{iy}' = \vec{h_i}\cdot\vec{y} = \sum_s h_i(x_s)
\end{equation}
is the number of training points whose coordinates are all
within $\epsilon$ of $x_i$, and
\begin{equation}
C_{ij}' = \vec{h_i}\cdot\vec{h_j} = \sum_s h_i(x_s)h_j(x_s)
\end{equation}
is the number of of points within $\epsilon$ of both $x_i$ and $x_j$.
Thus, the QA for binary optimization problem of Sec.~\ref{sec:classification}
with these coefficients corresponds the the special case of QIMS
with QUBO cost function
\begin{equation}
E({\bf z}) = \sum_{i,j} |B_i\cap B_j| z_iz_j
- \mu\sum_i |B_i|z_i + \lambda\sum_i z_i
\end{equation}
in which $\mu = 2$.
As we have done before, we can map this cost function
directly onto an equation of the form Eq.~\ref{eq:Ising}
in terms of Ising spins  $s_i = 1 - 2z_i$.
The resulting Ising equation can then be translated into a quantum problem
Hamiltonian $H_P$ (\ref{eq:HP-Ising}), which
can be employed by quantum annealing process described in Sec.~\ref{sec:QA}.

The QIMS algorithm uses $n$ qubits, where $n$ is the number of training
examples. In early implementations of quantum annealing machines, qubits
will be a precious commodity. The amount of training data can easily exceed
the number of qubits available. To address this issue, QIMS can be run in
batch mode. Suppose the quantum annealing machine has $m$ qubits. The most
obvious way to run QIMS in batch mode is to run QIMS separately on each
successive group of $m$ points, and take the union of those boxes. We
can do a lot better than that if after obtaining a set of $\epsilon$
boxes from previous runs, we check whether the next points are contained
in those boxes. Any point that is already covered, can be ignored.
We accumulate points not contained in those boxes until we reach $m$ points,
or run out of training points, and then run QIMS on these. There is yet
a further improvement we can make by ``growing'' each batch.
When we run QIMS on $m$ points, it returns $k$ boxes, where $k$ is usually
significantly less than $m$. If we retain the center of these boxes,
and add in $m-k$ new training points not covered by these boxes.
More specifically, we use $k$ qubits as indicator variables for the boxes we
already obtained, and the remaining $n-k$ qubits as indicator variables
for boxes centered at the new points.
We then run QIMS on this set of points to obtain a new set of boxes that may
or may not contain boxes found in earlier stages of processing this batch.
We can repeat this process until QIMS returns $m$ boxes. At this point,
we then go on to a new batch. In the batch growing step, we use all
points considered so far, not just the ones that are centers for
the current set of boxes, to compute the coefficients $|B_i|$ and
$|B_i\cap B_j|$. Batch mode for QIMS is summarized in Alg. \ref{alg:QIms-1}.
\begin{algorithm}
\caption{Pseudocode summarizing batch mode for QIMS}
\label{alg:QIms-1}
\While{unprocessed points remain}{
  $B$ = union of boxes from all completed batches\;
  \If{point $p$ not covered by $B$}{
    add $p$ to current batch (or start new batch)\;
  }
  \If{batchSize = $n$}{
    run QIMS on batch. Output will be $m \leq n$ $\epsilon$-boxes.\;
  }
  \If{$m = n$}{
    add this batch of boxes to $B$\;
    reset batchSize to $0$\;
  }
  \If{$m < n$}{
    reset batchSize to $m$\;
  }
}
\end{algorithm}

\section[Image Matching]{\label{sec:ImageProc} Data Fusion and Image Matching for Remote Sensing}

\begin{figure}
\centering
\includegraphics[width=0.75\textwidth]{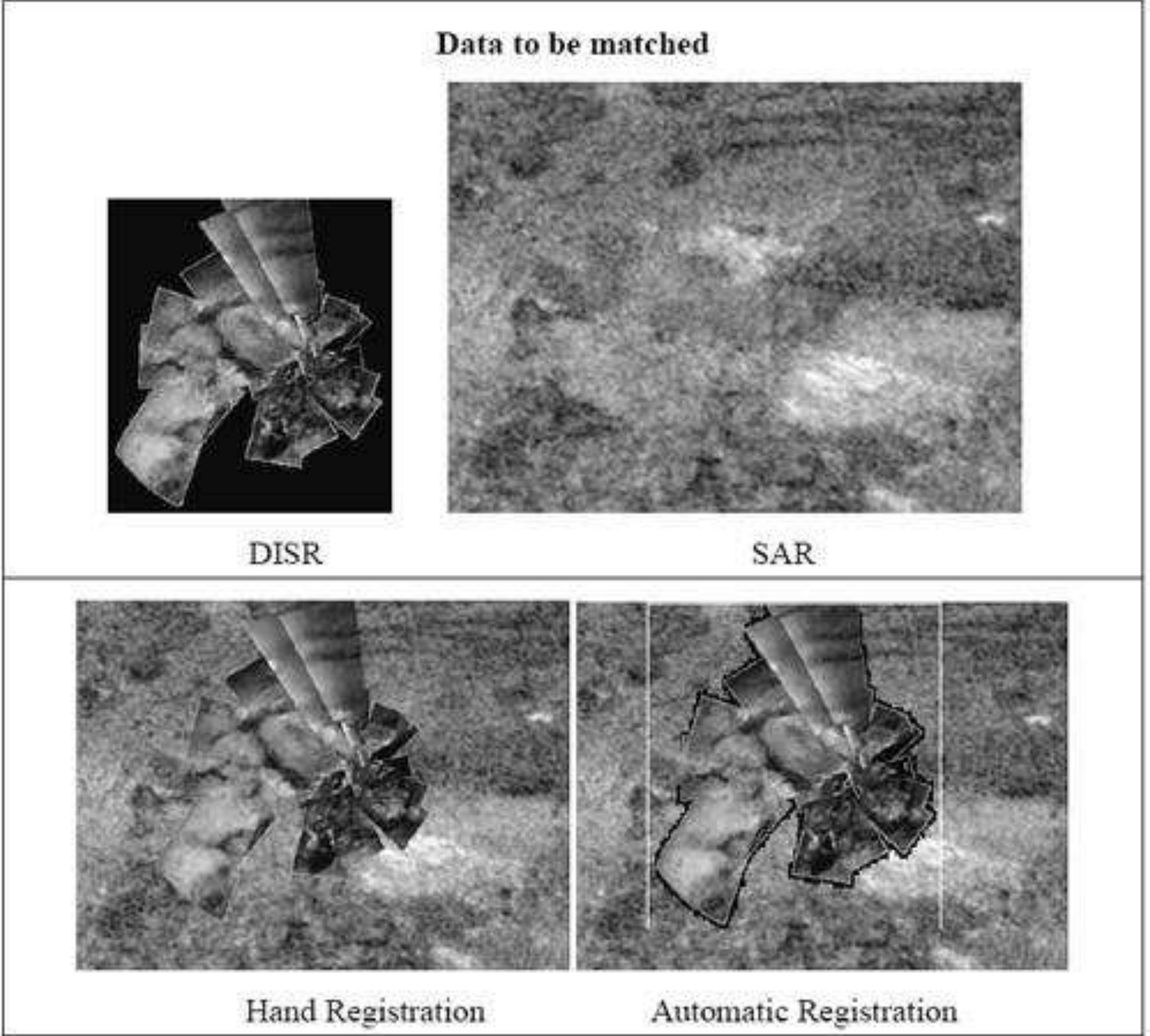}
\caption{A key component of imaging sensor fusion is image registration. This
figure shows Hand vs. Automatic registration of Huygens Descent Imager
Spectral Radiometer (DISR) image mosaic to Cassini Synthetic Aperture Radar
(SAR) \cite{FigForMMImReg}. The lower figure shows the DISR image
superimposed over the corresponding region of the SAR image, where,
on the left, the correspondence has been determined automatically and,
on the right, an improved correspondence has been determined by hand.}
\label{fig:ImRegDisrSar}
\end{figure}

Image registration is the first step in many image processing algorithms
such as stitching for image mosaicing and scene reconstruction
of planetary panoramas
\cite{MarsMosaic}, rover localization and mapping \cite{GeoMapping},
visual odometry and autonomous navigation \cite{PowerEfficRovers},
superresolution imaging of planetary terrain \cite{SuperRes}, and
photogrammetic measurements of lunar features \cite{PhotoGramLROC}.
Image registration underlies most video compression and coding schemes
\cite{Szeliski06} which are vital
for efficient communication between earth and space imaging devices.
Sensor fusion also frequently relies on image registration to align
images from different types of imaging devices.
The NASA/JPL Multi-modal Image Registration
and Mapping project was devoted to improved algorithms for registration of
and topographic mapping from different imaging devices such as
synthetic aperture radar (SAR) and visible/infrared (VIR) imaging
spectroscopy images from Cassini-Huygens's exploration of Titan and
descent imaging on the Huygens probe \cite{MMImReg}
 (see Figure \ref{fig:ImRegDisrSar}).

Image processing and machine vision are relatively mature areas,
with a variety of powerful techniques. Nevertheless,
machine vision lags behind human vision in many areas, and
many applications await better automated techniques.
While automated image registration tools are in widespread use, they make
sufficiently many errors that it is common practice to have human analysts
check the output and correct any errors made in the processing
\cite{PhotoGramLROC,GeoMapping}.
Greater accuracy and improved efficiency
are both desired.
The final section of Szeliski's tutorial on image alignment and stitching
includes a call for ``better machine learning
techniques for feature matching'' and to ``increase the
reliability of fully automated stitching algorithms''
\cite{Szeliski06}.
Thus, in spite of the maturity of the field, there remains significant
room for improvement in automated image registration algorithms.
In this section we describe the image alignment problem and
then describe Neven's mapping of this problem onto an optimization
problem that can be attacked with quantum annealing.

\subsection{\label{sec:ImageAl} Image Alignment}

The first step in aligning or registering images is to identify feature points.
A variety of methods exist for identifying feature points
in images and associating a {\it local descriptor} with each feature
point \cite{Lowe03,Ke04,Bay08,Liu09}. While we concentrate on
the case of points, the algorithm can also be applied to other
geometric objects with local descriptors, such as the features
used in the NASA/JPL Dynamic Landmarking project \cite{DynLand}.
The techniques we describe can be applied to images of various types, from
thermal images received from from IR cameras to
images from cameras detecting visual wavelengths.

\begin{figure}
\centering
\includegraphics[width=0.95\textwidth]{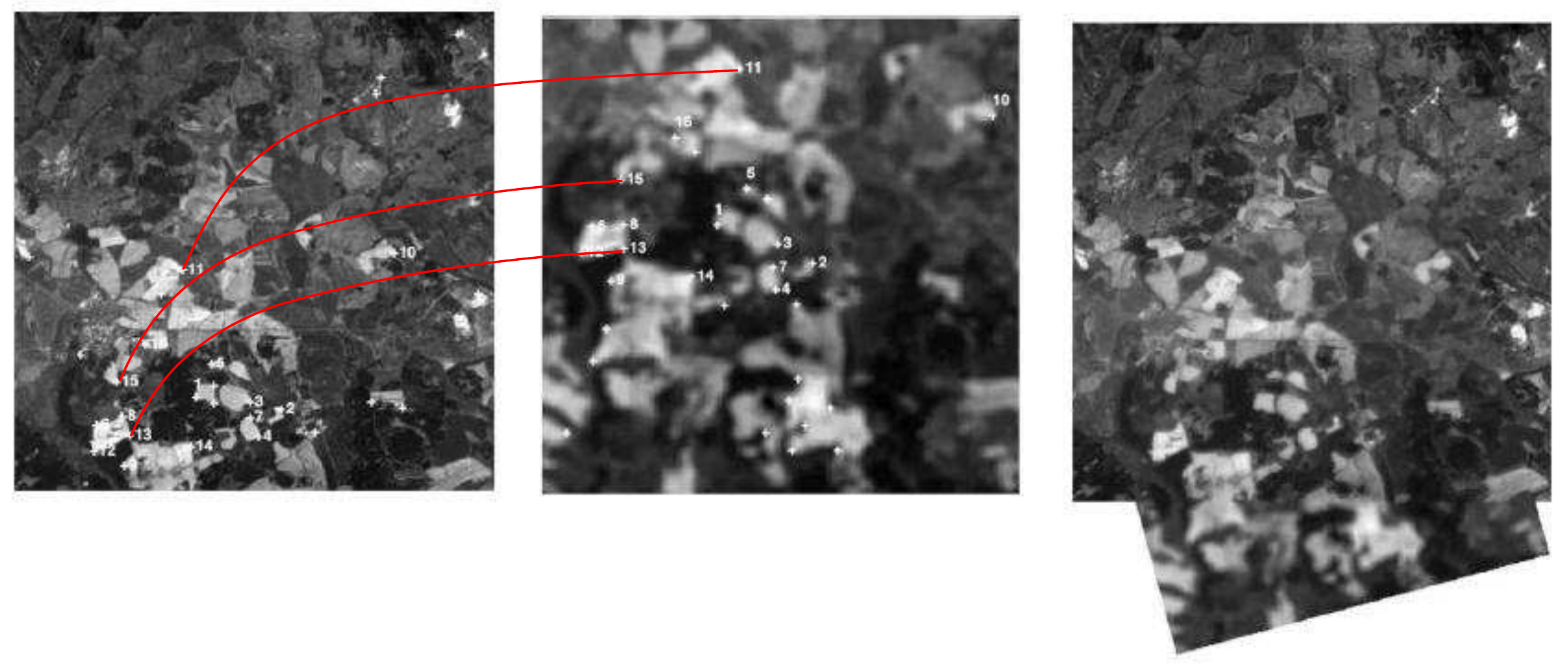}
\caption{Two satellite images control points were matched. Matches are denoted by the same numbers in both images. For illustrative purposes,  some of the matches are connected by red lines. The third figure on the right shows how the two images, once correctly registered, can be combined
into a mosaic (figure is re-plotted from Ref.~\cite{Zitova:2003}, Fig.~4).}
\label{fig:ImRegDiag}
\end{figure}

Images can be registered by finding pairs of points $(p_i, p_\alpha)$,
where $p_i$ is a feature point in the first image $I$ and $p_\alpha$
is a feature point in the second image $I'$, such that $p_i$ and $p_\alpha$
have similar local descriptors (see Figure \ref{fig:ImRegDiag}).
Often, however, multiple points have similar
descriptors. Geometric constraints can help determine which
is the correct set of matching points: frequently two pairs of points,
$(p_i, p_\alpha)$ and $(p_j, p_\beta)$, are not consistent with
a single registration between the two images, so only one of the pairs
should be included in the final set of matchers. The intuition is that
the best registration is the one that has the largest consistent set
of matches.

\subsection[Image Alignment and Quantum Annealing]{\label{sec:ImageMap} Mapping Image Alignment onto Quantum Annealing}

We present Neven's quantum annealing algorithm for finding a largest
consistent set of feature point matches between two images \cite{Neven}.
To turn the problem of finding the largest consistent set into a
binary operation problem, let $z_{i\alpha}$ be a binary variable
that determines whether the pair of points $(p_i, p_\alpha)$
should be included in the final set of matches. We
want a cost function that encourages geometric consistency and discourages
a single point in one image from being matched to two or more points in
the other image. Let $\{V_{i\alpha}\}$ be the set of all pairs
$\{(p_i, p_\alpha)\}$ viewed as vertices of a graph whose edges are
``conflicts.'' Specifically, there is an edge between vertex $V_{i\alpha}$
and $V_{j\beta}$ if the corresponding pairs are not geometrically
consistent with a single registration or if they match one point in one
image to two in another (i.e. $i=j$ or $\alpha = \beta$). Finding
the largest set of consistent matches is equivalent to finding
the {\it maximally independent set (MIS)} of vertices in the
conflict graph we have just constructed, where a maximally independent
set of vertices is the largest set of vertices that has no edges between
them. The problem of finding a largest maximally independent set,
the MIS problem is known to be NP-complete \cite{Garey79}, and thus
is generally attacked by heuristic algorithms. One approach is to
phrase it as a optimization problem over the binary variables $z_{i\alpha}$,
with QUBO energy function
\begin{equation}
E({\bf z}) = \sum_{i\alpha j\beta} C_{i\alpha,j\beta} z_{i\alpha}z_{j\beta}
\end{equation}
where, for $i\alpha\neq j\beta$, $C_{i\alpha,j\beta}=L$, where $L$ is the
number of vertices (the number of pairs $(p_i, p_\alpha)$ with
similar local descriptors), and $C_{i\alpha,i\alpha}=-1$ to encourage
the inclusion of vertices. This QUBO problem can be translated
into an Ising form \ref{eq:Ising} and encoded in the problem
Hamiltonian $H_P$ \ref{eq:HP-Ising} that can be used in quantum annealing.
When the non-zero interaction terms in $H_P$ match the connections
in the D-Wave quantum annealing implementation described in
Sec.~\ref{sec:QA-super}, the problem can be run on current hardware.

\section[Planning]{\label{sec:plan} Planning  of NASA operations: from ISS to Deep Space missions}

Automated planners have their origins in robotics and have been used
extensively in space applications. Early examples include
{\textsc{Optimum-AIV}} and {\textsc{PlanERS-1}} by European Space Agency,
support of ground teams for Voyager spacecraft and Hubble telescope
({\textsc{Spike}}, for scheduling observations), EO-1 observing sattellite,
ST-5 mission, and planning tool for Mars Express.
More recent example is {\textsc{Solar Array Constraint Engine}} (SACE), a
planner based on EUROPA framework \cite{sace}.

In keeping with the best practices the system consists of model-agnostic
kernel and a model describing various planning constraints using formal
language.
As the acronym suggests, constraint satisfaction is at the heart of the
planner. {\emph{Timelines}}, corresponding to state variables that are
functions of time and actions and subdivided into discrete intervals.
Continuous angles of rotary joints and gimbals are approximated by discrete
variables which limits the state space. Actions that can be performed at any
time are constrained by requirements of normal operating range as well as the
fact that any changes in orientation take a finite time. All these constraints
can be expressed using formal language. Since it is desirable to maximize
power output, the problem is properly reformulated as constrained
optimization.

Despite improving efficiency of hitherto manual operation, the design provides
an example of presently necessary tradeoffs. It sacrifices optimality for
speed as it uses a variant of a ``greedy'' algorithm, even though this
particular model can benefit from dynamical programming, at the expense
of having time
complexity scale as a {\emph{square}} of the size of state space in contrast  to
{\emph{linear}} scaling for greedy algorithm. Furthermore, dynamical
programming algorithm only works because constraints are ``local'' in time:
that is they either limit possible states in a given time interval due to
conflicts with scheduled operations or involve adjacent intervals (e.g.
insufficient time to reorient a panel). However, non-local constraints would
appear in more refined models: for example, thermal stress constraints depend
on entire histories. In this more general scenario time complexity increases
exponentially with the size of state space, which presents a challenge to
modern planners. The goal of ensuring the best use of resources (finding truly
optimal solutions) within the constraints of autonomous spacecraft (where vast
computational resources on the ground cannot be tapped) calls for radically
new approaches.

\subsection{\label{sec:plan-def} Planning Problem}

In a nutshell, the main ingredients of planning problem is a set of $N$
{\emph{propositions}} and $M$ {\emph{operators}}. A state at time $t$ is
determined by $N$ binary variables $\left\{ x_i \left( t \right) \right\}_{i =
1}^N$ where $x_i \left( t \right) = 1 \left( 0 \right)$ if proposition labeled
by $i$ is true(false). With each operation labeled by $j \in \left\{ 1, 2,
\ldots, M \right\}$ one associates two small sets of indices
$\mathcal{C}_j^{\left( + \right)}, \mathcal{C}_j^{\left( - \right)} \subset
\left\{ 1, 2, \ldots, N \right\}$ enumerating propositions that must be true
and false, respectively, {\emph{before}} the operation can be applied
(positive and negative {\emph{preconditions}}) as well as two sets
$\mathcal{E}_j^{\left( + \right)}, \mathcal{E}_j^{\left( - \right)} \subset
\left\{ 1, 2, \ldots, N \right\}$ (positive and negative {\emph{effects}})
which list propositions that will become true(false) after the operation is
performed. Given a complete specification of {\emph{initial state}} given by
$\mathcal{I}^{\left( + \right)} = \left\{ i \vert x_i \left( 0 \right) = 1
\right\}$ and $\mathcal{I}^{\left( - \right)} = \left\{ 1, 2, \ldots, N
\right\} \backslash \mathcal{I}^{\left(_{} + \right)}$, the planner aims to
find a sequence of operations that reaches a {\emph{goal state}} specified
using sets $\mathcal{G}^{\left( + \right)}, \mathcal{G}^{\left( - \right)}
\subset \left\{ 1, 2, \ldots N \right\}$ listing propositions that must be
true(false) at the end.

\begin{figure}
\includegraphics{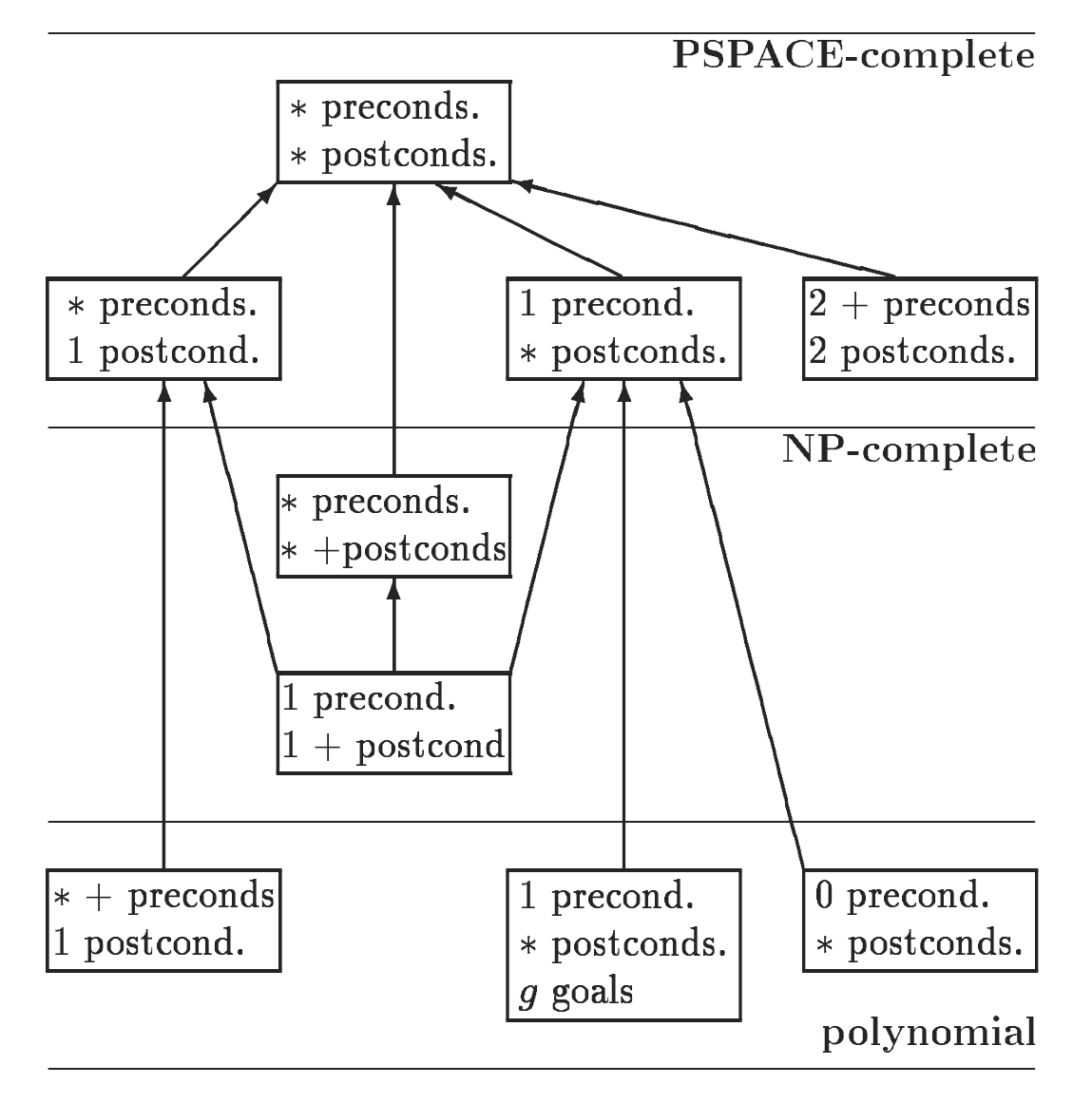}
\caption{\label{fig:plancomp}
Complexity of STRIPS planning problem (figure reprinted from Ref.~\cite{plancomp}).
Polymomial-time algorithms exist only if operations are restricted to (a) be
preconditioned on at most one proposition or (b) have a single effect and only
positive preconditions.
}
\end{figure}

In practice, inputs are expressed in less verbose form using notion of
{\emph{objects}}, {\emph{predicates}} and {\emph{actions}}. Predicates and
actions are merely templates for propositions and operations respectively
which (in PDDL notation) are written as ({\emph{predicate object$_1$
$\ldots$ object$_k$}}) or ({\emph{action object$_1$ object$_k$}}).
Best-performing algorithms \cite{satplan,graphplan} start off by transforming the problem
into fully propositional intermediate form presented above. Given $n$ objects,
$m$ predicates, and $\ell$ actions with $k$ representing the maximum arity of
predicates/actions, the number of propositions and actions is bounded: $N < m
\left( n + 1 \right)^k$ and $M < \ell \left( n + 1 \right)^k$ respectively.
Since $k$ is typically small (rarely more than 3), this step is polynomial in
time and memory requirements; the difficult part of the problem is finding the
actual plan. Indeed, the problem is quite hard (PSPACE-complete) unless
extraordinary (and unlikely to be encountered in practice) restrictions are
placed on it as illustrated in Fig.~\ref{fig:plancomp} reprinted from
Ref.~\cite{plancomp}.

\begin{figure}
\includegraphics[width=6in]{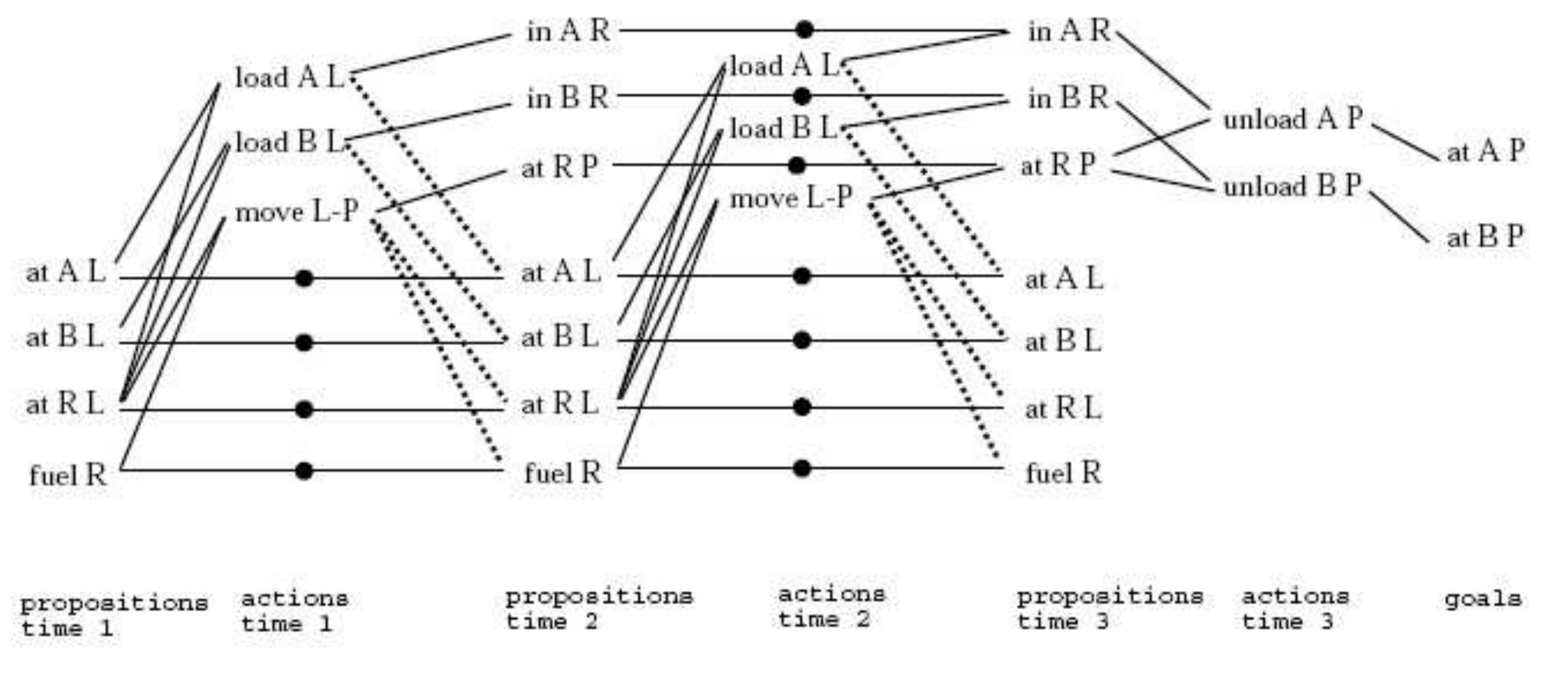}
\caption{\label{fig:graphplan}
An example of planning graph from Ref.~\cite{graphplan} with $L = 3$.
Columns alternate between propositions $\{ x_i^{(t)}\}$ and operations $\{ y_i^{(t)}\}$.
Propositions and operations were automatically generated from a model
containing objects \texttt{A}, \texttt{B}, \texttt{R}, \texttt{L},
\texttt{P}, predicates \texttt{at}, \texttt{fuel}, \texttt{in}, and
actions \texttt{load}, \texttt{move}, \texttt{unload}. The planning
graph roughly corresponds to the graph of interactions with solid lines
(positive preconditions/effects) indicating ferromagentic couplings and dashed
lines corresponding to negative preconditions/effects (antiferromagnetic
interactions). Many propositions and operations have been removed from the
full graph (similar optimization is possible in the approach we describe).
}
\end{figure}

One may distinguish \emph{sequential} and \emph{partial order} plans.
In the former the sequence of operations is completely ordered while
in the latter several operations can be completed in the same step.
Further extensions to STRIPS planning, such as typed objects, temporal
logic  constructs, operation costs (for preferential planning),
etc. are beyond the scope of present discussion but can be accommodated.
In the following section we describe a mapping of planning problem in the
propositional form onto quadratic unconstrained binary optimization,
inspired by {\textsc{GraphPlan}} algorithm (see Fig.~\ref{fig:graphplan}).
The intermediate form may be generated from PDDL definition using a classical
algorithm and quantum annealing could be applied to solve
Quadratic Unconstrained Binary Optimization (QUBO) problem.

\subsection[Mapping to QUBO]{\label{sec:plan-qubo} Mapping to Quadratic Unconstrained Binary Optimization (QUBO)}

Finding a plan of finite length $L$ is a constraint satisfaction problem which
can be encoded as an instance of quadratic binary optimization. We will need
$\left( L + 1 \right) \times N$ propositional variables $x_i^{\left( t
\right)}$ (note ``time'' index $t = 0, 1, \ldots, L$). In addition we will
need $L \times M$ binary variables $y_j^{\left( t \right)}$ indicating that
operation $j$ has been performed between the times of $\left( t - 1 \right)$
and $t$. The cost function will be built up from several components. First, we
need to ensure that a valid plan satisfies boundary conditions, which is
ensured by having the following contribution to the cost function:
\[ H_{\textrm{boundary}} = \sum_{i \in \mathcal{I}^{\left( + \right)}} \left( 1
   - x_i^{\left( 0 \right)} \right) + \sum_{i \in \mathcal{I}^{\left( -
   \right)}} x_i^{\left( 0 \right)} + \sum_{i \in \mathcal{G}^{\left( +
   \right)}} \left( 1 - x_i^{\left( L \right)} \right) + \sum_{i \in
   \mathcal{G}^{\left( - \right)}} x_i^{\left( L \right)} \]
This contribution is non-negative and equals zero if and only if states at
steps $t = 0$ and $t = L$ correspond to the initial state and the goal state
respectively.

The constraints on operations are due to preconditions encoded in the
quadratic term
\[ H_{\textrm{conditions}} = \sum_{t = 1}^L \sum_{j = 1}^M \left( \sum_{i \in
   \mathcal{C}_j^{\left( + \right)}} \left( 1 - x_i^{\left( t - 1 \right)}
   \right) y_j^{\left( t \right)} + \sum_{i \in \mathcal{C}_j^{\left( -
   \right)}} x_i^{\left( t - 1 \right)} y_j^{\left( t \right)} \right) \]
and a similar term for the effect of operations:
\[ H_{\textrm{effects}} = \sum_{t = 1}^L \sum_{j = 1}^M \left( \sum_{i \in
   \mathcal{E}_j^{\left( + \right)}} y_j^{\left( t \right)} \left( 1 -
   x_i^{\left( t \right)} \right) + \sum_{i \in \mathcal{E}_j^{\left( -
   \right)}} y_j^{\left( t \right)} x_i^{\left( t \right)} \right) \]
So far we have assumed that any operations may be executed in parallel.
Typical planning problems seek sequential plans. However, requiring
that $\sum_j y_j^{\left( t \right)} = 1$ at every step would lead to long
plans and be extraordinary wasteful in terms of resources. Instead we should
permit several operations at the same time for as long as they can be executed
sequentially in arbitrary order. This imposes an additional constraint
that operations are in conflict if positive preconditions of one
overlap with negative effects of another or vice versa. That such
operations cannot execute concurrently is ensured by the following term:
\[ H_{\textrm{conflicts}} = \sum_{t = 1}^L \sum_{i = 1}^N \left( \left( \sum_{j
   \vert i \in \mathcal{C}_j^{\left( + \right)}} y_j^{\left( t \right)} \right)
   \left( \sum_{j \vert i \in \mathcal{E}_j^{\left( - \right)}} y_j^{\left( t
   \right)} \right) + \left( \sum_{j \vert i \in \mathcal{C}_j^{\left( - \right)}}
   y_j^{\left( t \right)} \right) \left( \sum_{j \vert i \in \mathcal{E}_j^{\left(
   + \right)}} y_j^{\left( t \right)} \right) \right) \]
where the inner sums are restricted to operations $j$ such that proposition
$i$ belongs to sets $\mathcal{C}_j^{\left( + \right)}$, $\mathcal{E}_j^{\left(
- \right)}$, $\mathcal{C}_j^{\left( - \right)}$ and $\mathcal{E}_j^{\left( +
\right)}$. The net effect is that having conflicting operations $j$ and $j'$
in the same interval would contribute a positive penalty term proportional to
$\left| \mathcal{C}_j^{\left( + \right)} \cap \mathcal{E}_{j'}^{\left( -
\right)} \right| + \left| \mathcal{C}_j^{\left( - \right)} \cap
\mathcal{E}_{j'}^{\left( + \right)} \right|$.

The role of the last contribution is to ensure that the state is preserved
unless it appears in the effects set of any operations at a given time step.
We use
\[ H_{\textrm{no-op}} = - \varepsilon \sum_{t = 1}^L \sum_{i = 1}^N
   \left( 1 - 2 x_i^{\left( t - 1 \right)} \right) \left( 1 - 2 x_i^{\left( t
   \right)} \right) . \]
Unlike previous terms, it can be negative and it is a weak coupling favoring
$x_i^{\left( t - 1 \right)} = x_i^{\left( t \right)}$. The constant
$\varepsilon > 0$ should be small (e.g. $\varepsilon < \frac{1}{NL}$) so that
all other constraints are satisfied first.

The overall cost function is given by the sum of all contributions listed
above,
\[ H_{\textrm{plan}} = H_{\textrm{boundary}} + H_{\textrm{conditions}} +
   H_{\textrm{effects}} + H_{\textrm{conflicts}} + H_{\textrm{no-op}} . \]
Because this function is of QUBO form its minimum can be found using D-Wave quantum annealing procedure
described in Secs.~\ref{sec:QA}, \ref{sec:QA-super}.

The construction guarantees that the global minimum of this QUBO
corresponds to a valid plan of length $L$ if such plan exists. In such
a case the cost function is non-positive (all hard constraints are
satisfied) and additional processing (on a classical computer) is needed to verify
that no weak constraints involving no-ops are violated.

As with \textsc{GraphPlan}, increasing values of $L$, starting from an
initial guess, are tried until a valid plan is found. The subroutine verifying
the existence of a plan is the bottleneck (proving non-existence is
most difficult for classical computers) and may benefit from possible
speed-up with quantum adiabatic coprocessor.

The mapping presented above is not necessarily optimal in terms of the
number of qubits used. The decision to allow concurrent operations
prevents the length $L$ of shortest valid plan from becoming
inordinately large. In principle, all qubits corresponding to the
initial state $\{x_i(0)\}$ can be dispensed with altogether, replaced
by their known values. Similar substitutions can be performed for
all goal variables. Likewise, only operations with preconditions
consistent with the initial state are possible at $t=1$ so that the
set of $j$'s where $y_j(1)$ is guaranteed to be zero is known at
generation time. This constant propagation is continued for
$t=2,3,\ldots$ and backwards for $t=L,L-1,\ldots$; it can be done very
efficiently on a classical computer. The final graph looks
more like a tree rather than a lattice with the benefit of requiring
reduced number of qubits.

\section{\label{sec:diagnostics} Diagnostics and Recovery}

Onboard autonomy is of utmost importance to deep space exploration
where real time ground support is not available. A very important component
involves monitoring hardware modes, detecting anomalies, and isolating faults
as well as recovery and/or avoidance. Perhaps the best-known examples are
the {\textsc{Livingstone}} system and its immediate successor
{\textsc{Livingstone2}}. The former achieved prominence after being chosen as
part of the core autonomy architecture of {\textsc{Deep Space One}} probe
where it works in concert with higher-level planner and scheduler to enable
greater degree of robustness. This reflects an ongoing trend of departure from
bug-prone and costly monolithic systems to more flexible component-based
architectures with a generic core and easily reprogrammable models. As a
result, the system has been reused in numerous other applications including
EO-1 \cite{eo1liv} and flight experiments on X-34 and X-37 \cite{x37}
as well as various testbeds \cite{psa,rover}.

The central tasks of diagnostic tools such as {\textsc{Livingstone}} are closely
related {\emph{Mode Identification}} (MI) and {\emph{Mode Reconfiguration}}
(MR). The former continuously analyze input from sensors and known control
variables to identify a particular hardware mode and whether it deviates from
normal behavior. The latter attempts to adjust controls to achieve original
high-level goals even if undesirable transitions due to malfunction do take
place. The model is expressed in propositional logic, a property largely shared
with planning and scheduling; as a result many ideas presented in the previous
section can be reused for diagnostics and recovery. A subset of linear
temporal logic is used to define a transition system. Fully expanded, it
corresponds to non-deterministic finite automaton: in any given state in
addition to one nominal transition, there could be multiple low-probability
transitions indicating temporary or permanent failures. Since a combined input
of sensor measurements is typically insufficient to identify a particular
state uniquely, the software must use a detailed transition model and control
inputs in addition to sensor input to identify a trajectory in state space
with sufficient certainty and detect that one of faulty branches had been
taken. Practical tests showed that the detection is achievable with acceptable
delay. In mode reconfiguration, new values of control variables must be
generated to return to normal behavior; again testing on simplified model of
Cassini spacecraft showed successful recovery for any single failure. Thus
autonomous diagnostics and recovery has firmly established itself as
inseparable part of any viable space exploration program.

The pioneering work \cite{ds1} acknowledges that both MI and MR are essentially
combinatorial search/optimization problems and implementations utilized known
approaches that originated in that field, such as conflict-directed best first
search. The connection to optimization problems becomes self-evident once we
recognize that the goal of the mode identification problem is to find a trajectory
maximizing the posterior probability in accordance with Bayes' rule:
\[ P \left( X_n \vert O_n, \ldots, O_0 \right) \propto \sum_{\left\{ X_{n - 1},
   X_{n - 2}, \ldots \right\}} \mathcal{I} \left( O_n ; X_n \right) \times P
   \left( X_{n - 1} \rightarrow X_n \right) \times \mathcal{I} \left( O_{n -
   1} ; X_{n - 1} \right) \cdots \]
where $\mathcal{I} \left( O_i ; X_i \right) = 1 \left( 0 \right)$ if
observation $O_i$ is (is not) consistent with state $X_i$ and $P \left( X_i
\rightarrow X_j \right)$ are the transition probabilities. \ Model
reconfiguration may contain costs associated with performing specific
corrective actions, each using up precious resources. The overall cost
function is expected to have an additive form: sum of costs of individual
actions, which may facilitate mapping to quadratic optimization.

As with planning and scheduling, number of state variables can become quite
large, on the order of (\texttt{number of state variables}) $\times $ (
\texttt{number of timesteps within horizon}), but it is still bounded as a polynomial of problem
size. Yet the running time of the combinatorial optimization problem is known
to increase exponentially with problem size. Currently, reducing the number of
state variables and pruning branches that are explored are among strategies
that keep this running time within reasonable limits, but the relentless
exponential rise in time complexity will limit scalability of traditional
approach in future.

For illustration purposes, in the following sections we describe a simpler but
nevertheless important problem in diagnostics: the analysis of fault trees.
Due the more ``static'' nature of the problem, it requires fewer state
variables and can conceivably be implemented using current quantum hardware.

\subsection[ Fault tree analysis]{\label{sec:fault-tree} Fault tree analysis}

\begin{figure}
\includegraphics[width=5.0in]{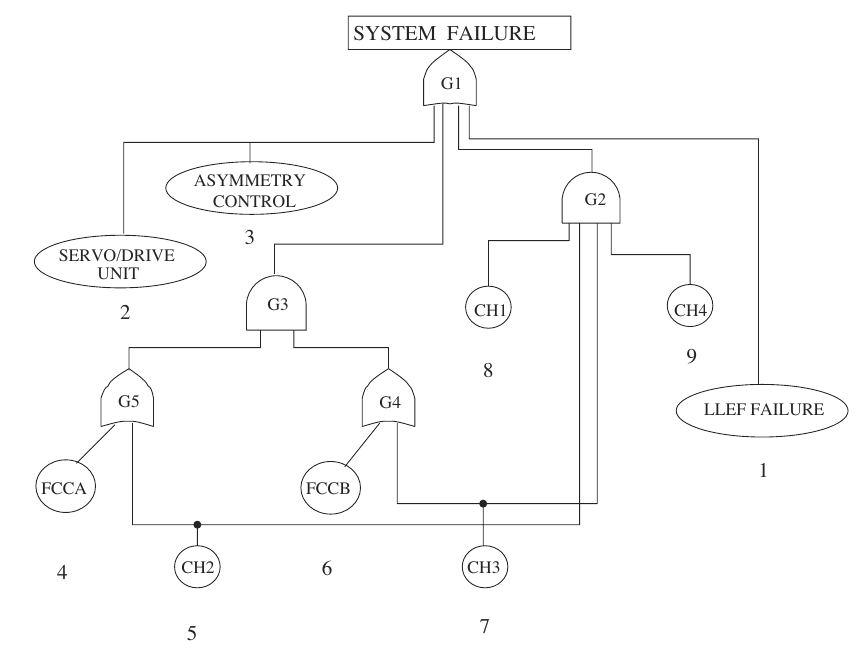}
\caption{\label{fig:faulttree}
A sample fault tree model of F18 flight control system.
Reprinted from Ref.~\cite{faulttree}
}
\end{figure}

Fault tree models are graphical representation of logical relationships
between events. It consists of a top event indicating (sub)system failure at the
root of the tree linked to more basic events (component failures) which are
the leaves of the tree by a combination of logic gates (internal nodes). The
fundamental problem in fault tree analysis is isolating the most likely cause
of malfunction: finding the most likely combination of basic events that would
lead to top event. Fig.~\ref{fig:faulttree} gives a simple example of the
fault tree; typical gates include logical AND, OR, or the majority (MAJ)
function, where e.g. at least 2 out of 3 subsystems must fail before a fault
can propagate further. One approach is finding the {\emph{minimum cut set}},
the smallest subset of basic events that would result in system failure. This
problem can be solved using {\emph{top-down}} approach either by hand or using
simple software for {\emph{pure trees}}. While originally conceived as such,
modern definition generalizes this to {\emph{directed acyclic graphs}} (DAG)
and some extensions allow loops \cite{loops}. When a model is a general DAG (i.e.
fanout of gates is allowed to be more than $1$) or basic events can appear
more than once, the problem becomes NP-complete \cite{mincut}. {\emph{Weighted}}
version of minimum cut set is more appropriate when {\emph{a priori}}
probabilities of individual component are known. But even computing the
probability of the top event given a vector of probabilities of basic events is
NP-hard in general case \cite{fault}.

\subsection{\label{sec:fault-tree-qubo} Mapping to QUBO}

The constrained optimization problem of the previous section can be easily mapped to
unconstrained quadratic optimization. We use binary variables $z_0, z_1,
\ldots, z_N$ where $z_i = 1$ indicates a particular event. {\emph{Top event}}
shall be represented by $z_0$ and a subset $\mathcal{B} \subset \left\{ 1, 2,
\ldots, N \right\}$ shall correspond to {\emph{basic events}}. All events are
interconnected through $M$ logical gates.

The main idea is encode relationships of the form $y = g \left( x_1, \ldots,
x_k \right)$ as a quadratic form having a value of $0$ if it is satisfied and
some positive value otherwise. For the 2-input AND gate we write
\[ H_{\textrm{AND}} \left( y, x_1, x_2 \right) = y + x_1 + x_2 + x_1 x_2 - 2
   y x_1 - 2 y x_2, \]
and for 2-input OR gate
\[ H_{\textrm{OR}} \left( y, x_1, x_2 \right) = 3 y + x_1 x_2 - 2 y x_1 - 2 y x_2 . \]
Gates with 3 or more inputs can be constructed by cascading 2-input gates. To
encode 3-input majority gate we use
\[ H_{\textrm{MAJ}} \left( y, x_1, x_2, x_3 \right) = 3 y - 2 y \left( x_1 + x_2
   + x_3 \right) + x_1 x_2 + x_1 x_3 + x_2 x_3 \]
and similar constructions can encode other gates, some possibly requiring use
of ancillary variables.

The complete cost function may be written in the following form:
\[ H_{\textrm{fault-tree}} = A \sum_{j = 1}^M H_{\textrm{gate}}^{\left( j  \right)} + B \left( 1 - z_0 \right) + \sum_{i \in \mathcal{B}} w_i z_i \]
where the first sum is over all gates, using an appropriate expression (e.g.
$H_{\rm AND}$, $H_{\rm OR}$, or $H_{\rm MAJ}$) with appropriate events $z_i$ substituted for $y, x_1,
\ldots, x_k$. For unweighed minimum cut we set $w_i = 1$ for all basic events
while choosing $A > 3 N$ and $B > 3 MA$ ensures that in the global minimum
$z_0 = 1$ and all gate constraints are satisfied. More generally, weights can
chosen in proportion to the logarithm of probabilities of basic events $w_i
\propto \log p_i$ if such information is available.

The above cost function is in QUBO form which can then
be translated into a problem  Hamiltonian $H_P$ (\ref{eq:HP-Ising}) implemented with D-Wave superconducting qubit hardware on Ising graph.
For this Hamiltonian  the optimal solution of QUBO is a ground state that
can be sought by running the QA algorithm described in Sec.~\ref{sec:QA}.
When the non-zero terms of $H_P$ match that of the D-Wave architecture,
the problem can be run on current D-Wave hardware of Sec.~\ref{sec:QA-super}.

\section[Autonomous Exploration]{\label{sec:uae} Unmanned Autonomous Exploration} As spacecraft venture farther from Earth, the decreasing chances of
human survivability and speed of light limited communication delays,
compel NASA to rely increasingly on unmanned autonomous systems.
Even today in safety critical situations on Earth many agencies are already using unmanned autonomous ground, sea, and aerial vehicles, i.e., UGVs, USVs, and UAVs to perform dangerous missions.
A representative example of the type of difficult computational problem
that arises in autonomous unmanned exploration is the assignment of tasks
among multiple UAVs designed to cooperate to achieve a collective mission
\cite{UAV2004, UAV2005a, UAV2006, UAV2009, UAV2010}. A feasible task
assignment must respect the physical limitations of the vehicles, such
as constraints on their turn radius, flight endurance, and altitude ceiling.

A simple model of a UAV flight dynamics is obtained by assuming the UAV is flying in a horizontal plane at constant speed $v$, has an onboard resource capacity (e.g., flight endurance due to a limited fuel supply) of $b$, and is constrained to bank at a certain maximum rate of turn $\omega^{\textrm{max}}$ \cite{RasmussenShimaBranchAndBound2006}. Thus, the dynamics of the $i$-th UAV in a group of UAVs can be described by the following set of differential equations:
\begin{eqnarray}
\label{eom1} \dot{x_{i}} &= &v_{i} \cos \theta_{i} \\ \label{eom2} \dot{y_{i}} &= &v_{i} \sin \theta_{i} \\ \label{eom3} \dot{\theta_{i}} &= &\omega_{i}^{\textrm{max}} u \\ \label{eom4}\dot{v_{i}} &= &0 \end{eqnarray} \noindent where $x_{i}$ and $y_{i}$ are the horizontal coordinates of the $i$-th UAV, $\theta_{i}$ is its azimuthal angle (i.e., its flight direction), $u = \{-1, 0, +1\}$ depending on whether the vehicle is banking left, flying straight, or banking right, and $\omega_{i}^{\textrm{max}}$ and $b_{i}$ are its limitations in terms of rate of turn and endurance. By fixing the $(x, y, \theta)$ coordinates of a UAV at a ``current'' location (e.g., the takeoff point, a way-station, or a target location) and a ``next'' location one can solve the aforementioned equations of motion to find a minimum distance trajectory, and hence (if flying at constant speed) a minimum flight time, connecting those coordinates while respecting the flight limitations of the UAV.

\subsection[Multi-UAV Task Assignment]{\label{sec:multi-uav} Multi-UAV Task Assignment as Combinatorial Optimization} To formalize the multi-UAV task assignment problem we need to introduce the basic parameters defining  the mission. Accordingly, let \begin{eqnarray} N_{T} &= &\textrm{the number of targets} \\ N_{U} &= &\textrm{the number of UAVs}\\ N_{t} &= &\textrm{the number of tasks per target} \\ N_{A} &= &\textrm{the number of task assignments to be made} \end{eqnarray} \noindent where $N_{A} = N_{T} \times N_{t}$ if the UAVs share an equal workload.
The multi-UAV task assignment problem can then be formalized as the problem of finding an assignment of tasks to UAVs such that all the tasks are performed on all the targets while minimizing some desirable objective function. Such an objective function could be, e.g., the cumulative flight times of all the UAVs. If the cumulative flight time is minimized and the UAVs fly at constant and equal speed this corresponds to minimizing the net fuel burn to accomplish the mission. Conservation of fuel is, of course, of special concern to NASA in remote environments be they on Earth or other planets.
To formalize the problem mathematically, we follow the prescription of Rasmussen and Shima \cite{RasmussenShimaBranchAndBound2006}.
Let us define a binary indicator variable, $z_{\ell, i, j} \in \{0, 1\}$,  such that:
\begin{equation}
z_{\ell, i, j} \in \{0, 1\} =\left[
\begin{array}{cc}
  1 & {\rm if \,at \,the \,}\,\ell^{\rm -th} \,{\rm assignment\, UAV} \, i\, {\rm performs\, a\, task\, on\, target} \, j \\
  0 & {\rm otherwise}
\end{array} \right. .
\end{equation}
\noindent We can then define a vector of such binary variables that represents the set of assignments of tasks to UAVs made to date. Specifically, let $\textbf{z}_{\ell} = \{z_{1,i,j}, z_{2,i,j}, \ldots, z_{\ell,i,j}\}$ be the set of assignments made up to (and including) the $\ell$-th assignment. This allows us to couch the optimization problem as that of finding the binary indicator vector, $\textbf{z} \equiv \textbf{z}_{N_{A}}$, which corresponds to a complete set of assignments of tasks to UAVs such that all the tasks have been performed on all the targets while respecting the flight characteristics of the UAVs. Formally, the cost function for optimization becomes
\begin{equation}  E(\textbf{z}) = \sum_{i=1}^{N_{U}} D_{i},
\end{equation}
\noindent where $D_{i}$ is the total distance traveled by the $i$-th UAV from the start of the mission to its termination. Note that the termination condition for the mission is not fixed. One could choose to terminate the mission at the locations of the last targets visited for each UAV, the origination points of the UAVs, or different landing sites.
To evaluate the distance traveled by the $i$-th UAV, $D_{i}$, we need to refer to the dynamical equations of motion (Eqns. \ref{eom1}-\ref{eom4}) that describe feasible fixed wing UAV trajectories. We can use these equations of motion to predict the quantity $d_{\ell, i, j}^{\textbf{z}_{\ell-1}}$, which is the incremental distance that would need to be traveled by UAV $i$ to perform a task on target $j$ if this were to be made the $\ell$-th task assignment given the prior history of assignments $\textbf{z}_{\ell - 1}$. Likewise, once engaged on the task we can define $r_{\ell, i, j}^{\textbf{z}_{\ell-1}}$ to be the resource required by UAV $i$ to perform the task on target $j$ as the $\ell$-th assignment. Combining the incremental distance flown and the increment resource consumed we can define the combinatorial optimization problem we must solve as:
\begin{equation}
E(\textbf{z}) = \sum_{\ell = 1}^{N_{A}} \sum_{i = 1}^{N_{U}} \sum_{j = 1}^{N_{T}} d_{\ell, i, j}^{\textbf{z}_{\ell-1}} z_{\ell, i, j} \end{equation} \noindent subject to the constraints \begin{equation} \sum_{i = 1}^{N_{U}} \sum_{j = 1}^{N_{T}} z_{\ell, i, j} = 1 \end{equation} \noindent  which requires that at the $\ell$-th assignment exactly one task on target $j$ is assigned to UAV $i$,  \begin{equation} \sum_{\ell = 1}^{N_{A}} \sum_{i = 1}^{N_{U}} z_{\ell, i, j} = N_{t}  \end{equation}  \noindent  which requires that each target has the requisite number of tasks performed upon it, and  \begin{equation} \sum_{\ell = 1}^{N_{A}} \sum_{j = 1}^{N_{T}} r_{\ell, i, j}^{\textbf{z}_{\ell-1}} z_{\ell, i, j} \le b_{i}  \end{equation}  \noindent which requires that the total resources expended by each UAV do not exceed its capacity. Finding an assignment vector $\textbf{z} \equiv \textbf{z}_{N_{A}}$ satisfying all these constraints will be equivalent to finding a set of assignments of tasks to UAVs such that all the tasks are accomplished in the least cumulative flight time, and hence for the least fuel consumed.

\subsection[Mapping  to QA]{Mapping UAV Task Assignment to Quantum Annealing} We can use a simplified version of the UAV task assignment problem to illustrate how this can be mapped to quantum annealing. Suppose that we have a single UAV that performs a single task on each target. For example, the UAV might simply orient and maintain a sensor on a target and record data as it flies over it. In typical NASA scenarios there may be a preferred azimuthal direction in which the overflight is to occur. For example, to maximize the visibility of a target to the sensor on the UAV one might prefer to approach each target in a direction that avoids the highest terrain surrounding it. In such a situation we can specialize the problem to the following. Let \begin{eqnarray} N_{T} &= &\textrm{the number of targets} \\ N_{U} &= &1 = \textrm{the number of UAVs}\\ N_{t} &= &1 = \textrm{the number of tasks per target} \\ N_{A} &= &N_{T} = \textrm{the number of task assignments to be made} \end{eqnarray} \noindent We further suppose that the targets are at known locations and are required to be traversed in specific directions. Hence, we parameterize each target by the triples $T_{1} \equiv (x_{1}, y_{1}, \theta_{1}), T_{2} \equiv (x_{2}, y_{2}, \theta_{2}), \ldots, T_{N} \equiv (x_{N}, y_{N}, \theta_{N})$. Given this target specification, we can pre-compute the $N^{2}$ incremental flight distances, $d_{i j}$, a UAV at target $T_{i}$ flying in the direction $\theta_{i}$ would need to incur in order to overfly a target $T_{j}$ in the direction $\theta_{j}$. Note that this is not the standard Euclidean distance between targets $T_{i}$ and $T_{j}$ as the UAVs are required to fly trajectories that respect their physically realizable flight characteristics. In particular, they cannot turn on a point but rather can only bank at some maximum rate of turn $\omega^{\textrm{max}}$. This constraint means that the UAVs fly trajectories built out of circular arcs (of some minimum feasible radius) and straight line segments. Although $d_{i j}$ might be unfamiliar it is feasible to compute it for any given pair target triples, $T_{i} \equiv (x_{i}, y_{i}, \theta_{i}), T_{j} \equiv (x_{j}, y_{j}, \theta_{j})$. Indeed, this metric has been consider by mathematicians before \cite{Dubins}.
Our simplified UAV task assignment problem can then be thought of as a traditional traveling salesman problem supplemented with a peculiar distance metric between cities. Once we recognize this we can map the UAV task assignment problem into a minimization problem over binary variables using a mapping similar to that proposed by Hopfield and Tank for mapping TSP to neural nets \cite{HopfieldTank}. Specifically, introduce the binary variable $z_{i,\alpha}$ defined as:
\begin{equation}
z_{i, \alpha} =
\begin{cases}
1 &\mbox{if target $i$ is the $\alpha$-th site visited on the tour} \\ 0 & \mbox{otherwise} \end{cases} \end{equation} \noindent As there are $N$ targets, there must therefore be $N^2$ such variables and the total length of a tour of all the targets is just:
\begin{equation}
 E(\textbf{z}) = \frac{1}{2} \sum_{i j \alpha} d_{i j} z_{i, \alpha} (z_{j, \alpha+1} + z_{j, \alpha - 1}) \end{equation} \noindent defined over the vector of $\textbf{z} \equiv \{z_{i, \alpha}\}$ binary variables. The intuition is that the products $z_{i, \alpha} z_{j, \alpha+1} \in \{0, 1\}$ and $z_{i, \alpha} z_{j, \alpha-1} \in \{0, 1\}$ serve as indicator functions as to whether or not a particular distance value, $d_{i j}$, should be included in the tour cost. The factor of $\frac{1}{2}$ prevents double counting a particular tour (as each tour may be traversed in two directions). We can then convert these indicator variables to Ising spins $\pm 1$ by making the transformation $z_{i,\alpha} \rightarrow \frac{1}{2}(s_{i,\alpha}+1)$, where each $s_{i,\alpha}$ can be either spin up ($+1$) or spin down ($-1$). The cost function for optimization can then be written in terms of the vector of these  spin variables ${\bf s}=(s_1,s_2,\ldots,s_N)$ as:
\begin{equation}
\label{tspLength}
 E(\textbf{s}) =  \sum_{i j \alpha} \frac{1}{4} d_{i j} + \frac{1}{8} d_{i j} (2 s_{i,\alpha} + s_{j,\alpha -1} + s_{j,\alpha +1}) + \frac{1}{8} d_{i j} s_{i,\alpha} (s_{j,\alpha -1} + s_{j,\alpha +1})
\end{equation}
\noindent \nopagebreak
 However, we must also impose two additional constraints that ensure each target is visited exactly once and all the targets are visited, namely,
\begin{eqnarray}
\label{tspOnce}
\sum_{\alpha} z_{i, \alpha} &= &\sum_{\alpha} \frac{1}{2} (s_{i,
  \alpha}+1) = 1  {\textrm{\footnotesize~~(the $i$ target is visited
    exactly once in the tour)}} \\ \label{tspAll} \sum_{i} z_{i,
  \alpha} &= &\sum_{i} \frac{1}{2} (s_{i, \alpha}+1) = 1
{\textrm{\footnotesize~~(only one target is visited at each position in the tour)}}
\end{eqnarray}
\noindent Hence, our overall binary function to be minimized, call it $E^{\prime}(\textbf{s})$, should be a weighted sum of the aforementioned contributions, namely, Eqns. (\ref{tspLength}),(\ref{tspOnce}), (\ref{tspAll}). The weighting factors chosen for the last two constraints need to be large enough to have influence and yet not so large as dominate the tour length component of the objective function. Thus, our UAV task assignment problem can be solved by minimizing the objective function:
\begin{eqnarray}
 E^{\prime}(\textbf{s}) &= &\frac{1}{4} d_{i j} + \frac{1}{8} d_{i j} (2 s_{i,\alpha} + s_{j,\alpha -1} + s_{j,\alpha +1}) + \frac{1}{8} d_{i j} s_{i,\alpha} (s_{j,\alpha -1} + s_{j,\alpha +1}) + \\ & &~~~~W_{1} \sum_{\alpha} \frac{1}{2} (s_{i, \alpha}+1) + W_{2} \sum_{i} \frac{1}{2} (s_{i,\alpha}+1)
\end{eqnarray}
\noindent where the weights $W_{1}$ and $W_{2}$ should be constants of order ${\cal O}(N \langle d_{i j} \rangle)$, which would give the three contributions to the objective function roughly equal weight. The cost function $E^\prime({\bf s})$ is in Ising form and its minimum can be sought using quantum annealing.
This process can be implemented with D-Wave machine described in Sec.~\ref{sec:QA-super} when the non-zero terms in Hamitonian derived from $E'$ match
the qubit connection architecture in the D-Wave hardware.

\section[Structured Learning]{\label{sec:strLearning} Structured Learning for Multiple
Label Classification}

A common first step in the automatic analysis of large and complex
data sets is the annotation of data with additional meta-data.
This meta-data may simply be a tag suggesting that a human take a
further look at potentially interesting structure, or the meta-data
may represent complex higher-level information inferred from the raw data.
Frequently there are structural dependencies between the labels.
For example, each category would have potentially dozens of
subcategories, leading to a very large number of potential
classifications.
Typically, the learning of appropriate meta-data annotation is done in a
supervised manner. Machine learning algorithms are trained from examples
annotated by humans. From the training examples,
structured learning algorithms learn to generalize the features that
lead to particular annotations.

A prototypical example
of a structured labeling task is optical character recognition (OCR)
which labels images of words with a sequence of letters. Each character
in the image could be labeled separately, but significantly better
performance is obtained by jointly labeling the characters in a word,
taking into account likely relationships between component letters
of words, and the relationship between characteristics of the
of the image and these labels.
This approach to character recognition is applicable to all NASA
projects involving scanned documents or images that include text.
Such documents form a substantial part of the information collected
by the NASA Engineering Network \cite{NEN,Topousis:2007} that captures
"lessons learned" from the NASA engineering community. Quick and accurate
interpretation of vast text and image data enables the highly targeted
retrieval and interpretation of the information collected over the long
history of engineering development from a broad variety of sources.

Much of the data returned from NASA missions has structure. The data
may be sequential, a set of images or sensor readings taken at
regular intervals within a small time window,
it may have spatial structure, a set of objects detected at various
distances from each other,
or there may be complex relationships between the various
properties assigned to an object, a set of interrelated characteristics
of an object.
For example, NASA may wish to assigm meta-data
to each image of a sequence of images taken by a rover, or to
events reported to the NASA Engineering Network (NEN) lessons learned
system \cite{NEN}.
NASA may wish to assign labels to each pixel in an image, such as
Fig.~\ref{fig:iceClass} or MODIS images as part of the Planetary Skin
project \cite{PlanetarySkin}, taking into account the spatial relationships
between the pixels.
Or NASA may wish to label terrain \cite{TerainClass},
using data from muliple sensors, with a
complex set of descriptors, such as whether smooth or rough, level
or sloped, sand, pebbles, soil, rock, or bedrock, traversable or
impassible, high or low ground clearance, hazardous or safe, where
there are interdependencies between these labels.

Many algorithms exist for the prediction of simple meta-data.
The simplest such predictions annotate raw data with an extra bit
(e.g., indicating whether an object is a rock or not).
Support vector machines (SVM) are a common and
effective approach for this class of problems.
In principle, more complex labeling with meta-data could be carried out
by applying an SVM to each label independently from all others. This
simplification, however, makes the problem more difficult than it really
is as there are relationships amongst meta-data elements
that can be exploited. Extensions
of support vector machines to this more complex problem have been
developed \cite{Taskar03,Tsochantaridis2005}, but exploiting this additional
structure creates difficult optimization problems. For a quantum
annealing approach we follow that of Bian et al.~\cite{Bian2011}.

\subsection{Mapping Optimal Labeling onto Quantum Annealing}

Let $X$ be the space of all possible objects ${\bf x}$ to be classified,
each represented by $N$ data values ${\bf x}=\{x_1,\ldots,x_N\}$,
that will be used to determine its
labeling. In other words, $X$ is an $N$ dimensional space.
Let $Z = Z_1 \times Z_2 \times \cdots \times Z_L$ be the space
of possible labelings. This general setting subsumes the binary
classification problem considered in section \ref{sec:classification} where there is a single label, and can capture a variety of
structured situations. In some cases, $Z_i$ and $Z_j$ consist of
the same set of labels, and an element ${\bf z}=\{z_1,\ldots,z_L\}\in Z$
is a sequence of those labels, or a spatially structured set of labels,
as in the case of labeling each pixel in an image.
For example, $X$ could be a set of images of handwritten words of length $L$,
where each image has $N$ pixels and the data values representing the
image are the pixel intensities. In this case, $Z_i$ is the alphabet,
and a label ${\bf z} = (z_1, z_2, ..., z_L)$ labels each of the $L$
handwritten characters in the word with a letter of the alphabet.
Each handwritten character could have been assigned a letter just
on the basis of its graphic characteristics alone, but jointly labeling
the characters of a word, taking into account statistical frequencies
for one letter following another for example, greatly increases the
accuracy of handwriting recognition.

As another simple example, each ${\bf x} \in X$ could represent a
collection of items, some of which are rocks, and
$Z_i$ could be $\{0,1\}$, where $z_i\in Z_i$ indicates whether the
$i^{\rm th}$ item is classified as a rock or not. In this case,
${\bf z}$ labels the collection ${\bf x} \in X$ with a length $L$
bit-string indicating which items in the collection are rocks or not.
Pixels in an image assigned the rock label are more likely to have neighboring pixels also assigned as rocks. Thus, improvements in labeling accuracy  can be obtained by exploiting such correlational information between labels. In other cases, $Z_i$ and $Z_j$ may contain different sets of labels.
For example, in the case of terrain,
$Z_1$ could be $\{{\rm smooth}, {\rm rough}\}$,
$Z_2$ could be $\{{\rm level}, {\rm sloped}\}$,
$Z_3$ could be $\{{\rm traversable}, {\rm impassible}\}$, etc.
We will restrict to the case where all $Z_i = \{0,1\}$, but
these values can be interpreted in these different ways.
In this case elements ${\bf z}$ are binary strings.

The goal is to learn a classification function $h: X \to Z$,
where $h$ typically comes from a parameterized family of functions
$\cal{H}$. Here, we consider a family consisting of linear
combinations of {\it basis functions}
or {\it features} $f_\alpha: X\times Z \to \mathbf{R}$, where
the family is parameterized by the weights ${\bf w} = \{w_\alpha\}$
determining the linear combination, ${\cal H} = \{h_{\bf w} \}$.
The functions $f_\alpha$, discussed in more detail shortly,
capture structural relationships between labels as well as relationships
between characteristics of objects in $X$ and the labels.
A hypothesis $h_{\bf w}$ maps an object ${\bf x}$ to an optimal
set of labels ${\bf z}_*$ for the weighting ${\bf w}$:
\begin{equation}
{\bf z}_{*}=h_{\bf w}({\bf x}) =  \underset{\bf z}{\operatorname{argmin}}\,\,{\cal E}({\bf x},{\bf z},{\bf w}),\label{eq:h}
\end{equation}\noindent
where
\begin{equation}
{\cal E}({\bf x},{\bf z},{\bf w})=\sum_\alpha w_\alpha f_\alpha({\bf x}, {\bf z}).\label{eq:Exz}
\end{equation}
\noindent
and  ${\bf w}=\{w_\alpha\}$. Thus, given a hypothesis specified by
a vector of weights ${\bf w}$, labeling is done by solving this binary
optimization problem to find the optimal assignment
of $0$ or $1$ to each $z_i$ in ${\bf z}$ given ${\bf x}$.

As mentioned earlier, common practice restricts to considering only
pairwise relationships between the labels. Furthermore, only
a sparse subset of these relationships may be considered. This
sparse subset can be represented by a sparse graph $G$, with a vertex
for each $Z_i$, and an edge $(i,j)\in E$ between those vertices
whose relationship we wish to capture. Thus, we restrict
to the case in which each basis function $f_\alpha$
is of the form $f_\alpha: X\times Z_i\times Z_j \to \mathbf{R}$ for some
pair $(i,j)\in E$.
There are various possibilities for the specific form of these basis
functions. Here, we consider a particularly simple form in which,
for each dimension $k$ of the $N$-dimensional domain $X$,
there is one basis function $f_\alpha$ for each
pair $(i,j)\in E$, and $f_{k, (i,j)}({\bf x},{\bf z}) = \psi_k(\mathbf{x}) z_iz_j$.
The feature function vector $\pmb{\psi}(\mathbf{x}) = [\psi_0(\mathbf{x}), \psi_1(\mathbf{x}), \cdots , \psi_F(\mathbf{x}) ]$ allows for non-linear dependence on $\mathbf{x}$. This freedom is exploited in the learning with kernels literature, where inputs $\mathbf{x}$ are mapped under $\pmb{\psi}$ into a feature space where a linear model is fit to data. Kernel methods rely indirectly on $\pmb{\psi}(\mathbf{x})$ through a similarity measure $K(\mathbf{x},\mathbf{x}') = \langle \pmb{\psi}(\mathbf{x}), \pmb{\psi}(\mathbf{x}')\rangle$ which measures the likeness of inputs $\mathbf{x}$ and $\mathbf{x}'$. Non-linear kernels such as $K(\mathbf{x},\mathbf{x}) = (1+\langle \mathbf{x},\mathbf{x}'\rangle)^2$ can be accommodated with appropriate definition of $\pmb{\psi}(\mathbf{x})$.\footnote{Kernels defining infinite-dimensional feature spaces may also be used using the methods of \cite{Rahimi2007}.} In this discussion, we confine ourselves to linear kernels where $\pmb{\psi}(\mathbf{x}) = [1,x_1,\cdots, x_N]$, but note that non-linear features are easy to incorporate.

The labeling process by a specific labeler $h_{\bf w}$ (\ref{eq:h}) can now
be phrased as a quadratic unconstrained binary optimization (QUBO)
with cost function ${\cal E}({\bf x},{\bf z},{\bf w})$
\begin{equation}
{\cal E}({\bf x},{\bf z},{\bf w})= \sum_{k=0}^{F}\sum_{(i,j )\in E} w_{k,(i,j)} \psi_k(\mathbf{x}) z_i z_j
\label{eqn: structQUBO}
\end{equation}
where ${\bf x}=\{x_0,x_1,\ldots,x_N\}$, and the number of $\mathbf{x}$ features is $F+1$.
To capture relationships between ${\bf x}$ and a single feature, we
allow $i=j$, which correspond to linear terms
because the $z_i$ are binary so $z_i^2 = z_i$.
Furthermore, for notational convenience,
we have allowed $k$ to take the value $0$ so that we can incorporate
terms that do not depend on $x$ by using the convention that $x_0 = 1$
for all $x$. As we have seen, this QUBO can be converted directly
to Ising form, as we did before,
by replacing the $z_i$ with spin variables $s_i = 1 - 2 z_i$. The
resulting Ising energy function can be translated into a problem
Hamiltonian $H_P$,
and an optimal or near optimal set of labels can be found using
quantum annealing.

\subsection[A Hybrid Approach to Learning the Weights]{A Hybrid Classical/Quantum Approach to Learning the Weights}

In this framework, structural learning is defined as learning the
parameters of the Ising model (vector of weights ${\bf w}$) so that
Ising minimization yields good labelings. The weights are learned from
a set of training examples
${\cal D} = \{{\bf x}_d,{\bf z}_d\}_{d=1}^{|{\cal D}|}$ each defined
by an object ${\bf x}_d$ and the assignment of label variables ${\bf z}_d$.
The optimal weighting ${\bf w}$  is determined by minimizing an
objective balancing the
two contributions. One contribution $R({\bf w})$ seeks to minimize the errors on the training set,
and the second contribution $\Omega({\bf w})$  ameliorates problems of overfitting by favoring \lq\lq simpler"
models. The best set of weights is then found from the solution of optimization problem as
\begin{equation}
\underset{\bf w}{\operatorname{argmin}}\,\,F({\bf w}),\quad F({\bf w})= \lambda\Omega({\bf w})+R({\bf w}),\label{eq:F}
\end{equation}
\noindent
where the $\lambda$ parameter balances the relative importance of the two contributions and it is customarily set by cross-validation procedure ~\cite{Bian2011}. Typically the regularization term
$\Omega({\bf w})$ favors simpler models having a smaller number of model parameters
\begin{equation}
\Omega({\bf w})= \frac{1}{2}\sum_{k=0}^{F}\sum_{(i,j )\in E} (w_{k,(i,j)})^2.\label{eq:Omega}
\end{equation}

To define the training set error $R({\bf w})$ we first quantify the cost of errors. For each training example $d$ the error can be defined as a Hamming distance $\Delta[{\bf z}_d,{\bf z}_*]$ between the assignment of label variables  ${\bf z}_d$ and the string of optimal labels ${\bf z}_*={\bf h}({\bf x},{\bf w})$ (\ref{eq:h}). Here
\begin{equation}
\Delta\left[{\bf z},{\bf z}^\prime\right]=\sum_{i=1}^{L}(z_i-z_
{i}^{\prime})^2=\sum_{i=1}^{L}(z_i+z_{i}^{\prime}-2 z_i z_{i}^{\prime})
\end{equation}
\noindent
Then the total error for the training set equals to the number of bits differing in the predictions and training data,
\begin{equation}
\sum_{d=1}^{|{\cal D}|}\Delta\left[{\bf z}_d,{\bf h}_{\bf w}({\bf x}_d)\right].\label{eq:R0}
\end{equation}
\noindent In general, this error is a discontinuous function of ${\bf w}$ because the bit string ${\bf h}_{\bf w}({\bf x})$ providing the minimum of the energy ${\cal E}({\bf x},{\bf z},{\bf w})$ with respect to ${\bf z}$ changes abruptly as the parameters ${\bf w}$ varies. In \cite{Bian2011} the training set error was defined via the following expression
\begin{equation}
R({\bf w})=\frac{1}{|{\cal D}|}\sum_{d=1}^{|{\cal D}|}\underset{\bf z}{\operatorname{max}}\left\{
\Delta\left[{\bf z}_d,{\bf z}\right]+{\cal E}({\bf x}_d,{\bf z}_d,{\bf w})-{\cal E}({\bf x}_d,{\bf z},{\bf w})\right\}.\label{eq:R}
\end{equation}
$R({\bf w})$ defined this way upper bounds the error in  Eq.~(\ref{eq:R0}) and
similar functions have been found to work well in practice \cite{Tsochantaridis2005}. With this error measure the cost function $F ({\bf w})$ (\ref{eq:F}) is the maximum of a set of convex quadratic functions and is thus strictly convex
so that it has a single minimum as a function of ${\bf w}$
(see \cite{Boyd2004}, p. 80).

Given a vector of weights $\mathbf{w}$, they used quantum annealing on the
Ising model associated with the QUBO of Eqn.~\ref{eqn: structQUBO}
to obtain a candidate labeling $\mathbf{z}_w^{(d)}$ for each $x^{(d)}$
in the training data. The Hamming distance between the predicted labels
$\mathbf{z}_w^{(d)}$ and the true labels $z^{(d)}$ is used to obtain a
measure of error. A classical optimization method based on subgradients
uses information from previously tested weights $\mathbf{w}$ to determine
new weights to try, and the process is iterated to find a
vector of weights $\hat{w}$ which minimizes Eq. \eqref{eq:F}.

\subsection[ D-Wave One Results]{\label{sec:structural-dwave}
D-Wave Results on Learning Structured Labels}

D-Wave has used quantum annealing on their D-Wave One processor to find labels for four different
types of data sets. Results on three of the data sets are reported in
D-Wave's preprint \cite{Bian2011}, the other on D-Wave's
website \cite{DWaveReuters}.
D-Wave ran this structured learning algorithm on four different data sets.

The first is a standard benchmark set, Scene, consisting of 1211 training
images and 1196 test images \cite{Boutell04}. The goal is to label each
image with a subset of the six tags:
Beach, Sunset, Fall Foliage, Field, Mountain, and Urban.
In this case, $z$ is a length $6$ binary string, with each bit
indicating whether or not each of the labels should be given
to the data point $x$. The graph of $G$ was chosen to be
the full bipartite graph between the vertices
Beach, Sunset, Fall Foliage on the one hand and Field, Mountain,
and Urban on the other. Each image was represented by $294$ real
values based on color histograms from subblocks of the image.
Thus, there are $294 + 1$ possible values for $k$ and $6 + 9$ pairs
for $(i,j)$, including the six terms with $i=j$ and the $9$ edges
in the bipartite graph. The results on this data set are shown
in Table \ref{DWaveExpResults}, and are compared with
results from an independent Support Vector Machine (SVM) approach,
a standard, state-of-the-art machine learning technique.
The second data set was the RCV1 subset of the Reuters corpus of labeled
news stories \cite{Lewis2004}. This data set has a significantly larger
number of labels, and more complex relationships between the labels, but
was set up in a similar way to Scene. The results,
also shown in Table \ref{DWaveExpResults}, show a more
significant improvement over independent SVM on this more complex data set.

To test this approach on a data set with highly complex, but
known, relationships, they created two synthetic data sets designated
MAX-3-SAT(I) and MAX-3-SAT(II).
Here, MAX-3-SAT is simply used to generate a data set with complex
relationships between the labels, and between the variables and
the features. The approach does not aim to solve MAX-3-SAT.
A 3-SAT problem is specified by a set of clauses all of which contain
three variables connected by {\sc or}s, and all clauses must be
satisfied. The MAX-3-SAT problem is an optimization problem
specified by clauses of the same form as for a 3-SAT problem together
with weights indicating the importance of satisfying the clause.

To generate a point in one of MAX-3-SAT data sets, a random
$20$-dimensional vector $x \in \mathbf{R}^{20}$, with each
entry $x_i$ chosen at random from $[0,10]$. The weights for
the clauses of a MAX-3-SAT problem instance are then determined from $x$,
in a manner we describe below. The label $z$ is then the bit
string giving the optimal solution to this problem. The problem instances
are small enough that a classical SAT solver can solve them to obtain $z$.

The first data set, MAX-3-SAT(I), consists of $34$ bit problems with a
restricted set of clauses, while the second set consists of 8 bit problems
in which all possible clauses could appear. In the first case, all
problems use the same 1500 clauses, all chosen to have variables
with exactly two edges between them, from a $72$ edge graph $G$.
The weights defining each MAX-3-SAT problem instance are linear
combinations of the $x_i$. More precisely, the weights are determined
by a fixed random $1500\times 20$ real valued matrix $V$, with entries chosen
at random from $[0,1]$, that maps $x$ to the $1500$ weights, one for
each clause. (Note that these weights are entirely unrelated to
the weights $w$ above that parameterize the possible labelers $h_w$.)
Thus, the matrix $V$ maps the $20$-dimension input vector to
a $1500$-dimensional vector that defines a MAX-3-SAT problem instance,
which is then solved to find the label vector $z$. This process, with
fixed $V$, was repeated 1600 times to generate 800 training data points
and 800 testing data points. The QUBO for the MAX-3-SAT (I) data set
has $N = 20$ and $34 + 72$ pairs $(i,j)$, from the $34$ pairs with $i=j$
and the $72$ edges in the graph $G$.
\renewcommand{\arraystretch}{0.8}
The second data set, MAX-3-SAT(II), consists of $8$ bit problems generated
in a similar manner as before, except this time all of the
$2^3 \times \left(\begin{array}{c}  8\\3 \end{array}\right) = 448$
possible clauses used, and the weights for the specific problem
instance are determined from $x$ by a fixed, random matrix $V$ of
dimension $448\times 20$. The edge graph is the $K_{4,4}$ graph
with $16$ edges, so the data set has $N=20$ and $8 + 16$ pairs $(i,j)$,
a total of $21\times 24 = 504$ components of the weight vector $w$.
They explored a different quantum/classical hybrid approach to learning
the weights in this problem.
The results for MAX-3-SAT(I) and MAX-3-SAT(II) are shown in
in Table \ref{DWaveExpResults}, illustrating even better performance
over independent SVM for these problems with highly complex
hidden relationships between the labels and the input.
\renewcommand{\arraystretch}{1.2}
It is known that quadratic and cubic kernels often work better for these
sorts of problems \cite{Taskar03}, so one early, attractive experiment
would be to run these problems using those kernels in both the quantum
and classical approaches.

\begin{table}
\begin{tabular}{|c|c|}
\hline
Dataset & Performance Gain in Labeling \\
        & Accuracy over Independent SVM \\\hline\hline
Scene & 14.4\% \\
Reuters & 15.4\% \\
MAX-3-SAT (I) & 28.3\%  \\
MAX-3-SAT (II) & 43.5\%  \\
\hline
\end{tabular}
\caption{Results for Quantum Annealing Experiments for Structured Learning
run on a DWave One machine. The same $\mathbf{x}$-dependent feature functions $\pmb{\psi}(\mathbf{x})$ are used in the SVM and structured classifier so that accuracy improvements are due entirely to exploitation of pairwise correlations in output labels captured by the quadratic label features.}
\label{DWaveExpResults}
\end{table}

\section{Probabilistic computing}

It is worth noting that hardware which relies on quantum mechanical processes at finite temperature is inherently stochastic. In many instances we can harness this inherent randomness to answer probabilistic queries. This probabilistic approach is at the core of Bayesian approaches to learning. Optimization-based approaches to learning offer only a single representative prediction. In contrast, probabilistic approaches provide a probability distribution over all possible predictions, and can thus provide more robust results and confidence estimates on these predictions. Here, we explore a probabilistic approach to structured classification which assigns a conditional probability distribution $P(\mathbf{z}|\mathbf{x})$ over possible labelings $\mathbf{z}$ associated with a given input $\mathbf{x}$.

Construction of a probabilistic model for $P(\mathbf{z}|\mathbf{x})$ proceeds similarly to the previous optimization-based approach. A parametric family of probabilistic models is posited and the best probability model is identified by optimizing a suitable criterion. The family we consider is
\begin{equation*}
P(\mathbf{z}|\mathbf{x},\mathbf{w}) = \frac{1}{Z(\mathbf{x},\mathbf{w})} \exp\bigl(-\mathcal{E}(\mathbf{x},\mathbf{z},\mathbf{w})\bigr)
\end{equation*}
where $Z(\mathbf{x},\mathbf{w})=\sum_{\mathbf{z}} \exp\bigl(-\mathcal{E}(\mathbf{x},\mathbf{z},\mathbf{w})\bigr)$ is the normalization of the distribution. Thus, we assume that the conditional probability is Boltzmann distributed according to the energy function $\mathcal{E}(\mathbf{x},\mathbf{z},\mathbf{w})$ defined as in Eq. \eqref{eq:Exz}. The energy function is parametric in $\mathbf{w}$, and a best $\mathbf{w}^\star$ must be determined.\footnote{Alternatively, we may integrate against the posterior distribution over $\vec{w}$, but we do not pursue that here.}

\paragraph{Identifying the best energy function}

One choice for $\mathbf{w}^\star$ can be determined by maximizing the conditional likelihood of the training data $\mathcal{D} = \{(\mathbf{x}_d,\mathbf{z}_d)\}_{d=1}^{|\mathcal{D}|}$. Rather than optimize the likelihood directly, it is more convenient to minimize the negative logarithm of the conditional likelihood. This defines the optimization problem determining $\mathbf{w}^\star$:
\begin{equation*}
\mathbf{w}^\star =  \text{arg} \, \max_{\vec{w}} \; \ln \prod_{(\mathbf{x}_d,\mathbf{z}_d)\in \mathcal{D}} P(\mathbf{z}_d|\mathbf{x}_d,\vec{w}) = \text{arg} \, \min_{\vec{w}} \sum_{(\mathbf{x}_d,\mathbf{z}_d)\in \mathcal{D}} \biggl\{ \mathcal{E}(\mathbf{x}_d,\mathbf{z}_d,\mathbf{w}) + \ln Z(\mathbf{x}_d,\mathbf{w})
\biggr\}.
\end{equation*}
For better generalization to points outside the training set we add a regularization term proportional to $\|\mathbf{w}\|^2$ to the objective. This minimization objective for $\mathbf{w}^\star$ is differentiable and convex. The gradient of the log likelihood with respect to $w_{k,(i,j)}$ at a given $\mathbf{w}$ is \begin{equation*}
\sum_{(\mathbf{x}_d,\mathbf{z}_d)\in \mathcal{D}} \biggl\{ \psi_k(\mathbf{x}_d) z_d(i) z_d(j) - \psi_k(\mathbf{x}_d) \mathbb{E}_{P(\mathbf{z}|\mathbf{x}_d,\mathbf{w})}\bigl(z_d(i) z_d(j) \bigr)\biggr\}.
\end{equation*}
This gradient can be evaluated by estimating the expectation of product of the $i$ and $j$ components of $\mathbf{z}_d$ at all training points. These expectations are approximated by Monte Carlo sampling from $P(\mathbf{z}|\mathbf{x}_d,\mathbf{w})$.

\paragraph{Boltzmann models}

\begin{figure}
\centering
\includegraphics[width=0.4\textwidth]{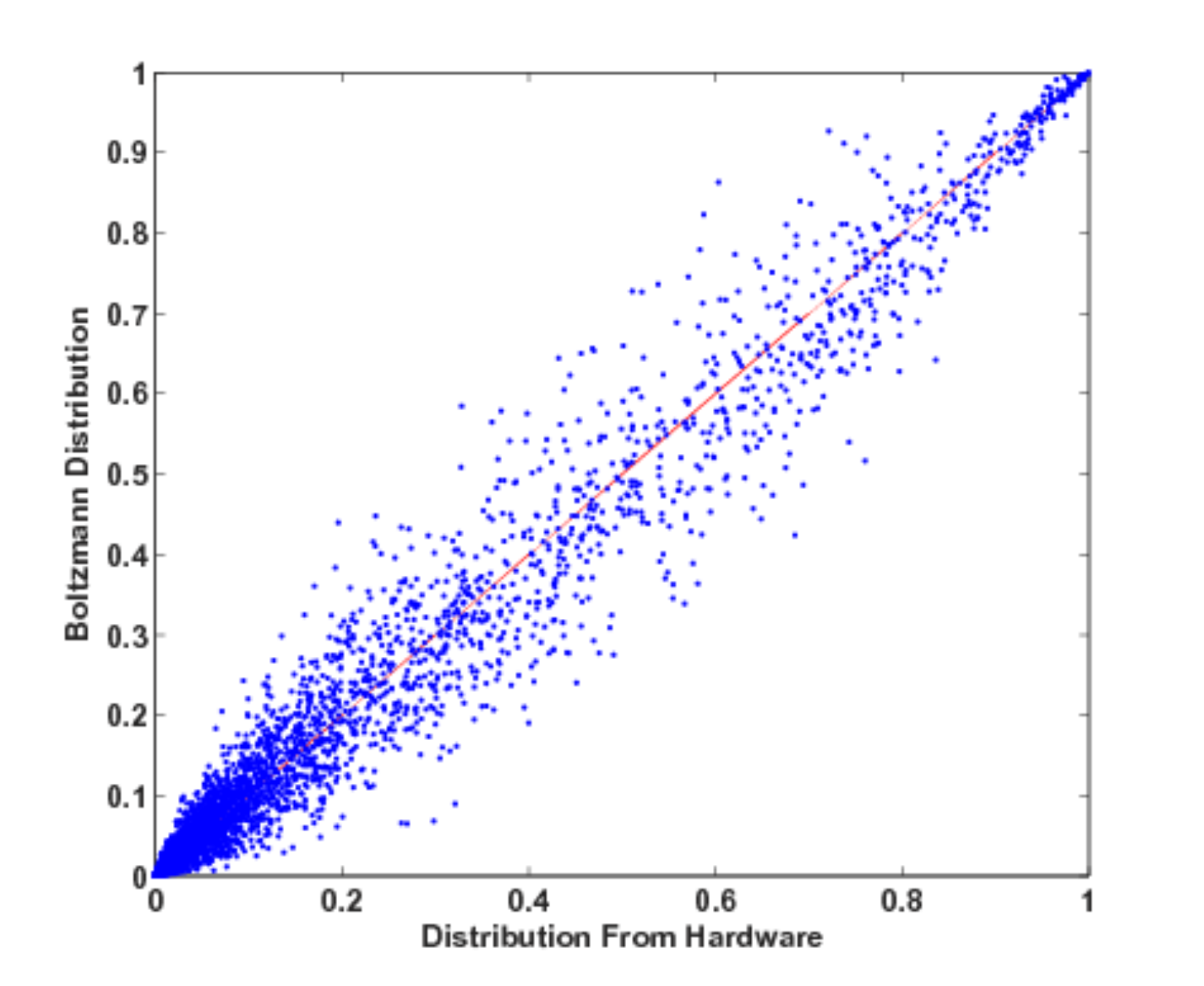}
\includegraphics[width=0.4\textwidth]{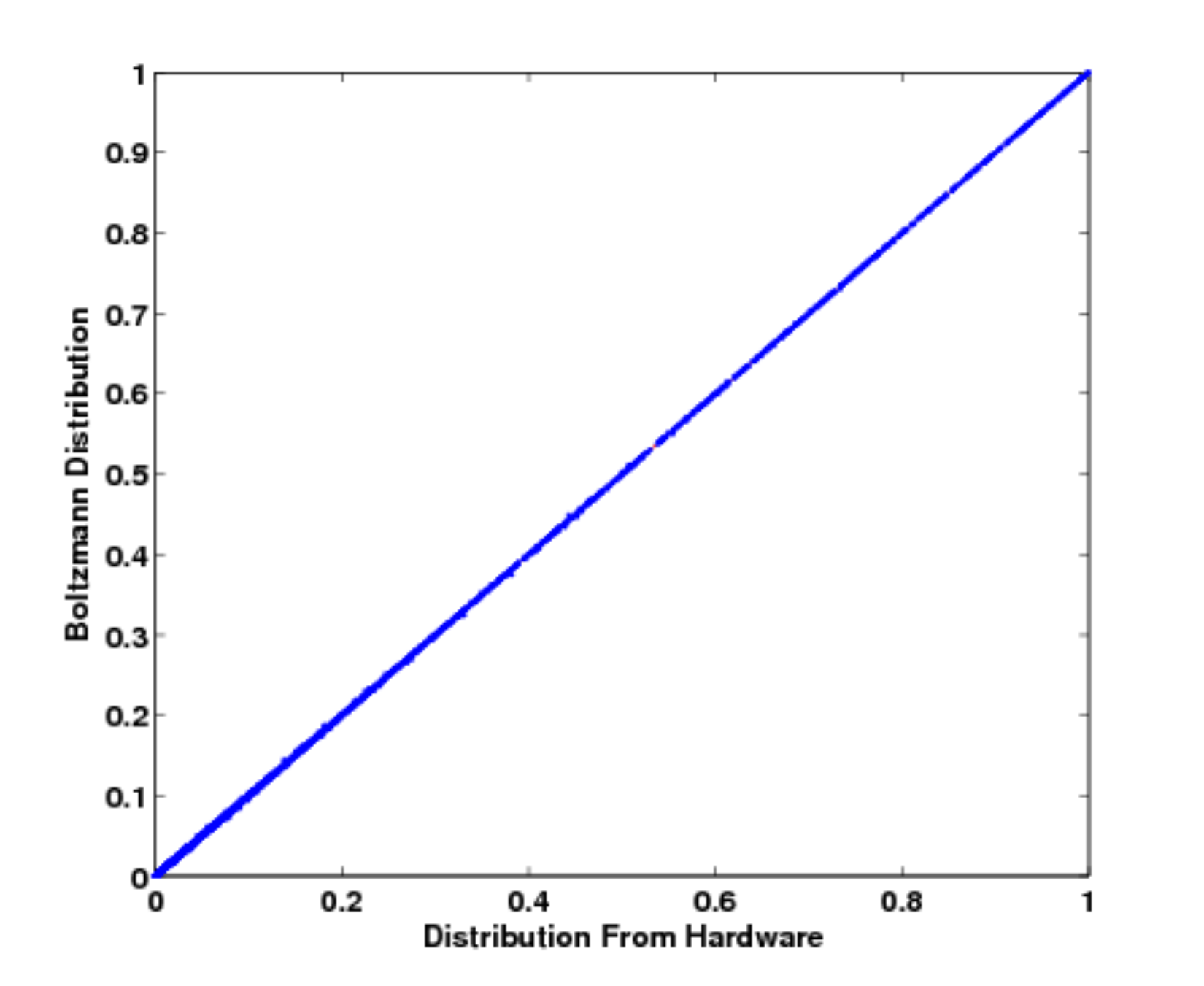}
\caption{A scatter plot comparing the empirical probabilities of observing a set of sampled configurations with a Boltzmann fit. The ideal fit would be a diagonal line.}
\label{fig:CRF}
\end{figure}

We have seen how there is a straightforward convex optimization approach to determine a good $\mathbf{w}^\star$ that relies on a Monte Carlo estimate of the gradient. However, here a complexity arises. A closed form description of the distribution determining the samples returned from the hardware is not possible (because we can't solve for the physics of large quantum systems coupled to an environment). For a given energy function $\mathcal{E}(\mathbf{x},\mathbf{z},\mathbf{w})$ the distribution over samples $\mathbf{z}$ returned from the quantum hardware is not exactly Boltzmann with energy function $\mathcal{E}$. However, a Boltzmann distribution at some perturbed energy function $\tilde{\mathcal{E}}$ can provide a good approximation to the samples returned from hardware. If we identify the perturbation to $\mathcal{E}$ then we can correct the required expectation using importance sampling. Reference \cite{Bian2009} provides further details. Figure \ref{fig:CRF} shows that the perturbed Boltzmann approximation to samples is typically very good. Fig. \ref{fig:CRF}(a) contrasts Boltzmann and empirical probabilities across a range of 8-variable Ising problems with randomly selected $\mathbf{h}$ and $\mathbf{J}$ values. Each point corresponds to one spin configuration sampled from an Ising problem at a given $\{\mathbf{h},\mathbf{J}\}$. The $x$ and $y$ coordinates in the scatter plot record the probability of seeing this configuration under a Boltzmann assumption and the empirical probability. We see in Fig. \ref{fig:CRF}(a) that these points do not lie exactly along the diagonal indicating a deviation from a Boltzmann distribution. However, in Fig. \ref{fig:CRF}(b) we see that if we allow for an effective $\{\tilde{\mathbf{h}},\tilde{\mathbf{J}}\}$ (corresponding to the perturbed energy $\tilde{\mathcal{E}}$) rather than the input $\{\mathbf{h},\mathbf{J}\}$ we obtain a much improved model. The effective parameters $\{\tilde{\mathbf{h}},\tilde{\mathbf{J}}\}$ are a function of $\{\mathbf{h},\mathbf{J}\}$, i.e. there is not a simple offset valid across all $\{\mathbf{h},\mathbf{J}\}$ that captures the effects of the quantum mechanical sampling. It should be noted that the good fits of \ref{fig:CRF}(b), are not simply an artifact of having a too many fitting parameters. If there are $N$ output labels there are roughly $4N$ $\{\tilde{\mathbf{h}},\tilde{\mathbf{J}}\}$ fitting degrees of freedom, but the Boltzmann fit provides good agreement on most of the $O(2^N)$ possible label configurations.
Fitting the effective $\{\tilde{\mathbf{h}},\tilde{\mathbf{J}}\}$ parameters can also be cast as a convex optimization problem, and consequently on any given problem we can efficiently build a quantitative model which provides a good approximation to the sampling distribution. In this way using importance sampling we can correct for the difference between $\{\tilde{\mathbf{h}},\tilde{\mathbf{J}}\}$ and the true $\{\mathbf{h},\mathbf{J}\}$. This allows for estimation of all quantities needed for the gradient and allows determination of the best model parameters $\mathbf{w}^\star$.

This application is particularly interesting as it is the first to show how quantum sampling (not just optimization) can be applied to solve inference problems. Further details and experimental results may be found in \cite{Bian2009}.

\subsection{\label{sec:blackbox}Monte Carlo Sampling for Solving non-Ising problems}

Here we comment on one important application of Monte Carlo sampling
that offers the potential to use quantum annealing to address a
broader class of discrete optimization problems. This use of
Monte Carlo sampling is under ongoing development, but because of
its potential, an implementation of this blackbox algorithm is available
for experimentation on the D-Wave hardware. Early experiments have
shown feasibility when applied to non-convex loss functions in binary
classification and in the optimization of radio-therapy treatment plans
for certain cancers (work in preparation). Here we outline the idea.
Because the algorithm described in this section is at an early stage
of development we maintained focus in the bulk of this report on the
more proven mappings to Ising models.

As witnessed by the many optimization problems that can be mapped to
Ising models, quantum annealing offers promise in diverse areas.
Nevertheless, the mapping into Ising form requires effort, and impedes
rapid prototyping. This section describes a hybrid classical-quantum
approach to minimizing any cost function $G$ even if it is not at
all obvious how to map it onto an Ising model. Furthermore, it can
be used with any hardware connectivity graph. The only requirement
is that the algorithm have access to the cost function value $G(s)$for any
input $s$. The algorithm will use only these values and does not need
access to a closed form expression for $G$ or any knowledge
about how $G$ might be inplemented. We emphasize this property by
saying that the algorithm has only ``blackbox'' or ``oracle'' access to $G$.
While this approach can be applied to any cost function $G$, not all
functions $G$ can be optimized effectively with this approach, but
this approach enables the development of more effective hybrid algorithms
relying on quantum annealing for general functions than has been possible
to date.

The general idea of behind the hybrid classical-quantum approach is
to iteratively improve the Ising parameters $\{h, J\}$ using information
from a population of candidate points $s$ and to find
an improved population by using quantum annealing to find low energy points
for the current values of $\{h, J\}$.
The general outline of the blackbox algorithm used to minimize an
arbitrary function $G(\mathbf{s})$ is described by the  pseudocode
given in Alg.\begin{algorithm}
 \caption{Overview of black-box minimization of a user supplied function $G({\bf s})$.}
\label{bbAlg}
\SetKwInOut{Require}{Require}\SetKwInOut{Ensure}{Ensure}
\Require{an oracle $G(\mathbf{s})$, an initial population of samples $P=\{\mathbf{s}_i\}_{i=1}^{|P|}$}
\Ensure{a candidate minimizer $\mathbf{s}^\star$ of $G$}
\While{not converged}{
  evaluate population: $\mathbf{G} \gets \{G(\mathbf{s}_i) \mid \mathbf{s}_i\in P \}$\;
  filter population: $[P,\mathbf{G}] \gets \text{filter}(P,G)$\;
  fit hardware-compatible model: $[\mathbf{h},\mathbf{J}] \gets \text{fit}(P,\mathbf{G})$\;
  sample new population: $P\gets \text{sample}(\mathbf{h},\mathbf{J})$\;
}
\Return $\text{bestInPopulation}(P)$.
\end{algorithm} \ref{bbAlg}.

The blackbox algorithm operates on a population $P$ of points or
configurations $s_i$.
For each configuration, the oracle $G$ is called to evaluate its objective.
In the filter step, the population is winnowed down by discarding those
configurations having high $G$ value. From the resulting population the
best-fitting hardware-structured model is learned using the population
$\{\mathbf{s}_i,G(\mathbf{s}_i)\}$ as a training set. The fitting
procedure automatically accommodates the restricted qubit connectivity
of Fig.~\ref{fig:architecture} giving a setting of $\mathbf{h}$ and
$\mathbf{J}$ values. The fitted function is then fed to the hardware to
obtain a set of samples which are approximate minimizers for the
learned $\mathbf{h}$ and $\mathbf{J}$ parameters. These basic steps
are iterated until some exit criterion is satisfied. This high-level
approach can be tailored in many ways, but all variants share the basic
structure outlined here. The approach offers no guarantees of solution
quality, but can be applied to any function $G$. The evaluate, filter, and
fit steps can be made very fast so that the bulk of the effort is
accomplished by quantum annealing in the sample step.

\section{\label{sec:conclusion} Concluding remarks}

\begin{figure}[!ht]
\hspace{-0.2in}
  \includegraphics[width=6.5in]{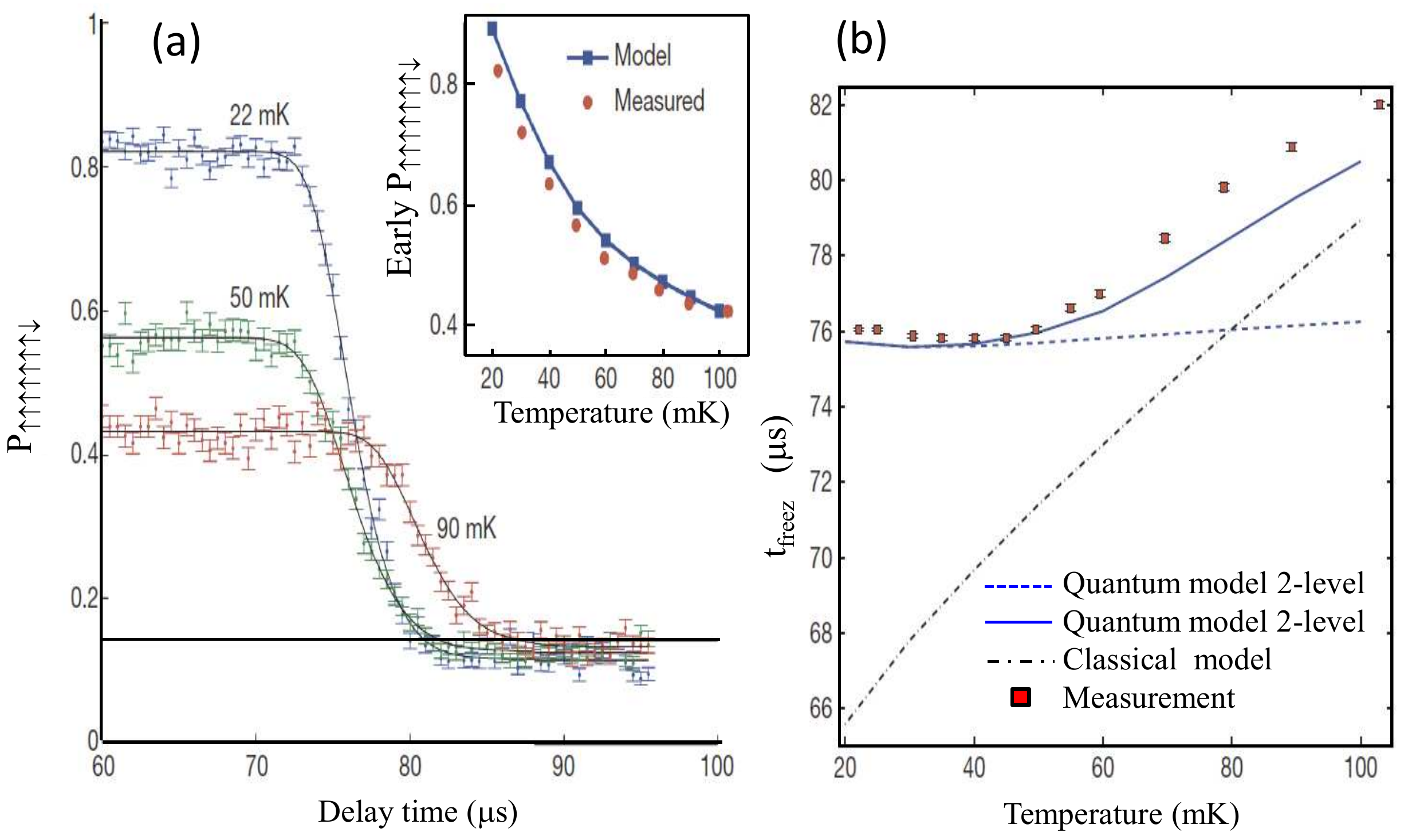}
  \caption{(a) Measured final ground-state probability, $P_{\uparrow\uparrow\uparrow\uparrow\uparrow\uparrow\uparrow\downarrow}$, in the eight-qubit chain versus delay time $t_{\rm d}$ for  $T=$22mK (blue), 50mK (green) and 90mK(red). The solid lines are the result of fits used to extract the freeze-out time, $t_{\rm freeze}$. Inset, measured and simulated (four-level quantum model) $T$ dependence of $P_{\uparrow\uparrow\uparrow\uparrow\uparrow\uparrow\uparrow\downarrow}$ for $t_{\rm d}\simeq 0$. (b) Measured $t_{\rm freeze}$ versus $T$ (red points). Also  simulated plots of $t_{\rm freeze}$ are shown  from two-level (dashed blue) and four-level (solid blue) quantum mechanical models and from a classical model of the qubits (black). \label{fig:Nature2011} }
\end{figure}

\subsection{Gaining insight into quantum annealing process}

In quantum annealing it is useful to think in terms of an `energy landscape' that is defined by the combinatorial optimization problem one wishes to solve. Each position in the landscape represents a spin configuration that is a potential solution to the problem (see Fig.~\ref{fig:1a}). Positions with lower energies represent better quality solutions. During quantum annealing the system Hamiltonian is transformed from the simple fully symmetric form
$H_D$ to a problem Hamiltonian $H_P$. This process could be described as the
``evolution'' of the landscape from a simple bowl shape to a
complex surface representing the problem under consideration.
At the end of the quantum annealing process, the original combinatorial optimization problem is fully
expressed in the hills and valleys of the energy landscape.

Conceptually, there are mechanisms that allow the QA processor
to move about the landscape in a way that favors low-energy positions, a type
of information processing. Classically, one can think of thermal excitations
allowing the processor to hop from site to site across only the surface of this landscape. Quantumly,
one can think of a probability amplitude wave that can flow \emph{through} the
landscape, and tunnel through barriers.
In any real quantum system, both modes of mobility will be present.
The challenge is to distinguish between
these two to show that quantum dynamics (mobility) are present and have
a dominant characteristic in typical operating conditions of the quantum annealing processor.

The experimental methods and results in \cite{Johnson2011} established just
such conditions in an 8-qubit unit cell of D-Wave's quantum processor.
Conceptually, one can imagine that over the course of an annealing period,
the energy landscape becomes more sloped and
there comes a point when the mechanisms of mobility can no longer overcome
the energy barriers between valleys. For some initial segment of time, the
processor can still move about the landscape. If we assumed that the
mechanisms for mobility were purely classical -- which is to say thermal --
then the period of mobility would get progressively shorter as processor
temperature decreases.

However, the experiments show that the mobility time becomes temperature
independent in the vicinity of the processor's operating temperature.
The temperatures in this vicinity are where the thermal mechanisms are
subordinate to quantum mechanical mechanisms for mobility in the energy
landscape. The results show that the processor circuitry does indeed
continue to cool through this temperature regime. The conclusions are
reinforced by a zero-free-parameter fit with a quantum mechanical model
of the processor and its environment using parameters that are
independently derived through primary measurements (Fig.~\ref{fig:Nature2011}).

\subsection{Overhead of the Ising Model Encoding}

The D-Wave processor implements a quantum annealing algorithm to
optimize Ising-spin configuration problems. All NP-complete problems
can be mapped onto an Ising model defined over a non-planar graph.
Therefore, as described above,
an overwhelming  majority of bottleneck computational problems in Space Exploration
can be couched in Ising model form. When one computer problem is mapped onto another
it often carries an overhead in terms of additional bits and algorithmic steps required to implement the mapping.
For NP-complete problems this overhead grows with the problem size no faster than polynomially, and is therefore regarded as cheap and tractable in classical computers. However quantum bits will remain a pricey commodity for the present and next generations of quantum computers such as the D-Wave quantum annealing machine. In trying to solve these  problems on quantum computer one has to carefully optimize the number of qubits involved.  In this context, the above overhead  can become very expensive and tricky to work around as illustrated below.

Consider, for example, mapping the Satisfiability problem, which is at the heart of NP-completeness, onto the Ising model. We will consider a version of Satisfiability called 3-SAT. An instance of 3-SAT  with $N$ bits $z_1,\ldots,z_N$ is defined by a list  of $M$ clauses with 3 bits in a clause and by the   assignment of bits in each clause that violates it. We define each clause {\cal C}  by  the triple of bits $(z_{i},z_{j},z_{k})$ appearing in  that  clause   together with the bit assignment  $(a_i,a_j,a_k)$ that violates the clause.   We define a clause energy as a function of bits  that it equals 1 for the  bit assignment that violates the clause, $z_i=a_i$, $z_j=a_j$, $z_k=a_k$, and  equals 0 for the seven other possible bit assignments
   \begin{equation}
E_{\rm C}= \prod_{r=i,j,k}[a_r z_{r}+(1-a_r)(1-z_{r})].\label{eq:sat}
  \end{equation}
\noindent
The total energy function of a 3-SAT instance is a  sum of the energies of all clauses, $E({\bf z})=\sum_{\cal C} E_{\cal C}$.
  It always equals the number of clauses violated by a bit assignment ${\bf z}$. Because the clause energy contains cubic terms of the form $z_i z_j z_k$  the model (\ref{eq:sat}) is not of QUBO (Ising) type and therefore cannot be solved directly (``natively'') on the D-Wave machine.
  To convert an instance of 3-SAT into QUBO form we will have to introduce a new (auxiliary) binary variable $u$ for each clause. The new energy function for a clause then becomes
\begin{eqnarray}
E_{\rm C}^{\rm QUBO}&=&1+5u-(1+3u)\sum_{r=i,j,k}\left(a_r (1-z_r)+(1-a_r) z_r\right)\label{eq:sat-QUBO}\\
&+&\sum_{\substack{
   r,r^\prime=i,j,k \\
   r\neq r^\prime }} \left(a_r (1-z_r)+(1-a_r) z_r\right)\left(a_{r^\prime} (1-z_{r^\prime})+(1-a_{r^\prime}) z_{r^\prime}\right).\nonumber
\end{eqnarray}
\noindent
The binary variable $u$ must always be set in a way to minimize the  clause energy
at fixed assignments of the bits $z_i,z_j,z_k$.  One can check, for example, that for the assignment of bits  $z_i=a_i$, $z_j=a_j$, $z_k=a_k$ that the minimum clause energy is 1 corresponding to $u=0$. For any other assignment of bits $z_i,z_j,z_k$ (and appropriately adjusted $u$) the minimum clause energy is 0. Therefore the function $E_{\rm C}^{\rm QUBO}$ encodes a clauses in an instance of 3-SAT similarly to
$E_{\rm C}$. The total energy function of 3-SAT instance is then a  sum of $E_{\rm C}^{\rm QUBO}$ of all clauses.
The full set of binary variables needed to encode an instance of 3-SAT this way contains $N+M$ variables $(z_1,\ldots,z_N,u_1,\ldots,u_M)$ where  $M$ is a number of clauses. It is clear from (\ref{eq:sat-QUBO}) that the cost function is now quadratic in the binary variables and therefore can be minimized using QA procedure on D-Wave machine.

We note however that the hardest instances of 3-SAT problems are those where the number of clauses $M$ is several times that of the number of variables $N$ (each represented by one bit). For the so-called uniform 3-SAT ensemble the hardest problems are near the Satisfiability threshold of $M/N\sim 4.26$. This means that  the number of binary variables needed to solve the Ising version of 3-SAT will be 4 to 5 times greater than that needed to solve 3-SAT in its original form. For example, if we are interested in instances of 3-SAT with few hundreds bits (smallest problem sizes where asymptotic complexity can be studied) we would  need to use superconducting  quantum annealing processor  with of the order of 1000 qubits. However building quantum bits is very difficult, at least at the present. Therefore,  if a practical problem can be represented \emph{natively} as an Ising model much larger problem instances can be investigated on the available hardware. Similarly, if the mapping is efficient, in that the computational cost of setting up the problem in Ising form is low, even small speed ups due to quantum annealing could be significant. For these reason, there is strong motivation to seek out efficient mappings of practical computational problems to Ising form.

We note in passing that the hybrid classical-quantum approach outlined in Sec.~\ref{sec:blackbox}
provides a means to apply quantum annealing even when it is unclear
how to map the original cost function onto an Ising model. This
heuristic \lq\lq blackbox'' approach, which iteratively tries to
approximate the original cost function through the iterative use of a
classical learning step and quantum sampling, risks losing vital
structure inherent in the original optimization problem. As a result
of this lost structure, this approach may not be able to find minima
that would be found using quantum annealing directly. Furthermore,
because of limitations of classical learning, even when there is a
good approximation or a direct mapping, this algorithm may not find
it. Therefore a direct mapping onto an Ising model is preferred
whenever it is possible to achieve such a mapping.

\section{Summary}

This paper described several difficult combinatorial optimization problems,
from four major system intelligence domains: data analysis and data fusion;
planning and scheduling; decision making; and distributed coordination.
For each problem, we showed how it can be phrased as a quadratic binary
optimization problem with cost function $E(z_1, z_2, ... z_N)$. Each
quadratic binary optimization problem, in turn, can be mapped to an
Ising energy function $E_{\rm Ising}(s_1,\ldots,s_N)$ via $s_i = 1-2z_i$.
In general, the lowest energy states, which
represent good solutions to the
original combinatorial optimization problem, can be sought via
Quantum Annealing. Many other problems can be similarly mapped.
Small versions of these problems can be run on current hardware, and
larger versions can be evaluated as quantum hardware advances.
We also described a hybrid classical/quantum approach to problems
for which a mapping onto an Ising model is model is more difficult.
This added applicability comes at the cost of repeated cycles of
quantum annealing, so it will be important to validate the efficacy of
this approach.

The strength and feasibility of quantum computation, and quantum annealing in
particular, is for difficult problems with small, compact solutions
that are hard to find. The output of the most famous quantum algorithm,
Shor's factoring algorithm, is just two numbers, but these numbers were
so hard to find that many cryptographic systems are based on
the difficulty of finding them. The second most famous algorithm,
Grover's search algorithm, outputs a single item from a large
collection of items, a needle in a haystack.

Similar to the above criteria, the Ising optimization problems
have three key properties: these optimizations are known to be
difficult, answers can be efficiently verified, and they share
the same form as other important problems. The output  from
a quantum annealing algorithm is a single, lowest energy state
or a handful of the suboptimal states sufficiently close in energy
to a global minimum.
A great many of NASA problems are of this form, requiring as output
only the best decision  (or a handful of close-to-best decisions)
given the data. These problems provide a rich testbed in which to
explore the power of quantum annealing.

Artificial intelligence (AI) problems are, almost by definition, difficult.
Once a problem becomes easy to solve every time by machine, people
no longer view that problem as needing intelligence to solve it.
Most AI problems are NP-complete or harder. They are currently
tackled via heuristic approaches, and the fields of artificial
intelligence and machine learning are constantly trying to improve
on current techniques. Even small improvements have significant
practical impact. While theory plays a role in these areas, it
is difficult to prove that one algorithm performs better than another.
Instead, most algorithms in artificial intelligence and machine
learning are compared by running them on a series of problem instances
to see how they perform. This empirical testing has not been possible
for quantum algorithms until very recently. Advances in quantum
annealing hardware means that we can begin what was
hitherto impossible: test quantum annealing algorithms on a
variety of AI problems.

This paper outlined several practical problems on which quantum annealing
approaches can be evaluated including problems in classification and
feature identification, clustering and pattern recognition,
anomaly detection, data fusion and image matching, planning
and scheduling, diagnostics, and optimal task assignment for
unmanned autonomous vehicles.
Up to now, understanding of the practical impact of quantum
computing has been limited to cases in which we can prove, mathematically,
that quantum algorithms outperform classical approaches. In practice,
however, many classical algorithms are used for which there is
empirical evidence, but no mathematical proof, that they outperform
other classical approaches. Such heuristics play a large role in the
practical application of computers to real world problems.
The near-term advances in quantum annealing hardware enable the
beginning of an entirely new approach to assessing the practical impact of
quantum algorithms by providing the first platforms on which
empirical testing of heuristic quantum algorithms can be performed.

\bibliographystyle{unsrt}
\bibliography{space-refs-v12}

\end{document}